\documentclass[aps,twocolumn,showkeys,showpacs,preprintnumbers,prd,superscriptaddress,nofootinbib]{revtex4-2}
\usepackage{amsmath}
\usepackage{amsfonts}
\usepackage{amssymb}
\usepackage{epstopdf}
\usepackage{natbib}
\usepackage{bm}     
\usepackage{booktabs}
\usepackage{dcolumn}   
\usepackage{mdwmath}   
\usepackage{booktabs}  
\usepackage{lipsum}
\usepackage{float}
\usepackage{multirow}
\usepackage{soul}     
\usepackage{verbatim}  
\usepackage{eqparbox}  
\usepackage{xcolor}
\usepackage{physics}  
\usepackage{url}       
\usepackage{csquotes}  
\usepackage{blindtext} 
\usepackage{rotating}  
\usepackage{adjustbox} 
\usepackage{hyperref}  
\usepackage{tabularx}  
\usepackage{array, booktabs, multirow, graphicx, caption, subcaption} 
\usepackage{caption}
\usepackage{orcidlink}

\usepackage[capitalize]{cleveref}
\usepackage{subcaption}
\hypersetup{
    linktoc=all,
    colorlinks=true,
    linkcolor=blue,
    citecolor=red,
    urlcolor=magenta
}

\definecolor{amethyst}{rgb}{0.6, 0.4, 0.8}
\definecolor{customblue}{rgb}{0.2, 0.2, 0.8}  

\begin{document}




\title{How Holographic is the Dark Energy? A Spline Nodal reconstruction approach}%

\author{Miguel A. Zapata\,\orcidlink{0009-0004-8575-4844}}
\email{miguel\_delacruz@icf.unam.mx}
\affiliation{Instituto de Ciencias F\'isicas, Universidad Nacional Aut\'onoma de M\'exico, 
Cuernavaca, Morelos, 62210, M\'exico}

\author{Gabriela Garcia-Arroyo\,\orcidlink{0000-0002-0599-7036}}
\email{arroyo@icf.unam.mx}
\affiliation{Instituto de Ciencias F\'isicas, Universidad Nacional Aut\'onoma de M\'exico, Cuernavaca, 
Morelos, 62210, M\'exico}

\author{Shahnawaz A. Adil\,\orcidlink{0000-0003-4999-7801}}
\email{shahnawaz@icf.unam.mx}
\affiliation{Instituto de Ciencias F\'isicas, Universidad Nacional Aut\'onoma de M\'exico, Cuernavaca, 
Morelos, 62210, M\'exico}

\author{J. Alberto Vazquez\,\orcidlink{0000-0002-7401-0864}}
\email{javazquez@icf.unam.mx}
\affiliation{Instituto de Ciencias F\'isicas, Universidad Nacional Aut\'onoma de M\'exico, Cuernavaca, 
Morelos, 62210, M\'exico}


\date{\today}


\begin{abstract}

In this work, we explore the generalized holographic dark energy (HDE) scenario. We relate the HDE density to the future-horizon scale via a non-parametric function, which is reconstructed via spline-based nodal interpolation. We perform a Bayesian analysis to assess the model consistency with current observations, including baryon acoustic oscillations (BAO) from the Dark Energy Spectroscopic Instrument (DESI) DR1, Type Ia supernovae (SNe Ia) from the Union3 and Pantheon+ compilations, and local measurements of the Hubble constant, $H_0$, from SH0ES. We show that under specific conditions, the model reduces to $\Lambda$CDM with one node.
We find strong statistical evidence against the standard HDE model, and in contrast, the reconstructed HDE model, with three nodes, provides a better fit to the data than the $\Lambda$CDM model, indicating a strong statistical preference for the reconstructed model. 

\end{abstract}

\maketitle




\section{Introduction}

The standard cosmological model, $\Lambda$CDM, has become a widely accepted framework for describing the evolution of the universe.
Its strength lies in its ability to consistently account for a broad array of cosmological and astrophysical observations. Central to this model is the concept of dark energy (DE)~\cite{Peebles:2002gy,Frieman:2008sn,Nojiri:2017ncd}, a component with negative pressure and positive energy density that drives the late-time accelerated expansion of the Universe. Since its discovery in 1998, DE has become a major paradigm in physics \cite{SupernovaSearchTeam:1998fmf,SupernovaCosmologyProject:1998vns}, and within the $\Lambda$CDM framework, it is interpreted as a cosmological constant,~$\Lambda$~\cite{Carroll:2000fy}.
However, the persistence of tensions within the standard model~\cite{Perivolaropoulos:2021jda,Abdalla:2022yfr, DiValentino:2020vvd,Nunes:2021ipq,DiValentino:2022fjm,DiValentino:2021izs,Vagnozzi:2023nrq,Vagnozzi:2019ezj}, along with the increasing limitations posed by systematic uncertainties in conventional cosmological probes, highlights the growing need to investigate alternative and independent observational approaches. 
In this context, recent results, i.e. from the Dark Energy Spectroscopic Instrument (DESI), offer compelling indications in favor of dynamical dark energy (DDE). Analyzes of the baryon acoustic oscillation (BAO) signal suggest a preference for DDE at a significance level of $3.1\sigma$, particularly when combined with complementary data sets~\cite{DESI:2024mwx,DESI:2024aqx,DESI:2024jis,DESI:2025zgx,DESI:2025fii}. 
Depending on the choice of the supernova data set in the joint analysis, this preference varies between $2.7\sigma$ and $4.2\sigma$. In particular, the DDE signal remains robust even when CMB data are excluded. 
Moreover, this evidence is not limited to DESI-BAO measurements alone, independent studies also report support 
for DDE without relying on DESI data~\cite{DES:2024jxu,DES:2025bxy}.

In particular, the cosmological constant is described by an equation of state (EoS) with value $\omega_{\rm{de}} = -1$, however, in order to explore extensions of the $\Lambda$CDM model, and hence introduce dynamics into the DE sector, a common approach is to assume a redshift-dependent equation of state, $\omega_{\rm{de}}=\omega_{\rm{de}}(z)$, thus allowing deviations from the constant $\Lambda$ scenario.
A widely adopted strategy for modeling this evolution is through EoS parameterizations, where the functional form of $\omega_{\rm{de}}$ is governed by free parameters that can be constrained using observational data. Although these parameterizations often show better statistical fits to the data, compared to $\Lambda$CDM, determining the most appropriate form among the many proposed remains challenging. 
Some of the most commonly used parametrization include the linear form~\cite{Akarsu:2015yea, cooray1999gravitational,ASTIER20018,PhysRevD.65.103512}, the logarithmic parametrization~\cite{Efstathiou:1999tm}, the Chevallier-Polarski-Linder (CPL) parametrization~\cite{chevallier:hal-00142125, CPL}, the Jassal-Bagla-Padmanabhan (JBP) parametrization~\cite{jassal}, and the Barboza-Alcaniz (BA) parametrization~\cite{Barboza_2008}.
On the other hand, beyond the phenomenological extensions of the $\Lambda$CDM model, there are other fundamental theoretical extensions, often motivated by high-energy physics frameworks. These include scalar field models such as Quintessence~\cite{quin1,quin2,quin3}, Phantom~\cite{phant1,phant2, Vazquez:2020ani}, K-essence~\cite{kessense1,kessense2,kessense3}, early DE scenarios~\cite{early1,early2,early3,early4}, hybrid models such as Quintom~\cite{Vazquez:2023kyx, quintom1,quintom2,quintom3} or interactions among scalar fields \cite{Garcia-Arroyo:2024tqq}. These approaches aim to provide a deeper understanding of the nature of DE from a field-theoretic perspective.

Recently, \textit{holographic dark energy} (HDE) has attracted considerable attention within the cosmological community. Its appeal lies in the holographic principle, which suggests a direct connection between DE and the event horizon of the observable universe. Within the HDE framework, numerous extensions of the standard model have been explored, incorporating various cosmological scenarios. These include models with spatial curvature~\cite{hdecurv}, interactions between dark matter (DM) and DE~\cite{zhang}, neutrino physics~\cite{Li_2013}, and perturbation theory~\cite{Li_2008}. Other notable extensions, considering cosmological holography, involve a time-varying gravitational constant~\cite{Jamil:2009sq,Lu_2010}, inflationary dynamics~\cite{chen}, Brans-Dicke theory~\cite{Gong_2004}, and scalar field analogs~\cite{Guberina_2005,PhysRevD.74.103505}.
Despite this broad landscape of modifications, several studies have shown that standard HDE models, in isolation, often do not provide a statistically better fit to observational data compared to $\Lambda$CDM~\cite{li2024comprehensivenumericalstudycategories,li2024revisitingholographicdarkenergy}. 
As a result, recent efforts have turned his focus on internal generalizations of HDE through the introduction of generalized entropy frameworks. These alternative entropies reduce to the Bekenstein-Hawking entropy in certain limits, thereby retaining consistency with the original proposal. Notable examples of holographic dark-energy models include Barrow (BHDE)~\cite{Saridakis:2020zol}, Tsallis (THDE)~\cite{TAVAYEF2018195}, and Kaniadakis (KHDE)~\cite{Drepanou_2022}, as well as recent generalizations of BHDE featuring a dynamically varying entropy exponent~\cite{Basilakos:2023seo}.

However, while parametric and fundamental approaches often better fit the observational data, they usually rely on an assumed functional form, which can introduce biases or lead to model-dependent interpretations that may not accurately capture the true behavior of the DE, for example \cite{escamilla_2025}. 
To address this limitation, non-parametric reconstruction techniques offer a more flexible and robust framework. When combined with current and upcoming observational datasets, these methods provide a powerful alternative to investigate the evolution of key cosmological quantities, such as $\omega_{\rm{de}}$ or the energy density~$\rho_{\rm{de}}$. 
A notable example is the methodology introduced in \cite{AlbertoVazquez:2012ofj, Hee:2016nho}, where the functional form is reconstructed using piecewise linear interpolation, cubic splines, or step functions. Recently, this approach has been successfully applied to constrain the EoS and other cosmological parameters using a variety of datasets~\cite{Escamilla_2023, Escamilla:2023shf}. 
In this work, we adopt a similar non-parametric reconstruction strategy, avoiding the imposition of a fixed parametric form and enabling the exploration of potential dynamics of DE in a data-driven manner.

The paper is structured as follows. In Section~\ref{sec:hde}, we introduce the background cosmology adopted throughout the study and outline the central concepts around our work. 
This is followed by Section~\ref{sec:data}, where we present the main framework for nodal reconstruction, together with the datasets and methodology used. In Section~\ref{sec:results}, we detail our findings and provide a comprehensive discussion of the results. Finally, Section~\ref{sec:conclusion} summarizes our conclusions.

\section{Holographic dark energy}\label{sec:hde}

In a spatially flat Friedmann-Lemaître-Robertson-Walker (FLRW) universe, the expansion dynamics is governed by the Friedmann equation:
\begin{equation}\label{eq:Hubble}
    H^2=\frac{8\pi G}{3}\left(\rho_{\rm{de}}+\rho_{m}\right),
\end{equation}
that relates the Hubble parameter $H$ with the matter content of the universe; $\rho_{\rm{de}}$ and $\rho_{{m}}$  represent the energy densities of DE and pressureless matter (baryons plus dark matter), respectively\footnote{This work is based on late time observations, and hence, at these epochs, radiation can be safely neglected.}.

\noindent 
In standard cosmology, densities evolve according to their conservation equations, and in particular the evolution of the DE is frequently described through the EoS parameter $\omega_{\rm{de}}$:
\begin{eqnarray}
    \dot{\rho}_{\rm{{de} }} +3H(1+\omega_{\rm{de}}) \rho_{\rm{de}}= 0.  \label{eq:conservation_de}
\end{eqnarray}
However, in this work, rather than prescribing a specific EoS form, we realize dynamical dark energy within the HDE framework; therefore, it is useful to rewrite the Friedmann Eq.~\ref{eq:Hubble} in terms of dimensionless parameter densities $(\Omega_i)$, i.e.:
\begin{equation}
    H^2 (z) = H_0^2 \left( \frac{\Omega_{m}(1+z)^{3}}{1 - \Omega_{\rm{de}}(z)}\right),
\end{equation}
%
where $\Omega_{m}$ is the density parameter of matter today and $\Omega_{\rm de}(z)$ is the density parameter of DE.
The HDE model is motivated by the holographic principle, which asserts that all information contained within a volume of space can be described by the information on its boundary. This fundamental idea was proposed by Gerard ’t Hooft in 1993~\cite{hooft} and since then has had profound implications in high-energy physics, particularly through the influential work of Juan Maldacena and Leonard Susskind~\cite{Maldacena_1999,susskind}.
Within the framework of effective field theory (EFT), a cubic region of size $L$ is interpreted in a cosmological context as a characteristic length scale of the universe with a UV cut-off denoted by $W$. 
The entropy within this volume scales as $S \sim L^3 W^3$~\cite{CHANG200614}. Drawing from his studies on black hole thermodynamics, Bekenstein proposed an upper bound for entropy in such systems, stating that $S \leq S_{\rm{BH}}$, where $S_{\rm{BH}}$ represents the Bekenstein-Hawking entropy \cite{Bekenstein1994}. 
This constraint, known as the Bekenstein entropy bound, reveals inconsistencies when applied to quantum field theory over large volumes.
To resolve this issue, Cohen, Kaplan, and Nelson (1999) proposed a stronger bound on the energy density~\cite{articlecohen}, leading to the condition:
\begin{equation}
\label{eq:entropy}
L^4 \rho_W \leq S_{\rm{BH}},
\end{equation}
where $\rho_W$ is the quantum zero point energy density associated with the UV cut-off. This inequality ensures that the total energy in a region of size $L$ does not exceed the mass of a black hole of the same size or, at most, saturates that bound. 

Building on the entropy bound and introducing a specific proposal for the characteristic length scale $L$, Miao Li proposed a model of DE~\cite{miao}. In this framework, if it is assumed that the vacuum energy density of the universe is governed by the holographic principle, then it can be interpreted as the origin of the accelerated expansion of the universe, that is, the dark energy. By saturating the inequality in Eq.~\ref{eq:entropy}, and considering that entropy scales with the area as $S \propto A \propto L^2$, the corresponding energy density of DE takes the form \cite{miao}:
\begin{equation}
\label{eq:density}
\rho_{\rm{de}} = 3 c^2 M_{\rm{p}}^2 L^{-2},
\end{equation}
where $c$ is a dimensionless holographic parameter and $M_{\rm{p}} = (8 \pi G)^{-1/2}$ is the reduced Planck mass.

The HDE model is therefore built on two key assumptions~\cite{Dai_2020}:
\begin{enumerate}
    \item The holographic principle can be applied to the entire universe, implying that the characteristic length $L$ must be connected to a cosmological horizon scale.
    \item DE originates from quantum vacuum energy.
\end{enumerate}

As a result, the formulation and dynamics of HDE models are critically dependent on the choice of this scale length~$L$.

\subsection{Future event horizon as the characteristic length scale}

The choice of the characteristic length scale $L$ is a nontrivial aspect and forms one of the foundational elements of the HDE framework, as it directly influences the model’s connection to the universe's expansion history. One of the first proposals identified $L$ with the Hubble radius, $L = H^{-1}$. 
However, this choice was shown to yield $\omega_{\rm{de}} = 0$, which corresponds to a non-accelerating universe~\cite{HSU200413}. Subsequently, the particle horizon was considered as an alternative~\cite{Fischler:1998st}, but this also resulted in $\omega_{\rm{de}} > -1/3$, again failing to produce an accelerated expansion today.
To address this issue, the future event horizon was proposed as a more suitable choice for $L$~\cite{miao}, defined as:
\begin{equation}
\label{eq:future_hoz}
R_{h} = a \int_{t}^{\infty} \frac{dt'}{a(t')} = a \int_{a}^{\infty} \frac{da'}{H a'^2}\, .
\end{equation}

Although various extensions and generalizations of the HDE model explore alternative horizon scales,  —such as the Granda–Oliveros model~\cite{Oliveros2022}— the present work adopts the future event horizon as the defining characteristic length scale.

Once $L=R_h$ is fixed, further extensions of HDE generalize the inequality Eq.~\ref{eq:entropy}, where different entropy formulations can be considered, recovering $S_{BH}$ in certain limits. This approach has been implemented in the Tsallis and Barrow models, which both HDE and $\Lambda$CDM are recovered at the appropriate limits. In this work, we assume that the exponent in the entropy can be modeled as a general non-parametric function of the cosmological parameters, where now the exponent contains a function of the scale 
factor,~$f(a)$:
\begin{equation}
\label{eq:entropy_general}
    S_{\rm{G}} = \gamma A^{2 + f(a)/2},
\end{equation}
where $\gamma$ is a constant with units, and the corresponding DE density can be obtained through Eq.~\ref{eq:density}, leading to:
\begin{equation}
\label{eq:density_general}
\rho_{\rm{de}} = \mathcal{K}L^{f(a)}\, .
\end{equation}
Note that the units of the constant $\mathcal{K}$ depend on the value of the function $f(a)$, and the units of $L$ are those of $H_0^{-1}$. To ensure that all relevant quantities have consistent dimensions, we define:
\begin{equation}
\mathcal{K} \equiv {3c^2M_{\rm{p}}^2H_{0}^{(f+2)}}\,,
\end{equation}
where $c$ is the new dimensionless holographic parameter. 

Taking into account the future event horizon (Eq.~\ref{eq:future_hoz}) as the characteristic length and the DE density given by Eq.~\ref{eq:density_general}, we obtain the following relation: 
\begin{equation}
    \int_{a}^{\infty} \frac{da'}{Ha'^2} = \frac{1}{a} \left(\frac{\mathcal{K}}{3 M_{\rm{p}}^2 H^2 \Omega_{\rm{de}}} \right)^{-\frac{1}{ f(a)}}\,,
\end{equation}
from which a differential equation for the DE density is given by: 
\begin{eqnarray}\label{eq:hor_general}
&&     \frac{1+z}{\Omega_{\rm{de}}  \left( 1 - \Omega_{\rm{de}}\right)} \frac{d \Omega_{\rm{de}}}{dz} =  3 \omega_{\rm{de}}(z),
\end{eqnarray}
with an associated EoS:
\begin{eqnarray} \label{eq:eos_gen}
  \omega_{\rm{de}}(z) &=&   - \frac{f(z) + 3 }{3} +  \frac{f(z) \sqrt{\Omega_{\rm{de}}}}{3c} \left[ \frac{Q(1-\Omega_{\rm{de}})}{\Omega_{\rm{de}}}\right]^{\frac{f+2}{2f}} \nonumber \\ 
  && -  \frac{(1+z)}{3f(z)}\frac{df}{dz}\ln{\left[ \frac{Q(1-\Omega_{\rm{de}})}{\Omega_{\rm{de}}}\right]} \, ,
\end{eqnarray}
where $Q=\frac{c^2(1+z)^{-3}}{\Omega_{m}}$. 

Note that the function $f(z)$ and its derivative directly influence the Eq.~\ref{eq:eos_gen}, and thereby the evolution of the dark energy density obtained through Eq.~\ref{eq:hor_general}. Looking at Eq.~\ref{eq:eos_gen}, when $f(z)$ is equal to a constant value, the last term vanishes, and the known models are recovered. In addition, in the limit of $f(z)\rightarrow0$, it reduces to an EoS $\omega_{\rm{de}}\sim- 1$, reproducing the scenario $\Lambda$CDM. 
For $f=-2$, the model reduces to standard HDE, where the equation of state becomes:
\begin{equation}
\omega_{\rm{de}} = -\frac{1}{3} - \frac{2}{3c} \sqrt{\Omega_{\rm{de}}}.
\end{equation}

Finally, the Barrow holographic model is obtained when $f = \Delta - 2$, which also corresponds to a constant $f$. However, it should be noted that our definition of the modified term $\mathcal{K}$ differs from that of~\cite{Saridakis:2020zol}, since we reformulate it using standard cosmological parameters and the parameter $c$. Allowing $f(z)$ to vary with redshift enables a broader class of models with richer dynamics, including smooth or abrupt departures from standard HDE. To explore this generality, we model this function using a nodal reconstruction framework.

\section{Methodology} \label{sec:data}
\begin{figure*}[!htbp]
    \centering
    \makebox[12cm][c]{
    \includegraphics[trim = 1mm  1mm 1mm 1mm, clip, width=6.0cm, height=4.cm]{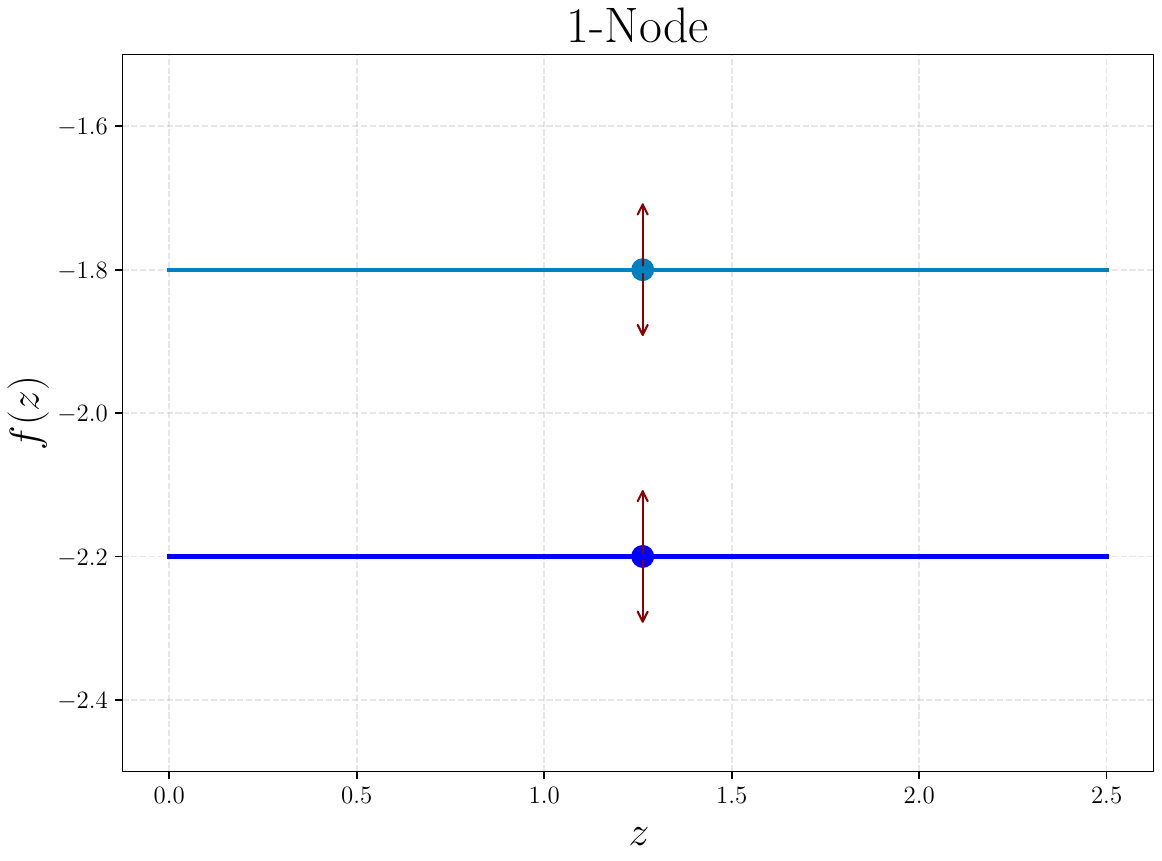}
    \includegraphics[trim = 1mm  1mm 1mm 1mm, clip, width=6.0cm, height=4.cm]{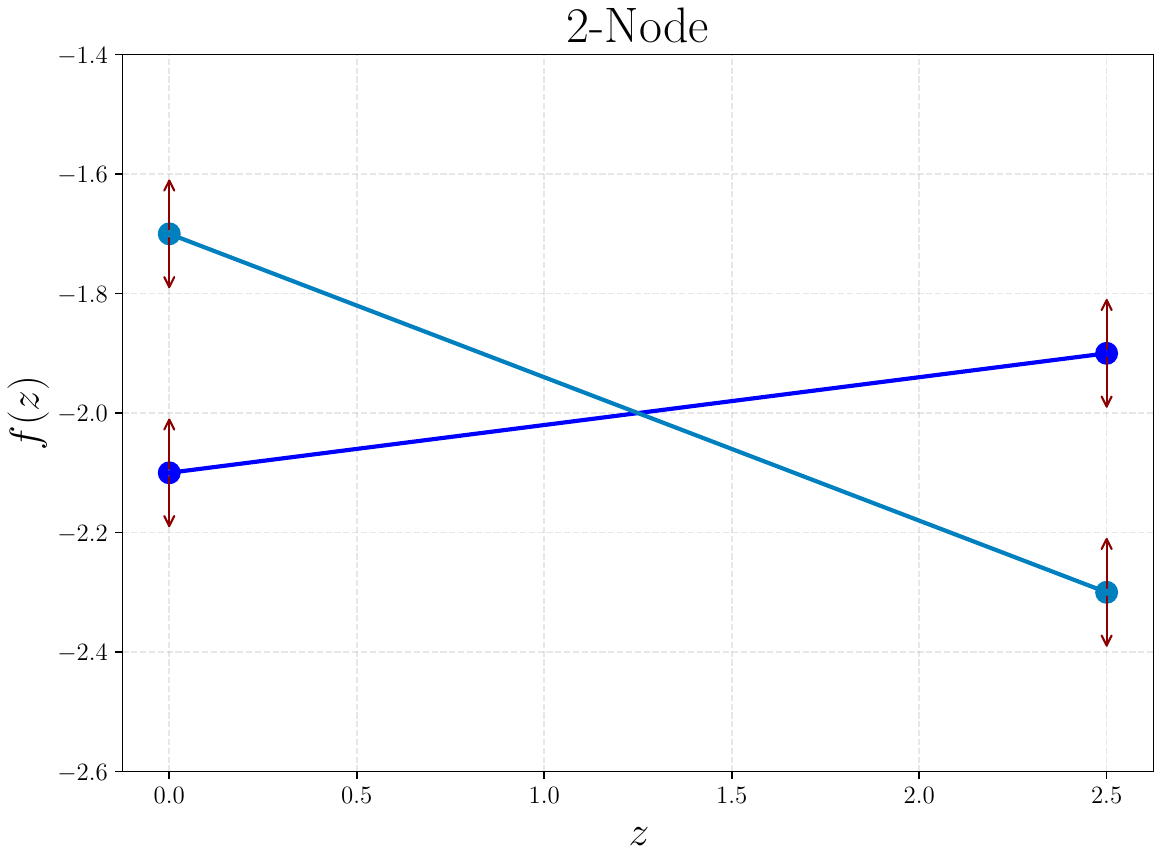}
    \includegraphics[trim = 1mm  1mm 1mm 1mm, clip, width=6.0cm, height=4.cm]{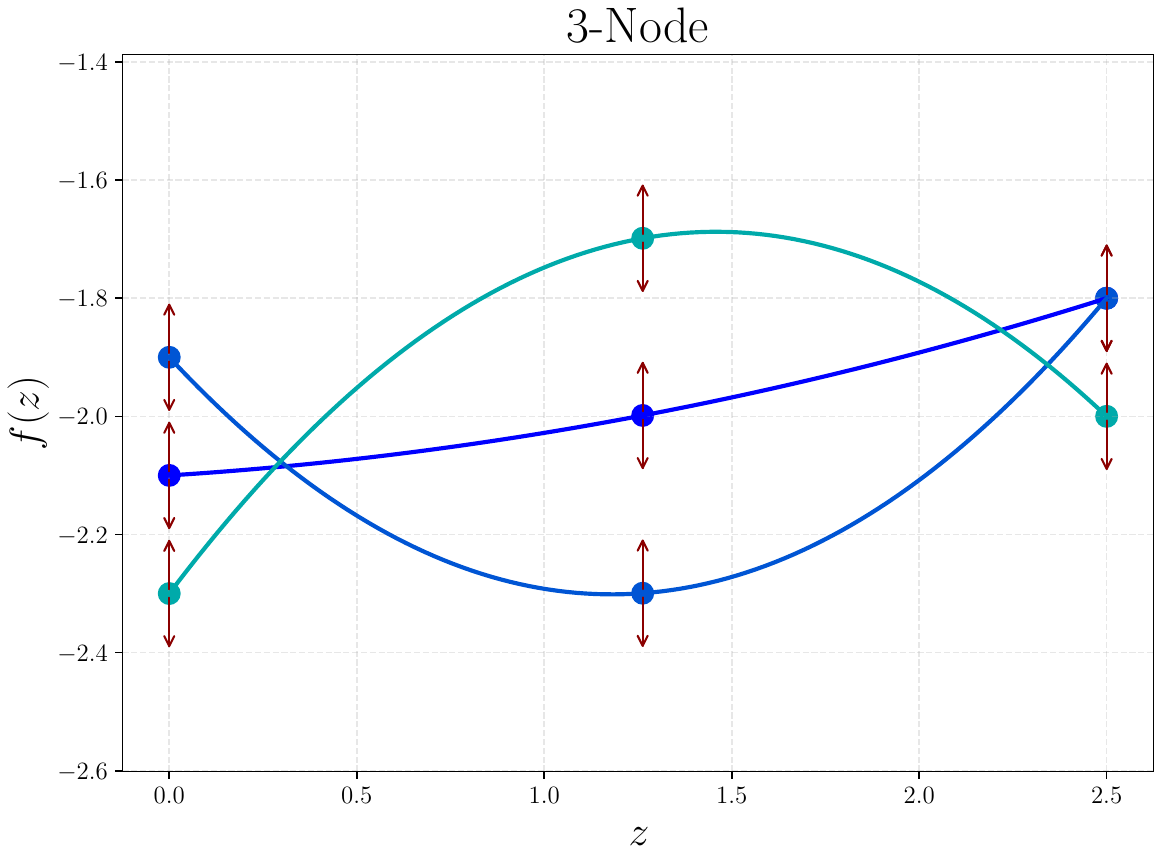}
    }
    \captionsetup{singlelinecheck=off, justification=raggedright} 
   \caption{Left and middle panels: Reconstruction of $f(z)$ using linear interpolation for 1-Node and 2-Node. Right: For the 3-Node case, a quadratic spline is employed. The implications of these reconstructions are further illustrated in Figure~\ref{fig:combined_omega_de}. Arrows indicate the directions in which the nodes of $f_i$ may move freely.}
    \label{fig:nodal}
\end{figure*}

\subsection{Nodal reconstruction framework}

\begin{figure*}[htb!]
    \centering
    \begin{subfigure}[t]{0.32\textwidth}
        \centering
        \includegraphics[trim=3mm 0mm 0mm 0mm, clip, width=\linewidth, height=4cm]{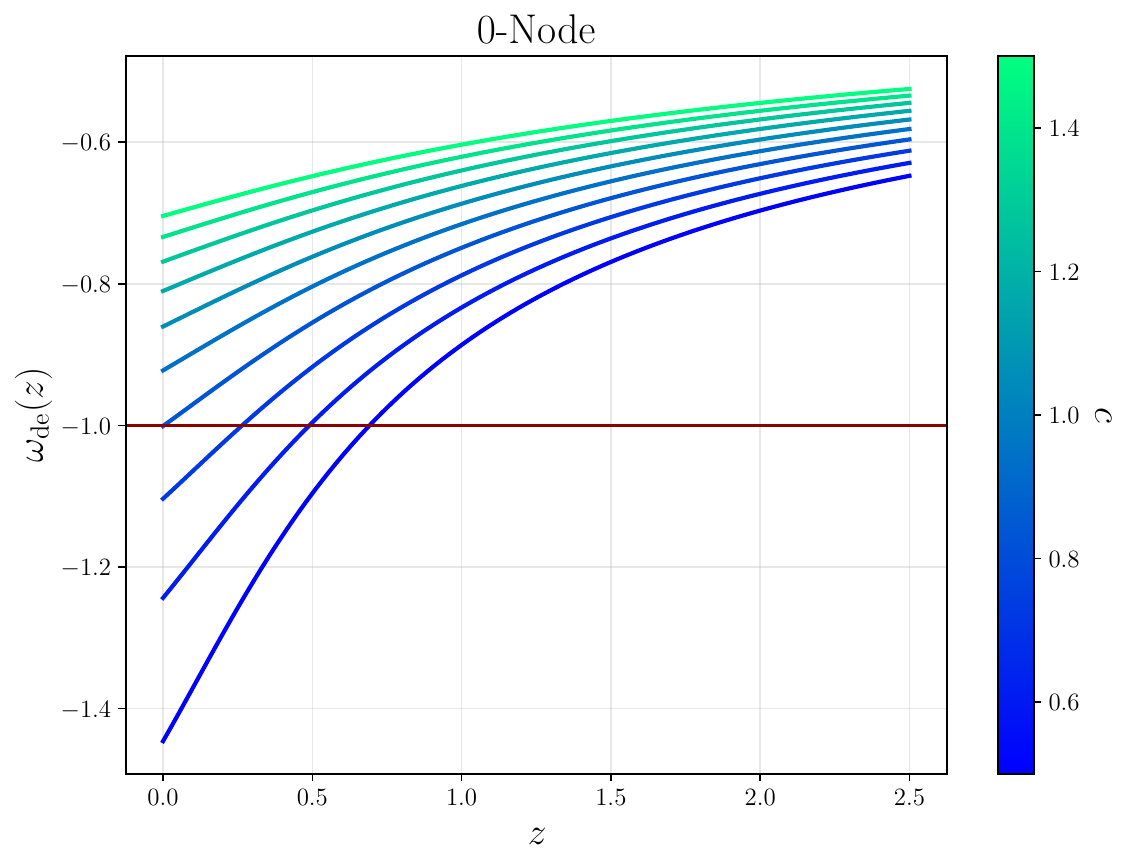}
        
        \caption{0-Node reconstruction.}
        \label{fig:0node}
        
    \end{subfigure}
    \hfill
    \begin{subfigure}[t]{0.32\textwidth}
        \centering
        \includegraphics[trim=3mm 0mm 0mm 0mm, clip, width=\linewidth, height=4cm]{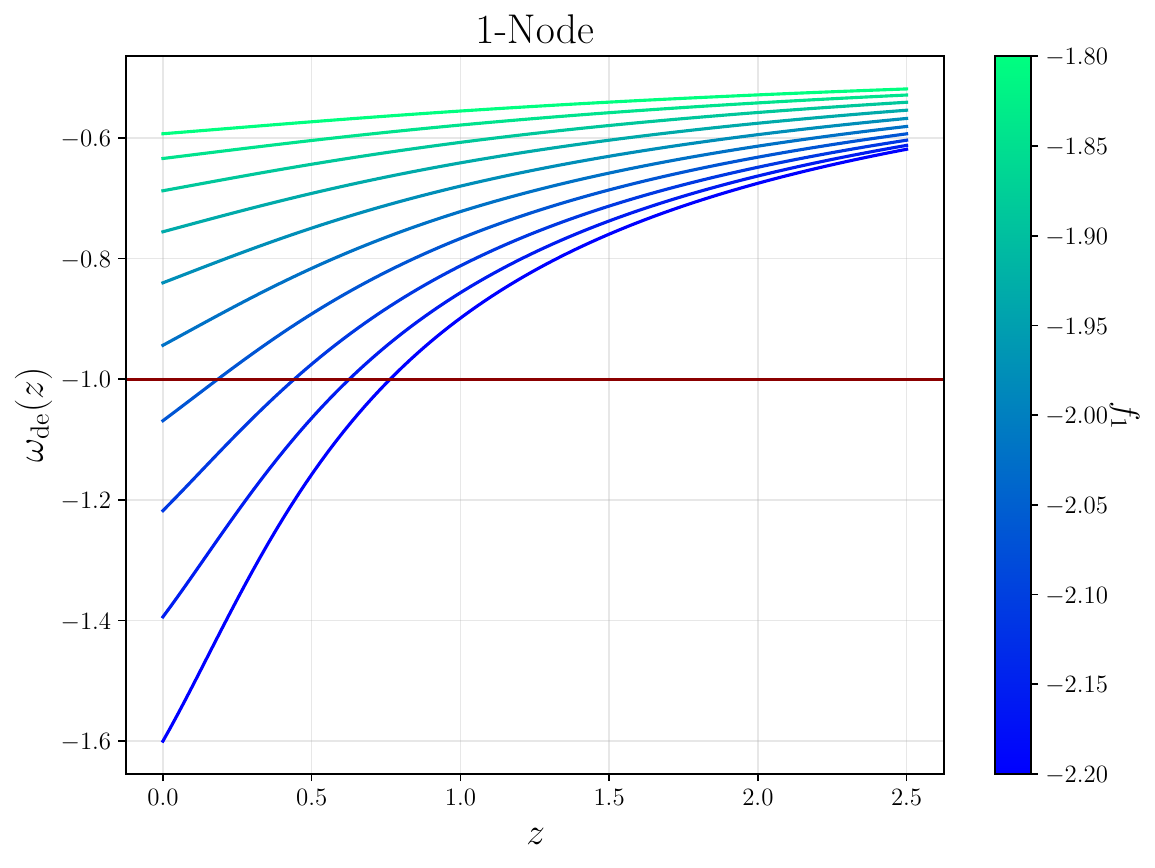}
        
        \caption{1-Node reconstruction.}
        \label{fig:1node}
        
    \end{subfigure}
    \hfill
    \begin{subfigure}[t]{0.32\textwidth}
        \centering
        \includegraphics[trim=3mm 0mm 1mm 1mm, clip, width=\linewidth, height=4cm]{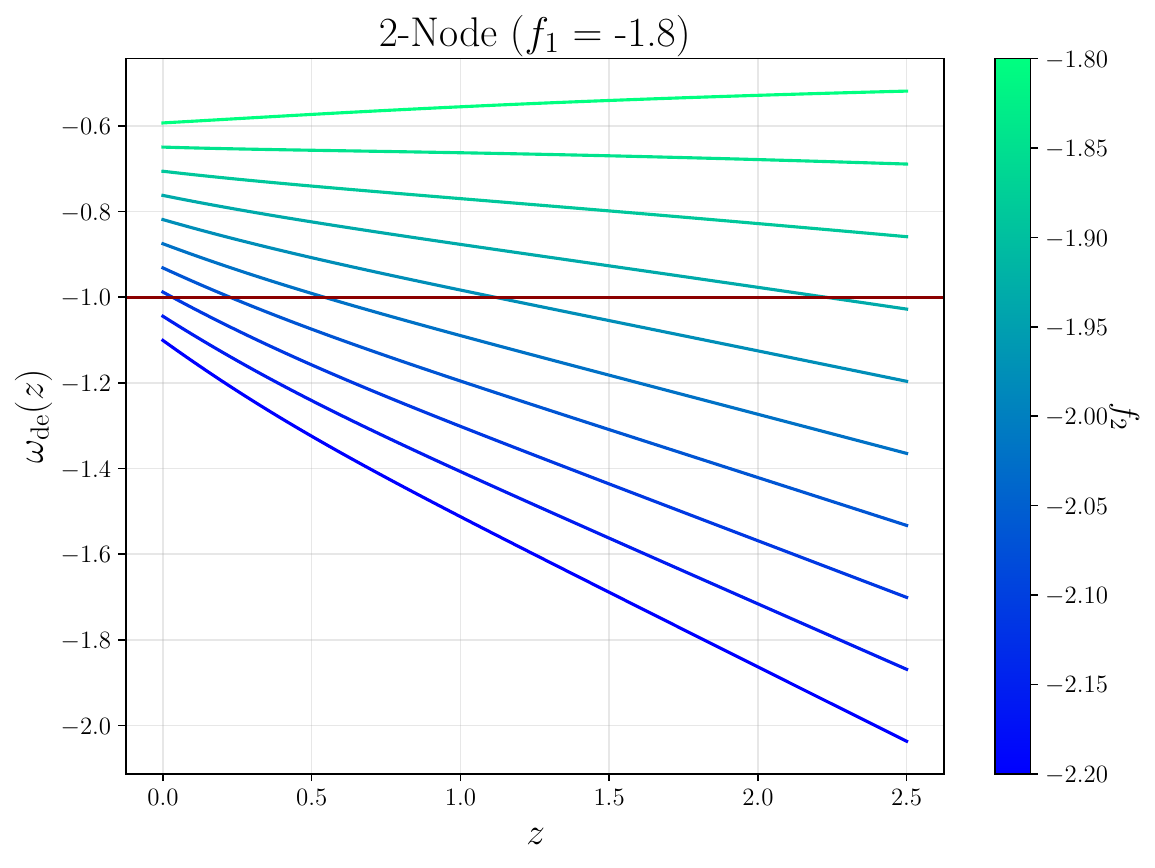}
        
        \caption{2-Node reconstruction: $f_1 \geq f_2$.}
        \label{fig:2node1}
        
    \end{subfigure}

    \vspace{0.4cm}

    \begin{subfigure}[t]{0.32\textwidth}
        \centering
        \includegraphics[trim=3mm 0mm 1mm 0mm, clip, width=\linewidth, height=4cm]{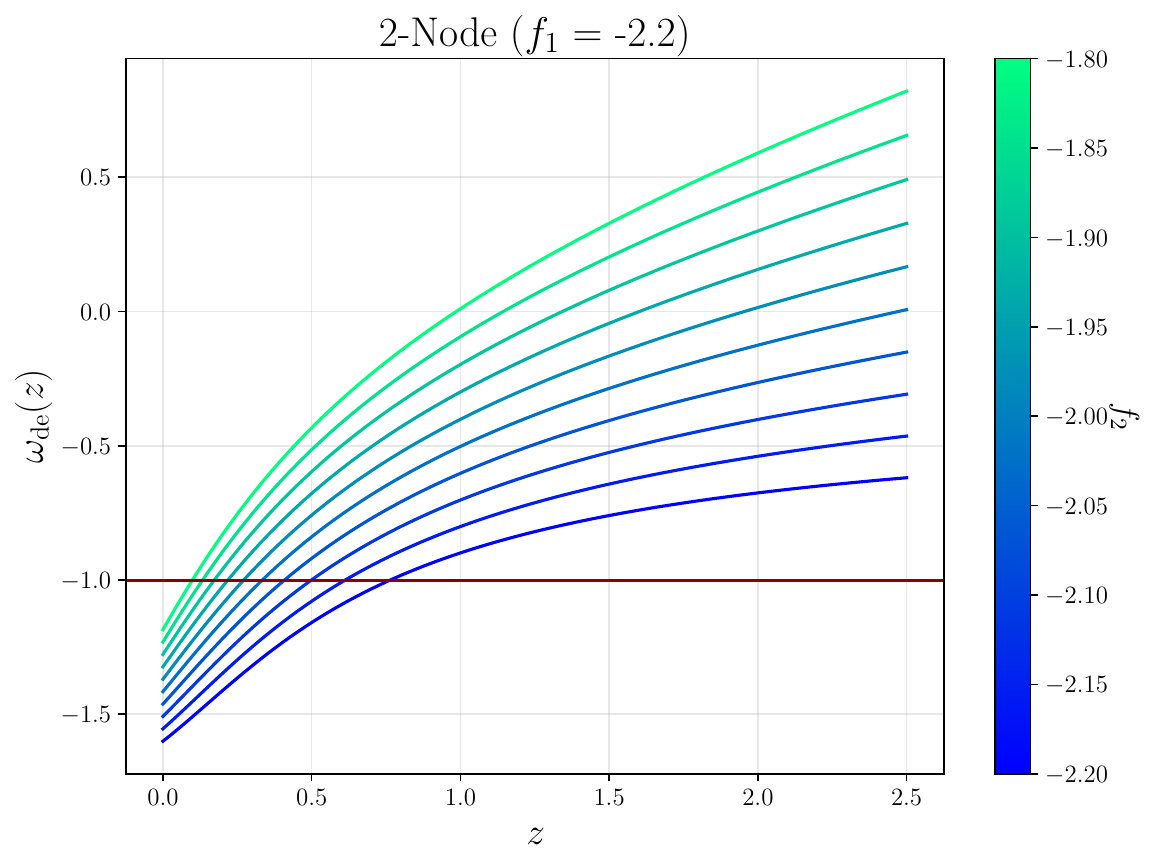}
        
        \caption{2-Node reconstruction: $f_1 \leq f_2$.}
        \label{fig:2node3}
        
    \end{subfigure}
    \hfill
    \begin{subfigure}[t]{0.32\textwidth}
        \centering
        \includegraphics[trim=3mm 0mm 1mm 0mm, clip, width=\linewidth, height=4cm]{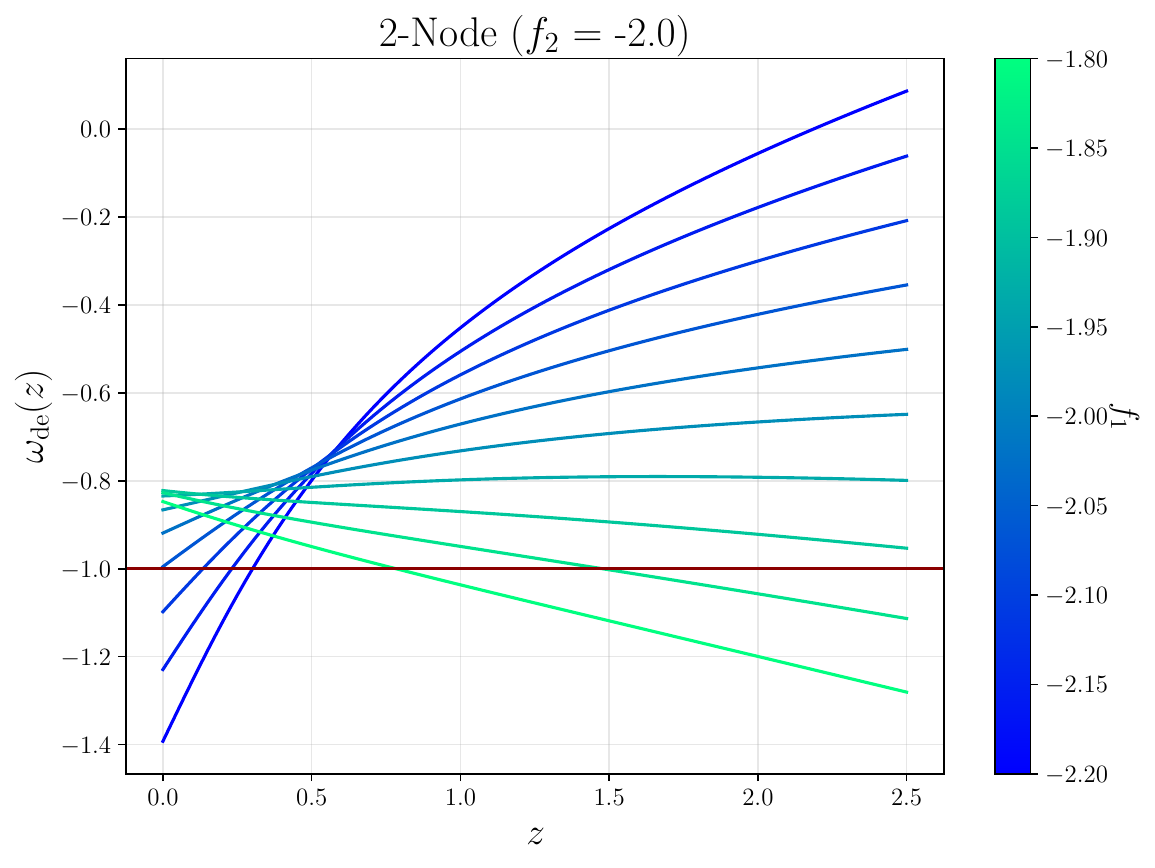}
        
        \caption{2-Node reconstruction: $f_2$ fixed and $f_1$ varying.}
        \label{fig:2node33}
        
    \end{subfigure}
    \hfill
    \begin{subfigure}[t]{0.32\textwidth}
        \centering
        \includegraphics[width=\linewidth, height=4cm]{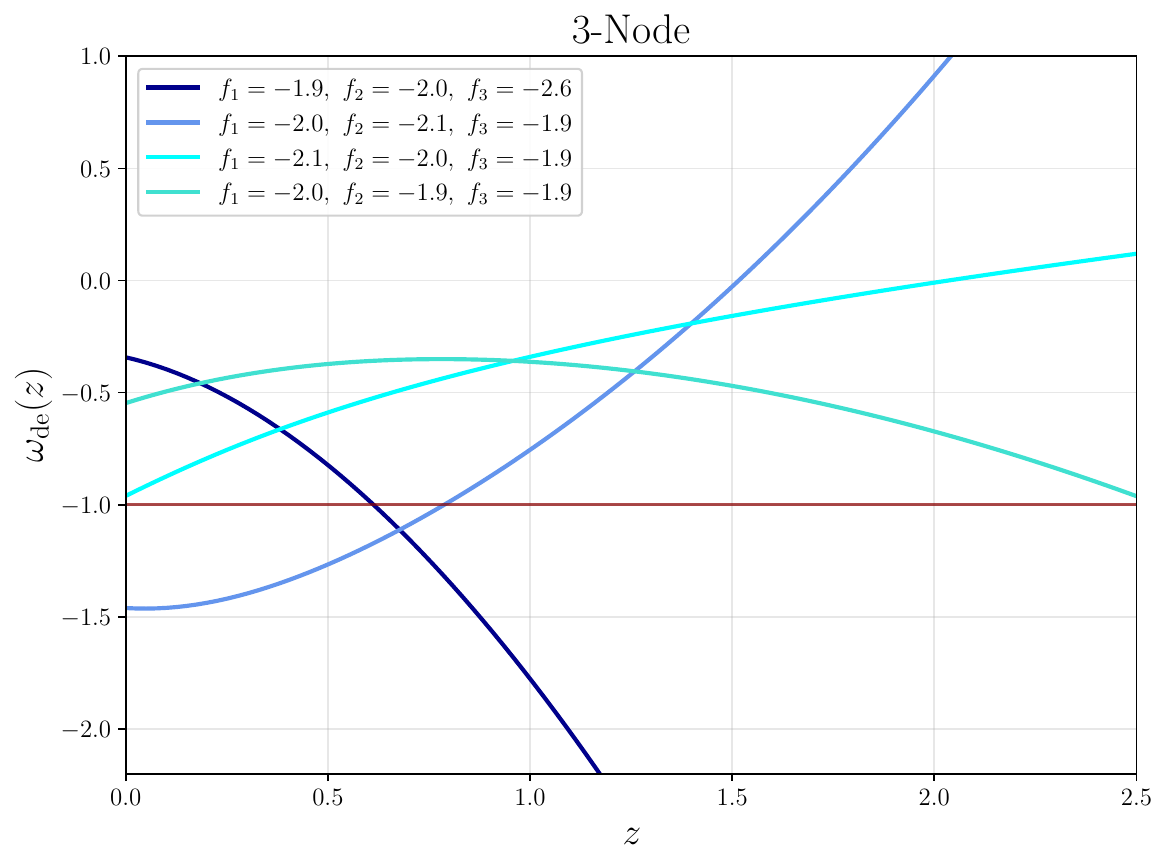}
        
        \caption{3-Node reconstruction.}
        \label{fig:3node}
        
    \end{subfigure}

    \captionsetup{singlelinecheck=off, justification=raggedright} 
    \caption{Examples of reconstructions of the $\omega_{\rm{de}}(z)$ with increasing node complexity, using fixed $c=1$, $\Omega_{m} = 0.30$, and $h = 0.68$. Top row: 0-Node, 1-Node ($c$ fixed), and 2-Node ($f_1$ and $c$ fixed), respectively. Bottom row: remaining 2-Node cases; one of them considers $f_2$ fixed and $f_1$ varying. Additionally, bottom right, 3-node reconstruction ($c$ fixed) using quadratic splines.}
    \label{fig:combined_omega_de}
\end{figure*}

In this work, we adopt a nodal approach to reconstruct the function $f(z)$, which generalizes several holographic DE models. This approach is not limited to any functional mechanism and has a broader perspective on the flexibility of the function form, extracted through a Bayesian mechanism. The reconstruction is based on interpolations through a discrete set of redshift nodes $\{z_i\}$ with the corresponding amplitudes $\{f_i =f(z_i)\}$. Then these nodes are connected using linear, quadratic, or higher-order splines.
In this work, as a proof of the concept, we explore linear and quadratic splines, with node positions uniformly spaced in redshift.
\begin{itemize}
    \item Linear interpolation connects pairs of consecutive nodes with straight segments. It is simple and preserves monotonicity between nodes, but may produce discontinuities in the first derivative.
    \item Quadratic spline interpolation uses piecewise parabolic curves that ensure the continuity of the function and its first derivative, offering a smoother reconstruction.
\end{itemize}
We consider reconstructions with 0, 1, 2, and 3 nodes, corresponding to increasing levels of flexibility in the modeling of the function $f(z)$. 
The redshift positions of the nodes are fixed and uniformly distributed in the range $z \in [0, 2.5]$, as shown in the horizontal axis of Figure~\ref{fig:nodal}.  
At the same time, their amplitudes $f_i$ may vary along the vertical axis as indicated by the vertical arrows.
These variations particularly affect the evolution of the associated $\omega_{\rm{de}}(z)$, as shown in Figure~\ref{fig:combined_omega_de} for the cases of 0 to 3 nodes. A key motivation for extending the reconstruction beyond the 2-node case comes from the features revealed in the DESI BAO and combined DESI+supernova datasets \cite{DESI:2024aqx}. The public DESI analyzes, particularly DESI indicate that the effective dark-energy equation of state (EoS) exhibits statistically significant departures from a simple linear evolution in scale factor \cite{DESI:2025zgx}.

These results display a quadratic-like bending in $\omega_{\rm de}(z)$, with mild but non-negligible curvature centered around intermediate redshifts ($z \sim 0.5$--$1$). Such curvature cannot be captured by linear interpolations or by 0--2 node constructions, which effectively impose a linear or nearly linear evolution on $f(z)$ and, consequently, on $\omega_{\rm de}(z)$. In particular:

\begin{itemize}
    \item The DESI reconstruction of $w(z)$ shows a characteristic turning behavior, first dipping below $-1$ and then gently returning toward $-1$ at higher redshift.
    \item This feature is mathematically equivalent to a second-order (quadratic) deviation from a monotonic trend, which inherently requires a representation with at least three nodes in order to capture the curvature, or ``bending,'' in a non-parametric way.
    \item With only two nodes, the reconstructed $f(z)$ is restricted to a single linear segment (or, for quadratic splines, a segment with insufficient flexibility), and thus lacks the degrees of freedom needed to follow the DESI-implied structure.
\end{itemize}

Thus, the 3-node reconstruction represents the minimal configuration capable of capturing the qualitative behavior favored by DESI. The ability of the 3-node (and higher-node) models to encode this curvature enables a more faithful, data-driven description of dark-energy evolution without imposing restrictive parametric assumptions on different analytic forms.

In light of these considerations, our analysis includes reconstructions up to the 3-node case. Future extensions---allowing adaptive node placement or additional nodes---are planned to more fully explore the rich structure suggested by the latest spectroscopic surveys.

Figure~\ref{fig:combined_omega_de}\subref{fig:0node} illustrates the implications of the 0-node case, which corresponds to the standard HDE scenario with $f = -2$. It is clear that varying the parameter $c$ significantly affects the behavior of $\omega_{\rm de}$. In particular, all values of $c$ lead to deviations from the constant $\omega_{\rm de} = -1$, characteristic of the $\Lambda$CDM model; lower values exhibit a transition from a phantom regime to a quintessence-like regime, while $c$ higher shifts the EoS above the phantom-divide line, preserving quintessence behavior throughout. 
Additionally, at high redshifts, all variations generally converge towards $\omega_{\rm{de}} = -1/3$.

In Figure~\ref{fig:combined_omega_de}\subref{fig:1node}, we examine the Barrow-like model with $f = f_1$ for fixed $c = 1$, and investigate how different values of $f_1$ affect the evolution of the EoS. It is evident that both $c$ and $f_1$ strongly influence the dynamics of $\omega_{\rm de}$. For example, in the case of 0-nodes (Figure~\ref{fig:combined_omega_de}\subref{fig:0node}), $f = -2$ and $c = 1$ produce a predominantly quintessence-like behavior. In contrast, in the one-node scenario, by fixing $c = 1$ and choosing more negative values of $f_1$ the EoS shifts downward to the phantom regime. 
In general, while the behavior resembles the case 0-node, the key difference is a more pronounced transition between the phantom and quintessence regimes, with increasingly phantom-like behavior for larger negative values $f_1$.

For the 2-node case, $f = f(z)$ is modeled as a piecewise linear function, where the slope can be positive or negative depending on the choice of amplitude values. Figures~\ref{fig:combined_omega_de}\subref{fig:2node1}, \subref{fig:2node3} and \subref{fig:2node33} illustrate various 2-node reconstructions. In particular, for \subref{fig:2node1} and \subref{fig:2node3}, $f_1$ is fixed at $-1.8$ and $-2.2$, respectively, while the second node $f_2$ varies. 
In contrast, in Figure~\ref{fig:combined_omega_de}\subref{fig:2node33}, $f_2$ is fixed to $-2$ and the first node $f_1$ is varied. This setup generates different linear profiles for $f(z)$, which in turn determine the evolution of $\omega_{\rm de}(z)$. 
In Figure~\ref{fig:combined_omega_de}\subref{fig:2node1}, where $f_1 \geq f_2$, the slopes of $f(z)$ are predominantly negative and $\omega_{\rm{de}}(z)$ tends to behave as a decreasing function, and as the slope becomes more pronounced, $\omega_{\rm{de}}(z)$ starts at smaller values today and decreases faster with redshift, often crossing into the phantom regime. 
In contrast, Figure~\ref{fig:combined_omega_de}\subref{fig:2node3},  corresponds to cases where $f_2 \geq f_1$, leading predominantly to positive slopes. In these cases, the EoS tends to increase with redshift, starting in the phantom region and evolving towards quintessence. 
As the slope steepens, the growth of $\omega_{\rm de}$ accelerates and the EoS enters the region $\omega_{\rm{de}} \ge 0$, significantly affecting the matter and radiation components of the total energy budget and making these configurations observationally disfavored.
In a complementary way, Figure~\ref{fig:combined_omega_de}\subref{fig:2node33} shows EoS trajectories that evolve monotonically from the phantom to the quintessence regime, and vice versa, similar to the behaviors observed in Figures~\ref{fig:combined_omega_de}\subref{fig:2node1} and \subref{fig:2node3}. 
This is expected, as varying $f_1$ with fixed $f_2$ allows for both negative and positive slopes.

A richer extension of the previous cases is the three-nodes scenario, where we employ a quadratic spline reconstruction to model $f(z)$. Figure~\ref{fig:combined_omega_de}\subref{fig:3node} illustrates the resulting behavior of the EoS for various combinations of node parameters $f_i$. For example, the configuration $f_1 = -2$, $f_2 = -2.1$, and $f_3 = -1.9$  is physically implausible, since EoS begins in the phantom regime, increases with redshift, and eventually exceeds $\omega_{\rm{de}} > 1$, which implies domination over matter and radiation at early times. In contrast, the combination $f_1 = -1.9$, $f_2 = -2$, and $f_3 = -2.6$ produces a rapidly decreasing EoS that quickly enters the phantom regime, primarily due to the low value of $f_3$, and may still be consistent with observational constraints.  
Moreover, we observe that other choices of $f_i$ lead to transitions in the EoS, which can evolve either from quintessence to phantom or vice versa. Interestingly, the evolution is no longer necessarily monotonic, highlighting the dynamic nature and flexibility of the 3-node reconstruction, allowing it to capture a wide range of dark energy scenarios. In this framework, the amplitudes $f_i$ are treated as free parameters that are constrained by cosmological observations through Bayesian inference, as detailed in the following sections.

\subsection{Datasets}

Following the nodal reconstruction framework outlined above, we employ late-time data to constrain the holographic and cosmological parameters.

\subsubsection{Type Ia Supernovae}

In our analysis, we use two SNe Ia datasets:
 \begin{enumerate}
        \item \textbf{Union3:} The updated `Union' compilation of 2,087 cosmologically useful SNe Ia from 24 datasets ('Union3'). These 2,087 SNe Ia are then compressed to 22 redshift bins. 
        In our analysis, we use these binned modulus distance observations of Union3 compilations \cite{UNION3}. 
        \item  \textbf{PantheonPlus\&SH0ES (PPS):} Supernova data from the Pantheon+ dataset, which provides the most recent measurements of the distance modulus \cite{Brout:2022vxf}. This data set comprises 1,701 light curves from 1,550 distinct SNe Ia, covering a redshift range of $0.0012 < z < 2.2614$ and with $\rm Cov_{SN}$, which includes statistical and systematic errors. The PPS sample incorporates the SH0ES Cepheid distances to calibrate the SN Ia absolute magnitude as discussed in equation (14) and (15) of section 2.3 in \cite{Brout:2022vxf}. 
    \end{enumerate}

\subsubsection{Baryon Acoustic Oscillations}

\begin{table}[ht!]
\begin{center}
\begin{tabular}{ccccc}
\toprule 
\midrule
\textbf{Tracer} & \textbf{$D_M / r_d$} & \textbf{$D_H / r_d$} & \textbf{$r$ or $D_V / r_d$} & \textbf{$z_{\text{eff}}$} \\
\midrule
BGS &  --- & --- & 7.93 $\pm$ 0.15  & 0.295  \\
\multicolumn{5}{c}{\dotfill}\\
LRG1 & 13.62 $\pm$ 0.25 & 20.98 $\pm$ 0.61 & $-0.445$ &  0.510 \\
LRG2 & 16.85 $\pm$ 0.32 & 20.08 $\pm$ 0.60 & $-0.420$ &  0.706 \\
\multicolumn{5}{c}{\dotfill}\\
LRG3+ELG1 & 21.71 $\pm$ 0.28 & 17.88 $\pm$ 0.35 & $-0.389$ &  0.930 \\
\multicolumn{5}{c}{\dotfill}\\
ELG2 & 27.79 $\pm$ 0.69 & 13.82 $\pm$ 0.42 & $-0.444$ &  1.317 \\
\multicolumn{5}{c}{\dotfill}\\
QSO & --- & --- & 26.07 $\pm$ 0.67 &  1.491  \\
Lya QSO & 39.71 $\pm$ 0.94 & 8.52 $\pm$ 0.17 & $-0.477$ & 2.330  \\
\hline 
\end{tabular}
\end{center}
\caption{DESI DR1 BAO measurements \cite{DESI:2024mwx}.}
\label{tab:DESI}
\end{table}

Throughout the analysis, we incorporate the DESI BAO measurements \cite{DESI:2024mwx} listed in \cref{tab:DESI}. These are based on four different classes of extragalactic targets which are: the Bright Galaxy Sample (\textbf{BGS}) over 5.5 million reliable redshifts were measured for BGS targets, covering a redshift range of $0.1 < z < 0 .4$, the Luminous Red Galaxy sample (\textbf{LRG}) in the ranges $0.4 < z < 0.6 $ and $0.6 < z < 0.8 $, the Emission Line Galaxy sample (\textbf{ELG}) in the range $1.1 < z < 1.6$,  the combined LRG and ELG sample (\textbf{LRG+ELG}) in  $0.8 < z < 1.1$ range, the Quasar sample (\textbf{QSO}) in $  0.8 < z < 2.14$ range and finally the Lyman-$\alpha$ forest sample (\textbf{Ly}-$\alpha$), in the range $1.77 < z < 4.16$.

In addition to the BAO and SNe Ia data, we incorporate the SH0ES measurement~\cite{Riess:2021jrx} of the local Hubble constant as a Gaussian likelihood:
$$
H_0 = 73.04 \pm 1.04 \, \text{km}\,\text{s}^{-1}\,\text{Mpc}^{-1}.
$$

Therefore, two combinations of data sets are considered: DESI+SH0ES+Union3 and DESI+PPS.
In addition to these combinations, we impose a Gaussian likelihood on $ \Omega_b h^2 $, as it is tightly constrained by the combined Planck and BBN measurements~\cite{Cooke:2017cwo}.
For both, we perform a Bayesian inference of the model parameters where we include standard cosmological parameters —such as the matter density $ \Omega_m $, the physical baryon density $ \Omega_b h^2 $, and the Hubble parameter $ h $—as well as model-specific holographic parameters: the dimensionless parameter $ c $, and the nodal amplitudes $ f_i $ used in the reconstruction of the function $ f(z) $.

\subsection{Bayesian inference}

We perform our statistical analysis within the framework of Bayesian inference~\cite{Padilla:2019mgi}, a widely adopted approach for parameter estimation in cosmology. Bayesian inference offers a systematic approach to updating the probability of a model as new observational data become available. In this framework, the posterior probability distribution of the model parameters $\theta$ is obtained by combining the likelihood function with the prior probability distribution. It is given by:
\begin{equation}
    \mathcal{P}(\theta|D,M) = \frac{\mathcal{L}(D|\theta,M)\mathcal{\pi}(\theta)}{\mathcal{Z}(D|M)},
\end{equation}
where $D$ are the observational data, $\theta$ is the vector parameter of our model $M$. From this expression, we can obtain the posterior probability distribution $\mathcal{P} (\theta | D,M)$, (which represents the updated belief about the parameter $\theta$ after observing the data $D$) if the likelihood $\mathcal{L}(D|\theta, M) $ and the prior $\mathcal{\pi} (\theta|M )$ are known. The $\mathcal{Z} (D|M)$ is known as the \textit{Bayesian evidence}. In addition to normalizing the posterior distribution, it is key to compare models, since it is the total probability of the observed data under a given model.

\begin{table}[!htbp]
    \centering
    \begin{tabular}{c c}
    \hline 
    \hline
    Parameter & Prior \\
    \hline
    \textbf{Cosmological} & \\
     $\Omega_{m}$    & $\mathcal{U}\left(0.1,\,0.5 \right)  $ \\
      $\Omega_{b}h^2 $    & $\mathcal{U}\left(0.02,\, 0.025\right) $ \\
      $h$ & $\mathcal{U}\left(0.4,\,0.9 \right) $\\
      \hline 
      \textbf{Holographic} & \\
      $c$ & $\mathcal{U}\left( 0,\,3 \right)$\\
      $f_i$ & $\mathcal{U}\left( -3.5,\,0\right)$ \\
      \hline
      \hline
    \end{tabular}
    \caption{Uniform prior ranges for cosmological, holographic, and CPL parameters.} 
    \label{tab:prior}
\end{table}

Here, we assume uniform (flat) priors for all parameters, as listed in \cref{tab:prior}, which means that all values are initially considered equally probable\footnote{For comparison with our reconstruction approach, we used the CPL parameterization, with prior ranges $\mathcal{U}\left( -2.0,\,0.0 \right)$ for the $w_0$ parameter and $\mathcal{U}\left( -2.0,\,2.0\right)$ for $w_a$.}.
Under this assumption, the posterior becomes directly proportional to the likelihood, whereas for a Gaussian distribution, the likelihood is approximated to
\begin{equation}
-2\log{\cal L} (D|\theta,M) \propto \chi^2(D|\theta,M),
\end{equation}
where $\chi^2(D|\theta,M)$ denotes the chi-squared statistic, quantifying the discrepancy between observed data and model predictions for a given parameter set $\theta$ and model $M$. 
The definition of the statistic $\chi^2$ depends on the nature of the data under consideration. For correlated data points, the $\chi^2$ function incorporates the full covariance matrix and is expressed as:
\begin{equation}
    \chi^2 = \Delta \alpha^T \, \mathrm{Cov}^{-1} \, \Delta \alpha,
\end{equation}
where $\Delta \alpha \equiv \alpha_{\rm th} - \alpha_{\rm obs}$, and $\mathrm{Cov}^{-1}$ is the inverse of the covariance matrix. The total $\chi^2$ for independent datasets is given by the sum of individual contributions:
\begin{equation}
    \chi^2 =  \chi^2_{\rm BAO} + \chi^2_{\rm SN} + \chi^2_{\rm SH0ES}.
\end{equation}

This statistical estimator provides a robust tool for evaluating the consistency between datasets and their compatibility within the parameter space of a given theoretical model. By examining the best-fit values and associated uncertainties, one can reveal potential discrepancies that may point to systematic errors, dataset tensions, or the need to review the underlying cosmological framework.
To assess which theoretical scenario is the most favored by the data, it is necessary to introduce a model comparison criterion~\cite{Liddle:2007fy}.
Here, we employ two complementary criteria to evaluate the model performance: the p-value criteria based on the simple difference $\Delta \chi^2=\chi_{\Lambda \rm{CDM}}^2 - \chi^2_{i}$ and the corresponding extra degrees of freedom (dof) for the respective models; and the Bayes factor $\mathcal{B}_{\Lambda {\rm{CDM}},\; i} = \ln{({\mathcal{Z}_{\Lambda \rm{CDM}}} /\mathcal{Z}_{i })}$, which is computed using \texttt{dynesty} \cite{Speagle_2020}, a dynamic nested sampling package to estimate Bayesian posteriors and evidence incorporated into \texttt{SimpleMC}\footnote{\href{https://github.com/ja-vazquez/SimpleMC?tab=readme-ov-file}{SimpleMC GitHub}}~\cite{BOSS:2014hhw}, a package for the estimation of cosmological parameters and the comparison of the model using Bayesian inference, optimization, and machine learning algorithms. The p-value quantifies the relative goodness-of-fit, while $\Delta\ln{\mathcal{Z}}$ naturally penalizes the added complexity and uncertainty of the parameters.  Both metrics are reported in Table \ref{tab:aic}. The Bayes factor is interpreted using the Jeffreys’ scale, while the p-values are calculated using \texttt{Scipy.stats} for different scenarios.

\begin{table}[!htbp]
\centering
\begin{tabular*}{0.47\textwidth}{@{\extracolsep{\fill}}cccc}
\toprule 
\midrule
\multicolumn{1}{c}{\textbf{p-value}} & 
\multicolumn{1}{c}{Interpretation} & 
\multicolumn{1}{c}{\textbf{$\Delta \ln {\mathcal Z}$}} & 
\multicolumn{1}{c}{\textbf{Evidence Result}} \\ 
\midrule
$> 0.32$              & Insignificant          & $< 1.0$        & Insignificant \\
$0.10 - 0.32$         & Weak                   & $1.0 - 2.5$    & Significant  \\
$0.01 - 0.10$       & Moderate               & $2.5 - 5.0$    & Strong  \\
$< 0.01$              & Strong                 & $> 5.0$        & Decisive \\
\midrule
\bottomrule
\end{tabular*}
\caption{
Model comparison metrics between $\Lambda$CDM and $i$-node HDE models, including the p-value with corresponding dof and the logarithm of the Bayes factor $B_{\Lambda\mathrm{CDM},i} = \ln(Z_{\Lambda\mathrm{CDM}} / Z_i)$. While $\Delta \ln Z$ is interpreted via Jeffreys’ scale~\cite{Burnham,Trotta_2008},  the p-value ranges follow the standard frequentist interpretation based on Wilks’ theorem \cite{wilks1938,Liddle:2007fy,Trotta:2008qt}}

\label{tab:aic}
\end{table}

\section{Results and discussion } \label{sec:results}

\begin{table*}[ht]
\centering 
\begin{tabular*}{0.85\textwidth}{@{\extracolsep{\fill}}c c c c c c c }
\toprule 
\midrule
\multicolumn{1}{c}{\textbf{Model}} &  \multicolumn{1}{c}{$\Lambda$\textbf{CDM}} & \multicolumn{1}{c}{\textbf{CPL}} &\multicolumn{1}{c}{\textbf{0-Node}} & \multicolumn{1}{c}{\textbf{1-Node}} & \multicolumn{1}{c}{\textbf{2-Node}}  & \multicolumn{1}{c}{\textbf{3-Node} } \\
\midrule
 \multicolumn{7}{l}{\textbf{DESI+SH0ES+Union3}} \\
{$\Omega_m$} &
 {$0.31 \pm 0.01$} &
 {$0.34 \pm 0.01$}&
{$0.29 \pm 0.01$} &
{$ 0.31 \pm 0.01$} & 
{$0.33 \pm 0.01$}&
{$0.34 \pm 0.01$}
\\

{$H_0 \,[\mathrm{km/s/Mpc}]$}&
{$70.13 \pm 0.69$ }&

{$71.26 \pm 0.82$} &
{$70.64 \pm 0.84$ } &
{$69.88 \pm 0.62$} &
{$71.40 \pm 0.79  $ }& 
{$71.60 \pm 0.77$ }\\
\multicolumn{7}{c}{\dotfill}\\
{$f_1$} & 0
&
- &%
$\rm{-}2$ &
{$> \rm{-}0.14        $}&
{$\rm{-}1.49_{-0.38}^{+0.24}$} &
{$\rm{-} 1.71 _{-0.54}^{+0.32}$}\\

{$f_2$} & 0
&
-&
$\rm{-}2$ &
 - &
{$\rm{-}1.76_{-0.47}^{+0.28}$ }&
{$\rm{-}1.91_{-0.56}^{+0.30}$}\\

{$f_3$} & 0 &
- &
$\rm{-}2 $ &
 - &
- &
{$\rm{-} 2.51_{-0.79}^{+0.48}$}\\

{$c$} & - &
- &
{$0.69 \pm 0.04$} &
{$1.52 \pm 0.86$ }&
{$1.77_{-0.67}^{+0.95}$ }&
{$1.50 \pm0.80$}\\

\multicolumn{7}{c}{\dotfill}\\
{$  \chi^2$} &
{56.23} &

{43.7}& 
{65.55}&
{55.75}&
{45.46 }& 
{38.45} \\

$\Delta \chi^2_{\Lambda \rm \text{CDM,i-Node}}$ &
0 &

{12.53} &
{$\rm- 9.32$} &
{ 0.48}&
{10.77}& 
{17.78}\\

${p-value}$ &
{0} &
{0.0019}&
{1}&
 {0.7866}&
{0.013} &
{0.0014}\\

$-\ln{\mathcal{Z}}$ &
{$  33.72 \mp 0.14$} &
{$ 31.47 \mp 0.19$}&
{$41.80 \mp 0.19$} &
{$37.39 \mp 0.19$ }&    
{$33.47 \mp 0.20$} &
{$32.60 \mp 0.23$}\\

${\mathcal{B}_{\rm \Lambda \text{CDM,i-Node}}}$ &
- &
{$\rm -2.25 \pm 0.24$ }&
{$8.08 \pm 0.24$ } &
{$3.67 \pm 0.24$} &
{$\rm - 0.25 \pm 0.24$} & 
{$-1.12 \pm 27$ }\\ 
\midrule
\multicolumn{6}{l}{\textbf{DESI+PPS}} \\
{$\Omega_m$} &
{$0.31 \pm 0.01$} &
{$0.33 \pm 0.01$}&
{$0.30 \pm 0.01$} &
{$0.30 \pm 0.01$} & 
{$0.33 \pm 0.01$} &
{$0.34\pm 0.01$ }\\

{$H_0 \,[\mathrm{km/s/Mpc}]$} &
{$70.15 \pm 0.60$} &
{$71.27 \pm 1.30$}&

{$70.05 \pm 0.79$} &
{$71.90 \pm 0.90$} &
{$71.26 \pm 0.75$} & 
{$71.61 \pm 0.75$}\\
\multicolumn{7}{c}{\dotfill}\\

{$f_1$} &
0 &%
- &
$\rm -2$& 
{ $> -0.15                  $}&
{$\rm -1.49_{-0.40}^{+0.20}$} &
{$- 1.77_{-0.53    }^{+0.31}$}\\

{$f_2$} &
0&
- &%
$\rm -2$ &
 - &
{$\rm - 1.75_{-0.50}^{+0.27}$} &
{$ \rm - 1.94_{-0.55}^{+0.28}$}\\

{$f_3$} &
0& 
- &%
$\rm -2$ &
 - &
- &
{$ \rm - 2.45_{-0.78}^{+0.48}$}\\

{$c$} &
- &
- &
{$ 0.73 \pm 0.04$} &
{$1.50 \pm 0.87$}  &
{$1.77^{+0.95}_{-0.65}$} &
{$1.51_{-0.91}^{+0.78} $}\\

\multicolumn{7}{c}{\dotfill}\\

$\chi^2$ &
{1585.75} &

{1573.48} &
{1601.08}&
{1585.20} &
{1574.43} &
{1568.82}\\
$\Delta \chi^2_{\Lambda \text{CDM,i-Node}}$ &
0 &
{12.27} &
{-$15.33$}&
{0.55} &
{11.32} &
{16.93}\\

${p-value}$ &
{0} &
{0.0022}&
{1}&
{0.7596} &
{0.0101} &
{0.002}\\
$\rm-\ln{\mathcal{Z}}$ &
{$803.01 \mp 0.20$} &
{$801.21 \mp 0.24$} &

{$813.54 \mp  0.23$} &
{$806.31 \mp 0.23$} &    
{$ 801.85 \mp 0.24$} &
{$ 802.13 \mp 0.27 $}\\

${\mathcal{B}_{ \Lambda         \text{CDM,i-Node}}}$ &
- &
{ $-1.80 \pm 0.31$}&
{$10.53 \pm 0.30$}&
{$3.30 \pm 0.30$} &
{$-1.16 \pm 0.31$}& 
{$-0.88 \pm 0.34$}\\
\midrule \bottomrule
\end{tabular*}
\captionsetup{singlelinecheck=off, justification=raggedright}
\caption{Mean values for cosmological parameters for the $\Lambda$CDM model, CPL parametrization and 0-3-Node HDE reconstructions,  obtained from the combined DESI+SH0ES+Union3  dataset (upper panel) and DESI+PPS dataset (lower panel). Quoted uncertainties are at $68\%$ confidence limits whereas lower bounds are calculated at 95~\% confidence level. Across all models and dataset combinations, the baryon density is consistently constrained to $100\Omega_b h^2=2.22 \pm 0.04$. } 
\label{tab:results}
\end{table*}

\begin{figure*}[!t]
    \includegraphics[width=.48\textwidth]{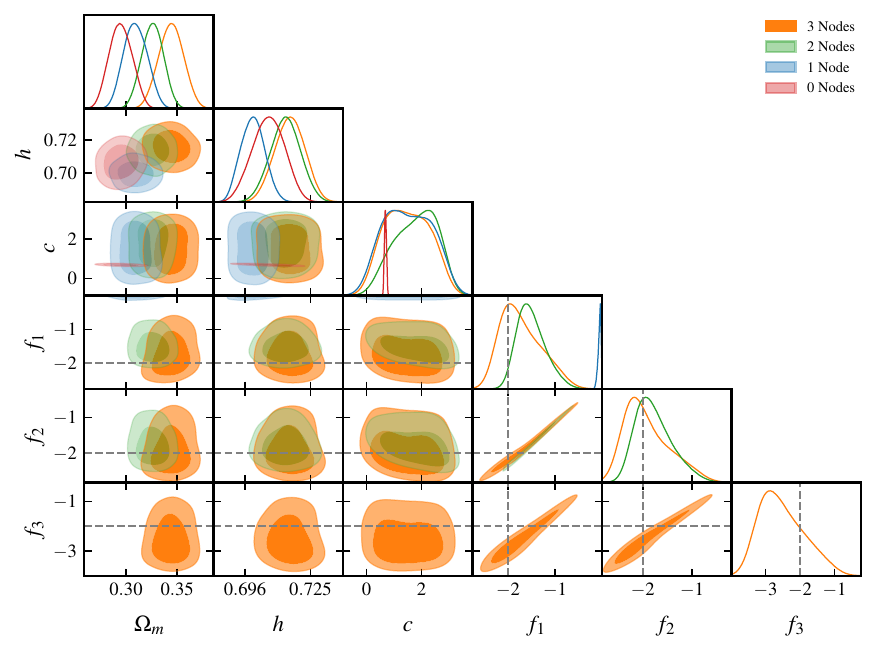}
    \includegraphics[width=.48\textwidth]{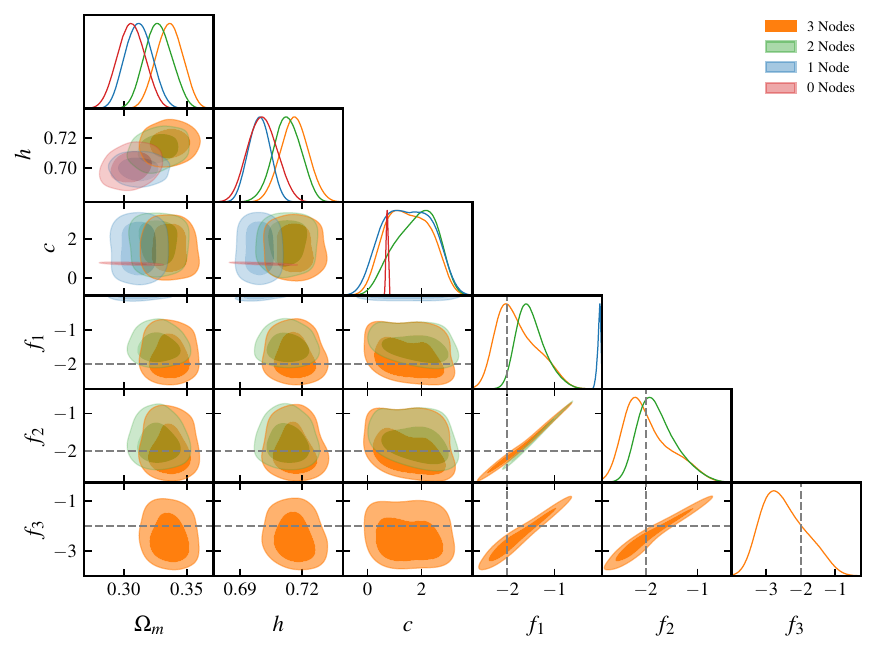}
    \captionsetup{singlelinecheck=off, justification=raggedright} 
    \caption{Triangle plot for marginalized posterior distributions using the DESI+SH0ES+Union3  (left) and DESI+PPS data combination (right) at 1-$\sigma$ and 2-$\sigma$ confidence level. The dashed lines represent the case of standard HDE, i.e. the case with all nodes fixed to $f_i = -2$. }
    \label{fig:triangle_both}
\end{figure*}

\begin{figure*}[!htbp]
    \centering
    \makebox[\textwidth][c]{
    
    \includegraphics[trim=1mm 0mm 0mm 0mm, clip, width=0.35\textwidth]{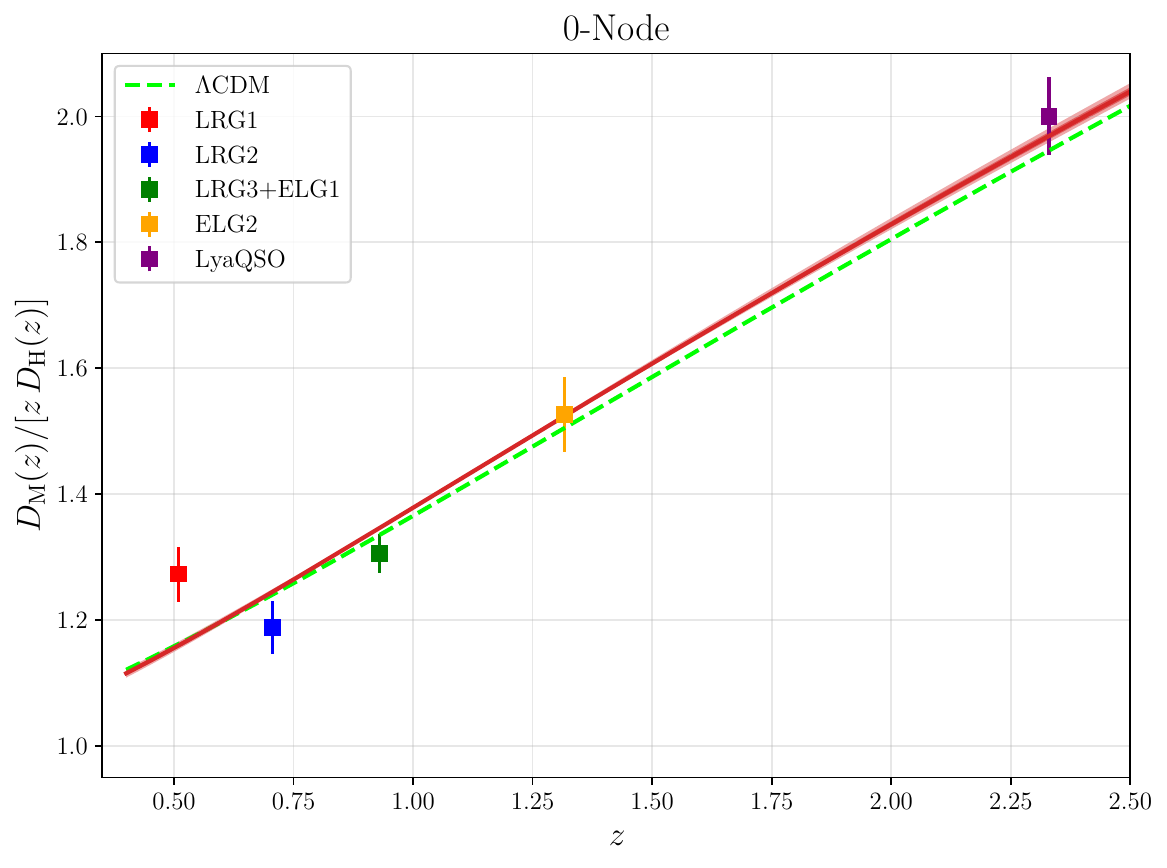}
   
    \includegraphics[trim=1mm 0mm 0mm 0mm, clip, width=0.35\textwidth]{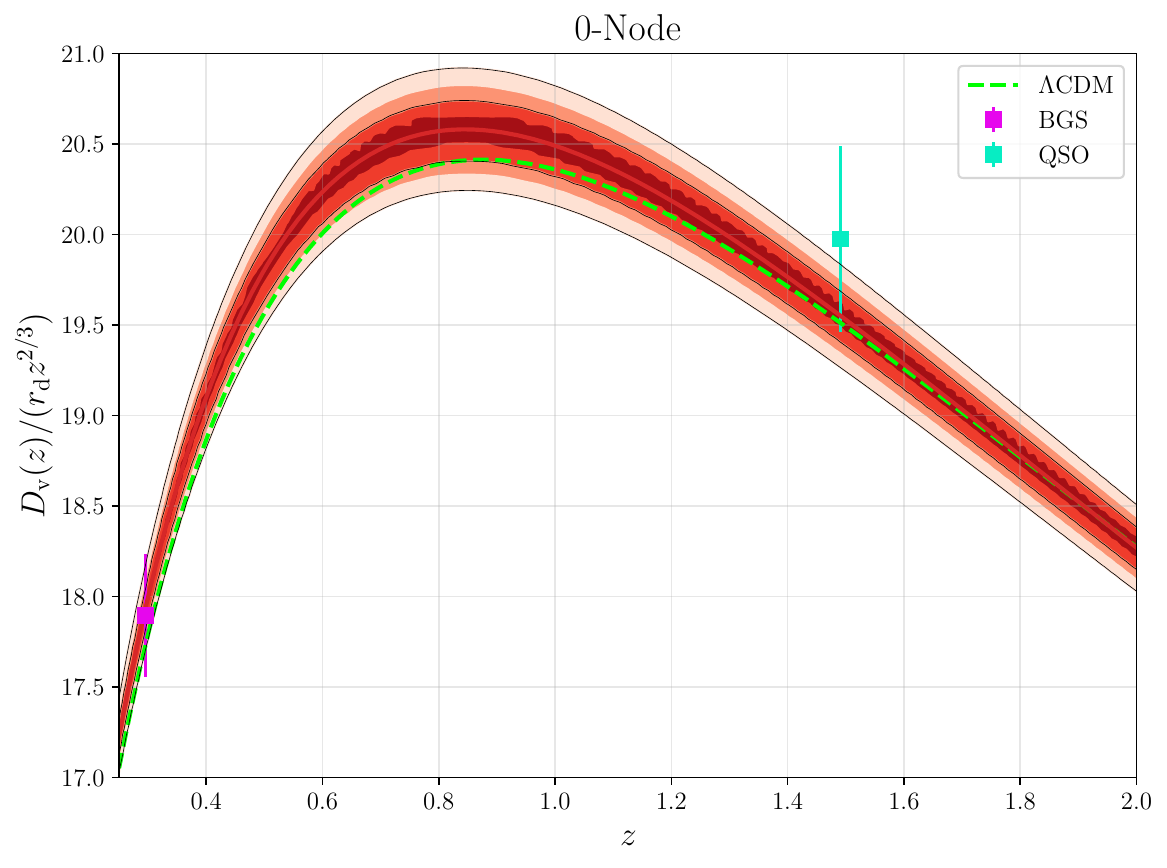}
     \includegraphics[trim=1mm 0mm 0mm 0mm, clip, width=0.33\textwidth]{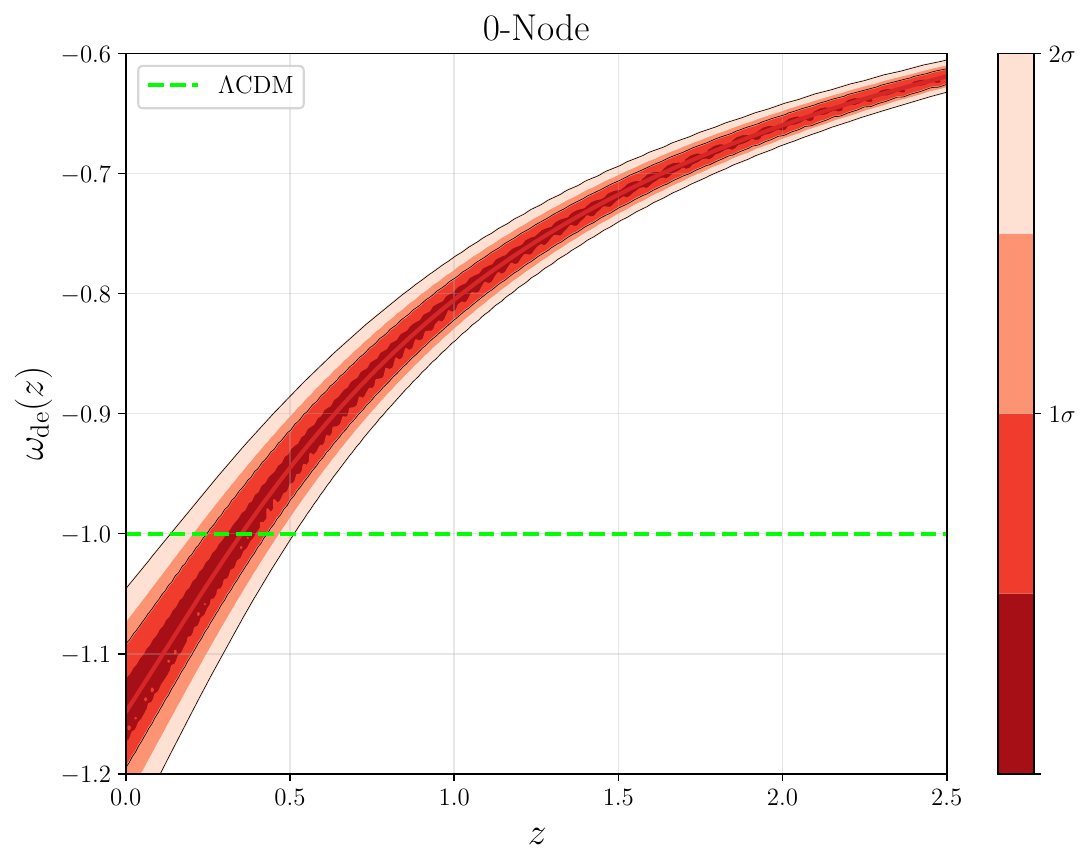}

    }
    \makebox[\textwidth][c]{
    
    \includegraphics[trim=1mm 0mm 0mm 0mm, clip, width=0.35\textwidth]{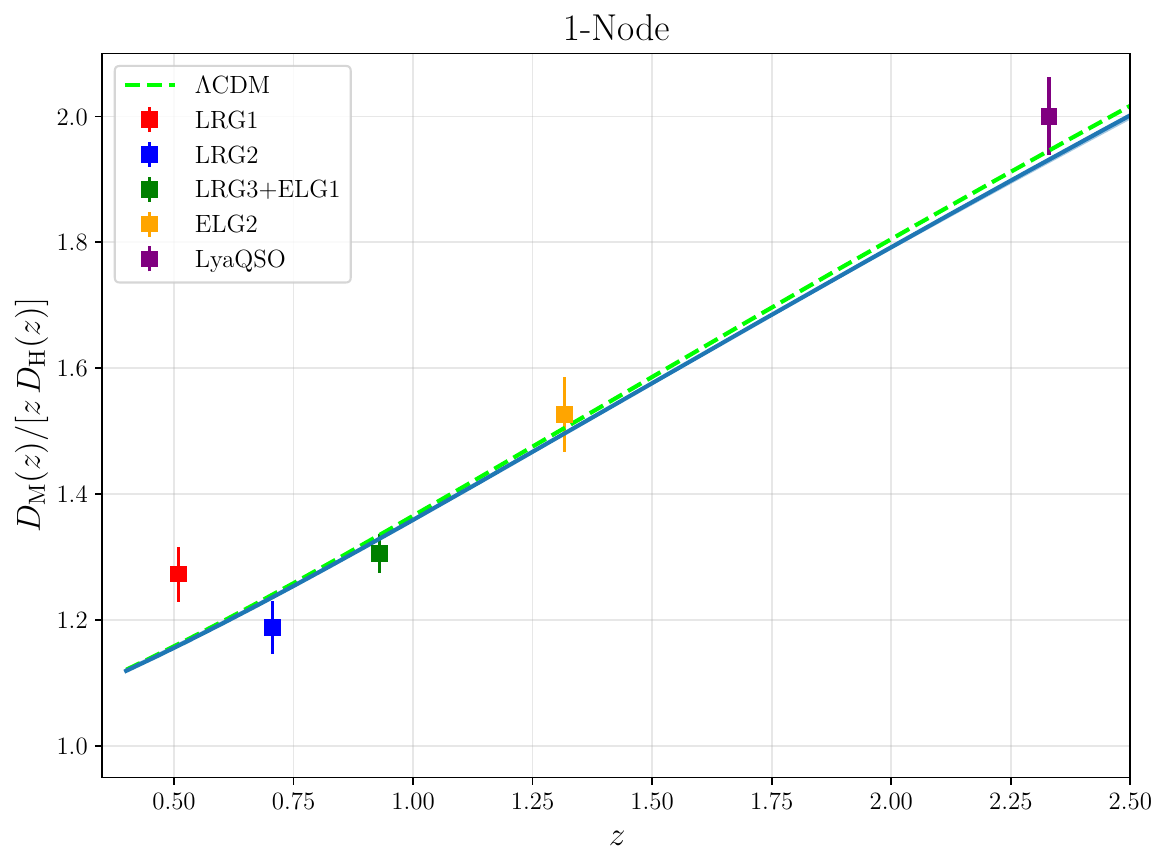}
    \includegraphics[trim=1mm 0mm 0mm 0mm, clip, width=0.35\textwidth]{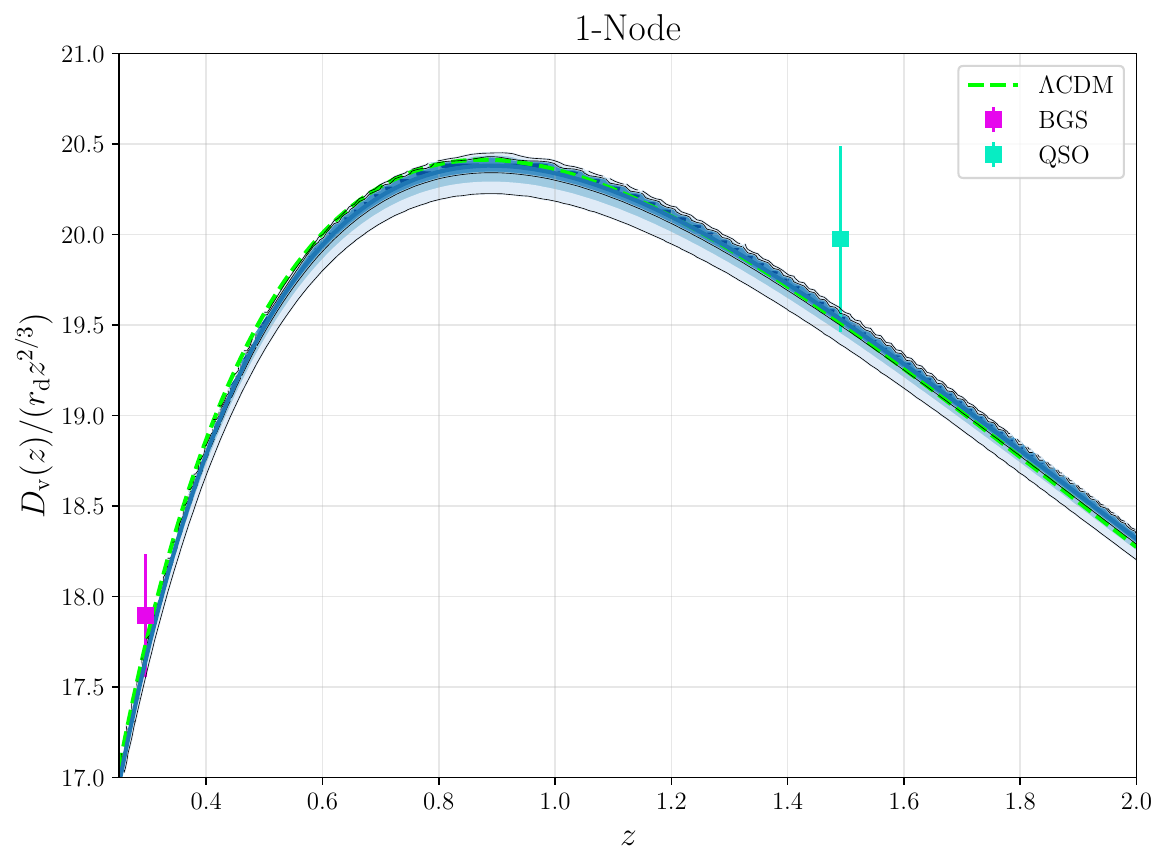}

    \includegraphics[trim=1mm 0mm 0mm 0mm, clip, width=0.33\textwidth]{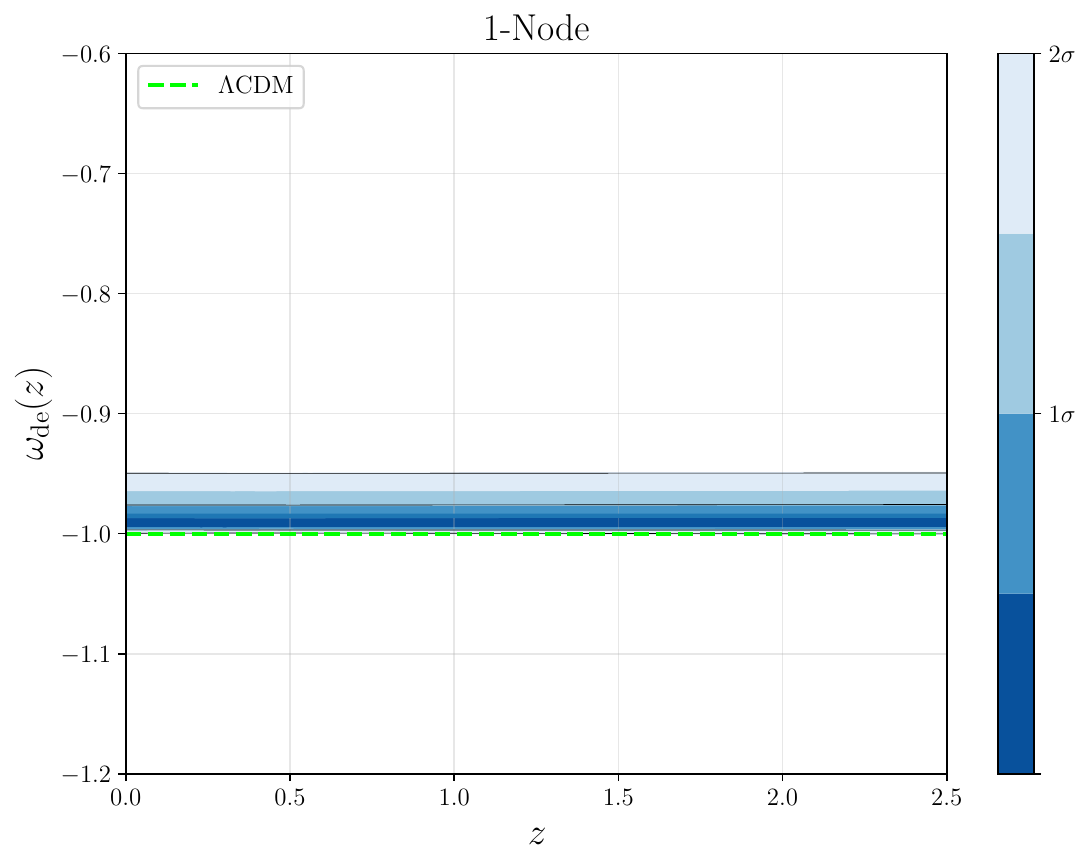}

    }
    \makebox[\textwidth][c]{
    
    \includegraphics[trim=1mm 0mm 0mm 0mm, clip, width=0.35\textwidth]{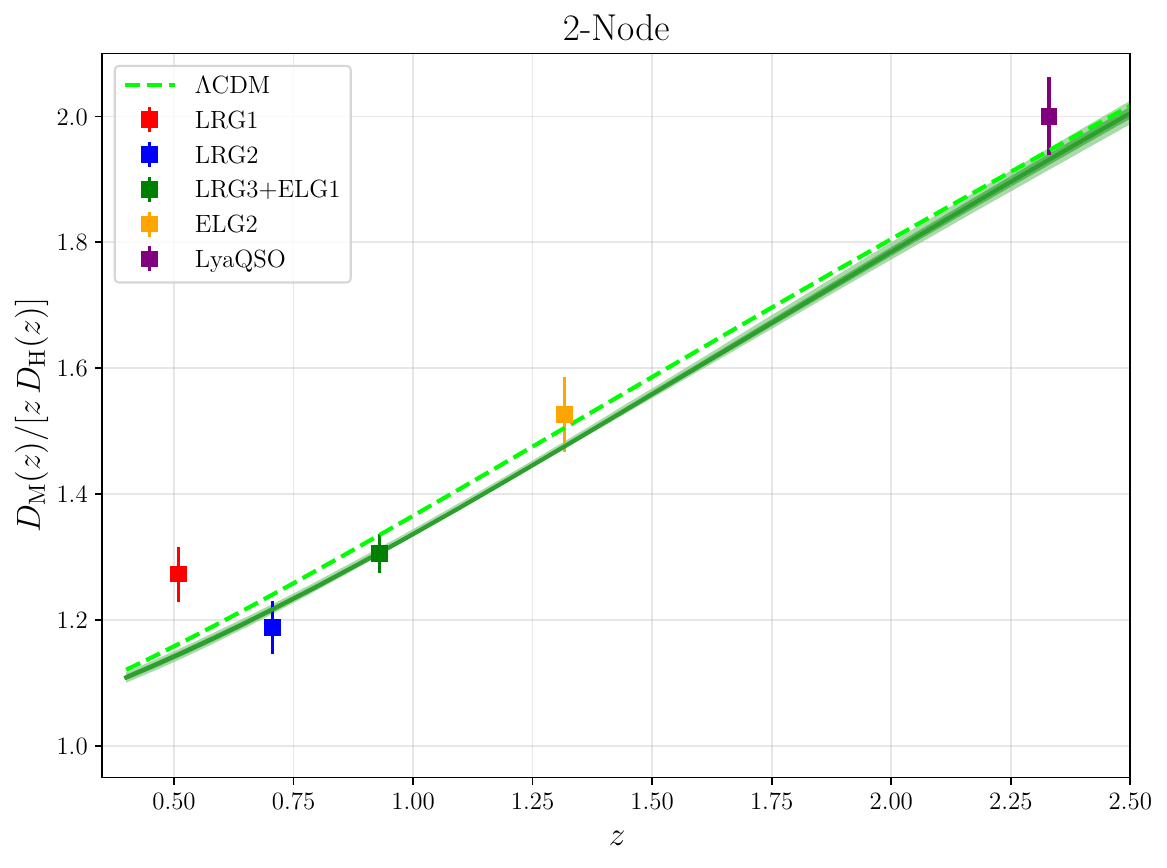}
    \includegraphics[trim=1mm 0mm 0mm 0mm, clip, width=0.35\textwidth]{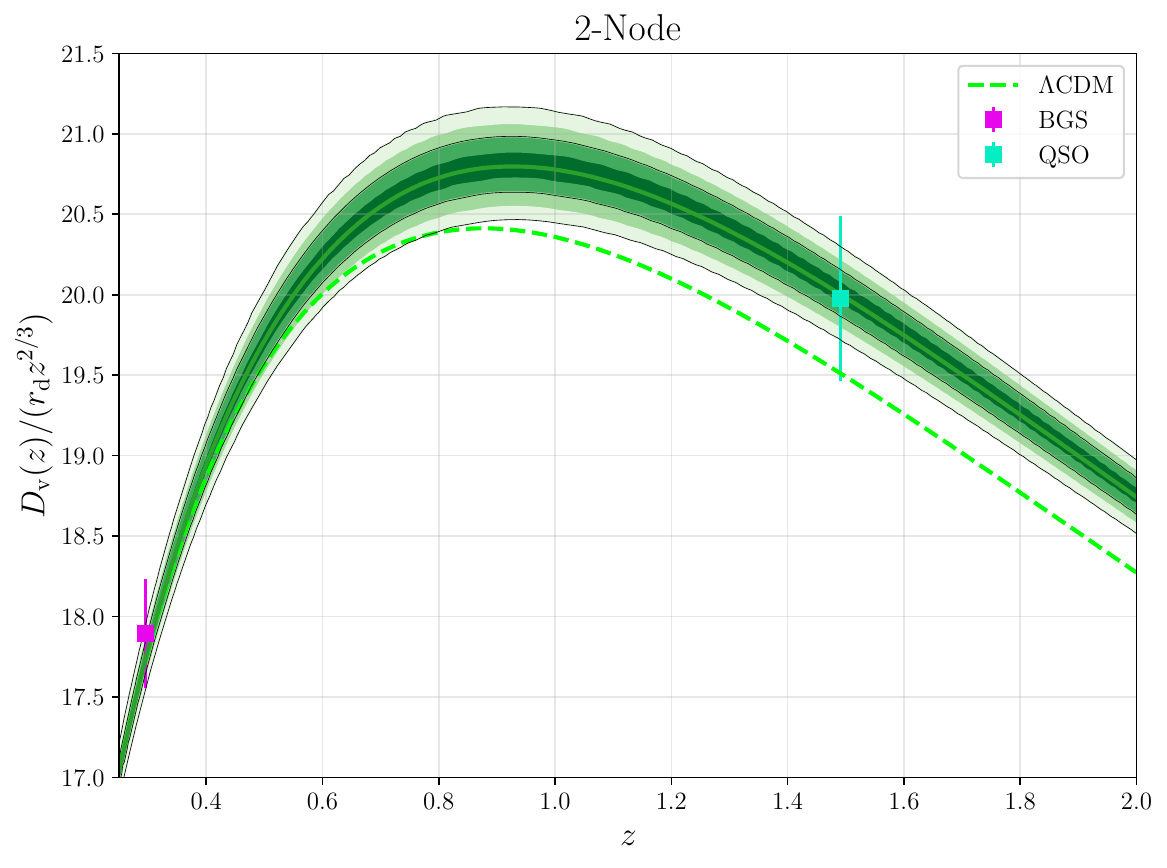}
    \includegraphics[trim=1mm 0mm 0mm 0mm, clip, width=0.33\textwidth]{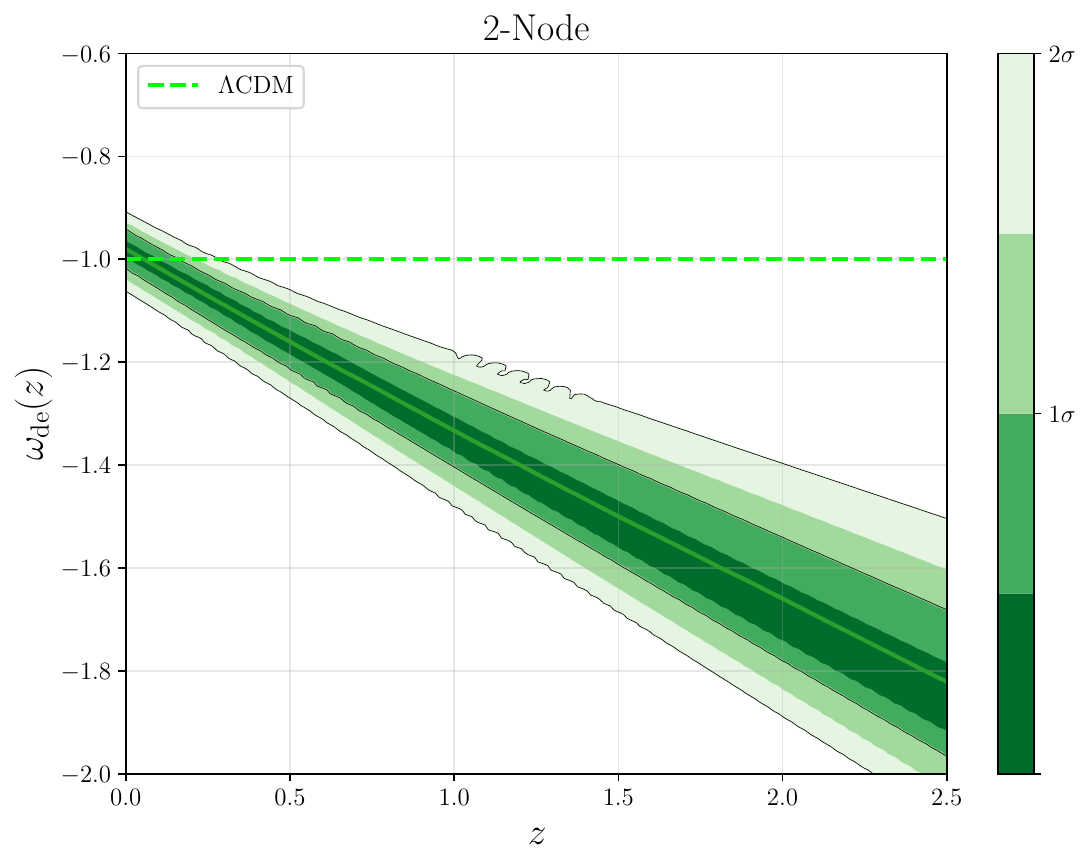}
    }

    \makebox[\textwidth][c]{
    
    \includegraphics[trim=1mm 0mm 0mm 0mm, clip, width=0.35\textwidth]{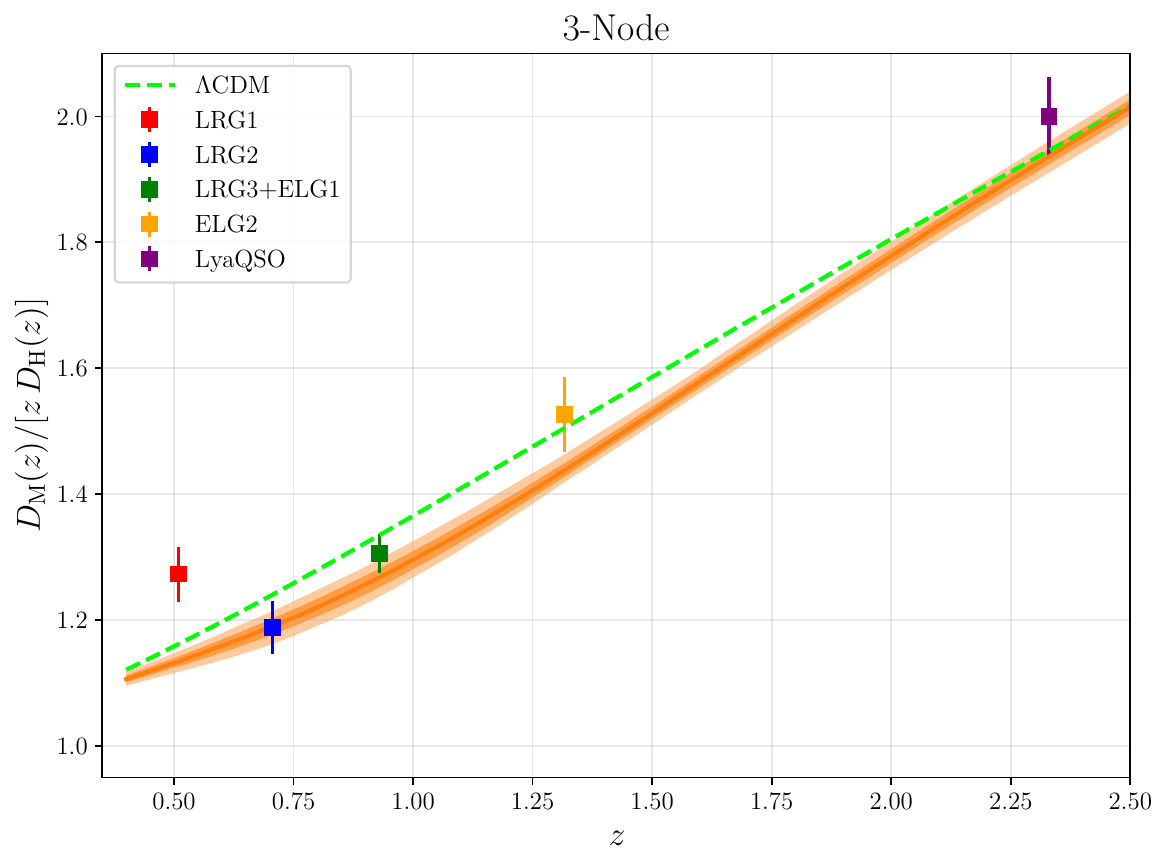}
    \includegraphics[trim=1mm 0mm 0mm 0mm, clip, width=0.35\textwidth]{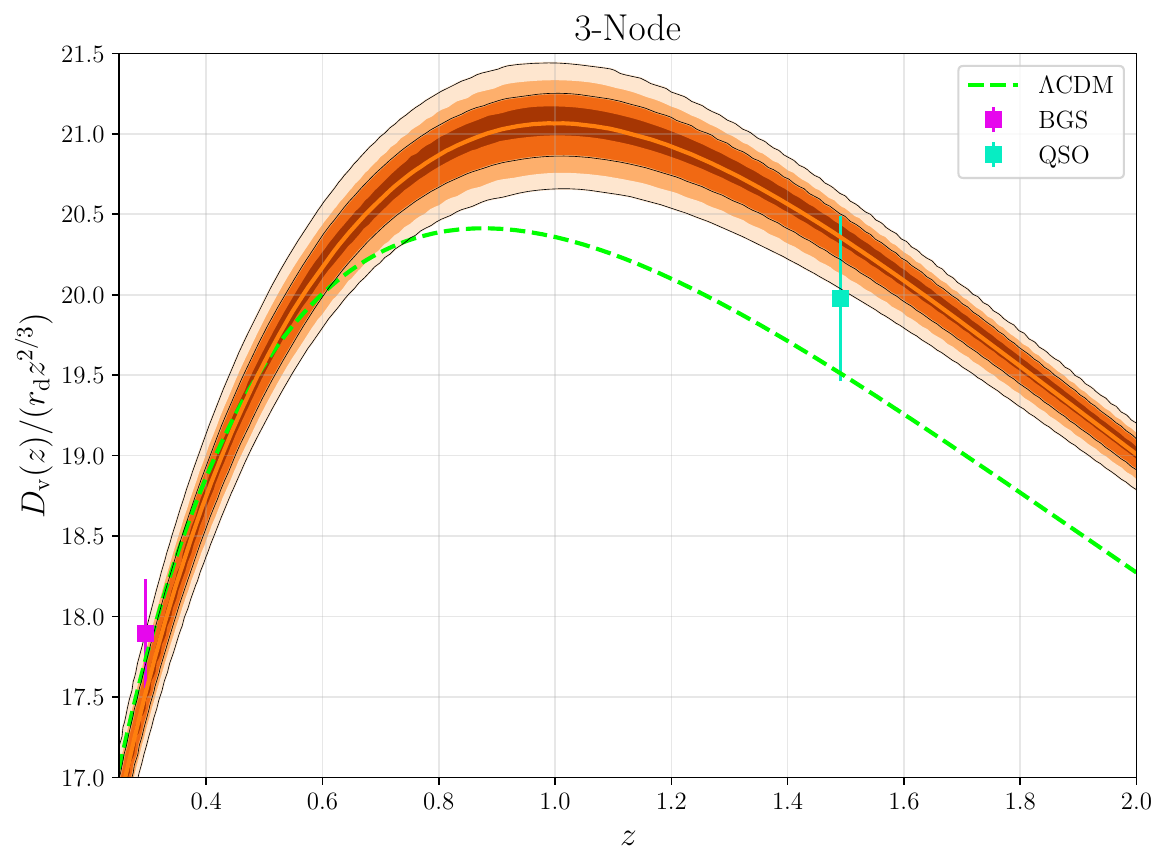}
    \includegraphics[trim=1mm 0mm 0mm 0mm, clip, width=0.33\textwidth]{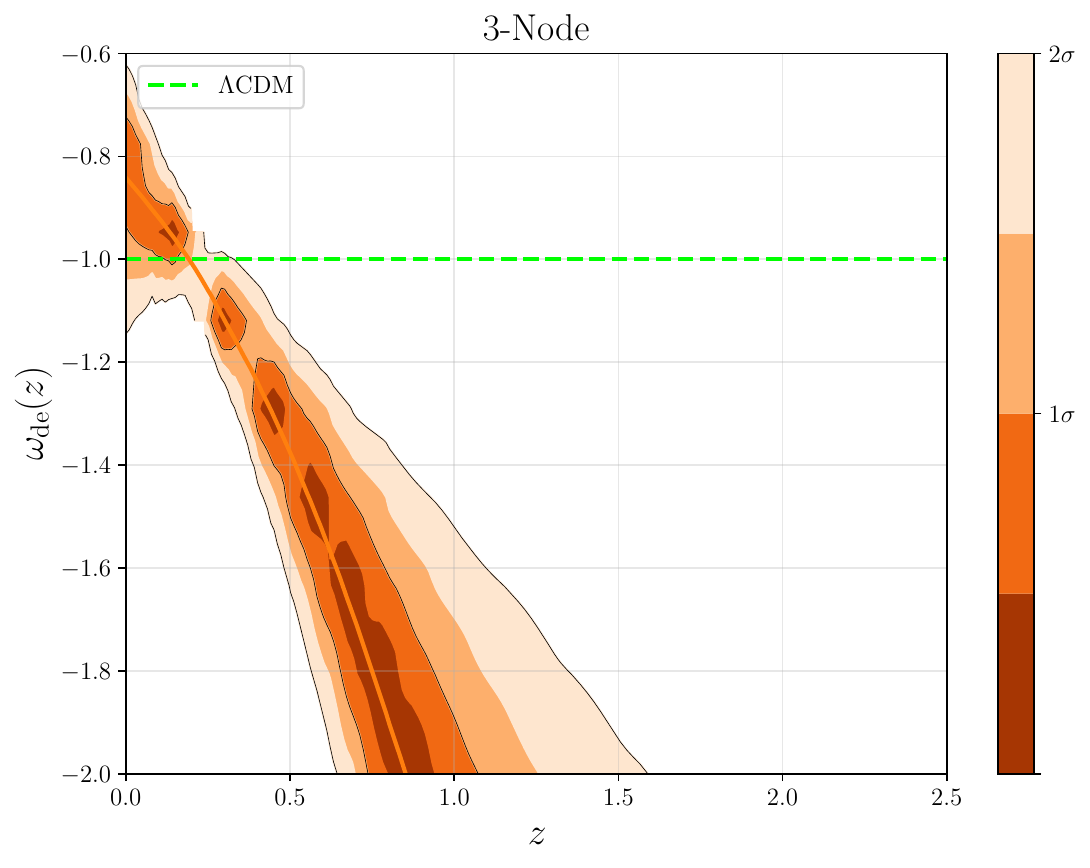}
    }
    
     \captionsetup{singlelinecheck=off, justification=raggedright} 
    \caption{{2-$\sigma$} functional posterior probability of $D_{\rm{M}}(z)/[z\,D_{\rm{H}}(z)]$ (left column), \(D_{\rm v}(z)/(r_{\rm d} z^{2/3})\) (middle column) and $\omega_{\rm{de}}(z)$ (right column) from 0-Node to 3-Node cases, from top to bottom, using DESI+SH0ES+Union3 data. {Shaded regions indicate different confidence levels, as indicated in the color bar on the right, and DESI data (Table~\ref{tab:DESI}) are included as reference points for comparison.} }
    \label{fig:distances_unions}
\end{figure*}

\begin{figure*}[!htbp]
    \centering

    \makebox[\textwidth][c]{
    \includegraphics[trim=1mm 0mm 0mm 0mm, clip, width=0.35\textwidth]{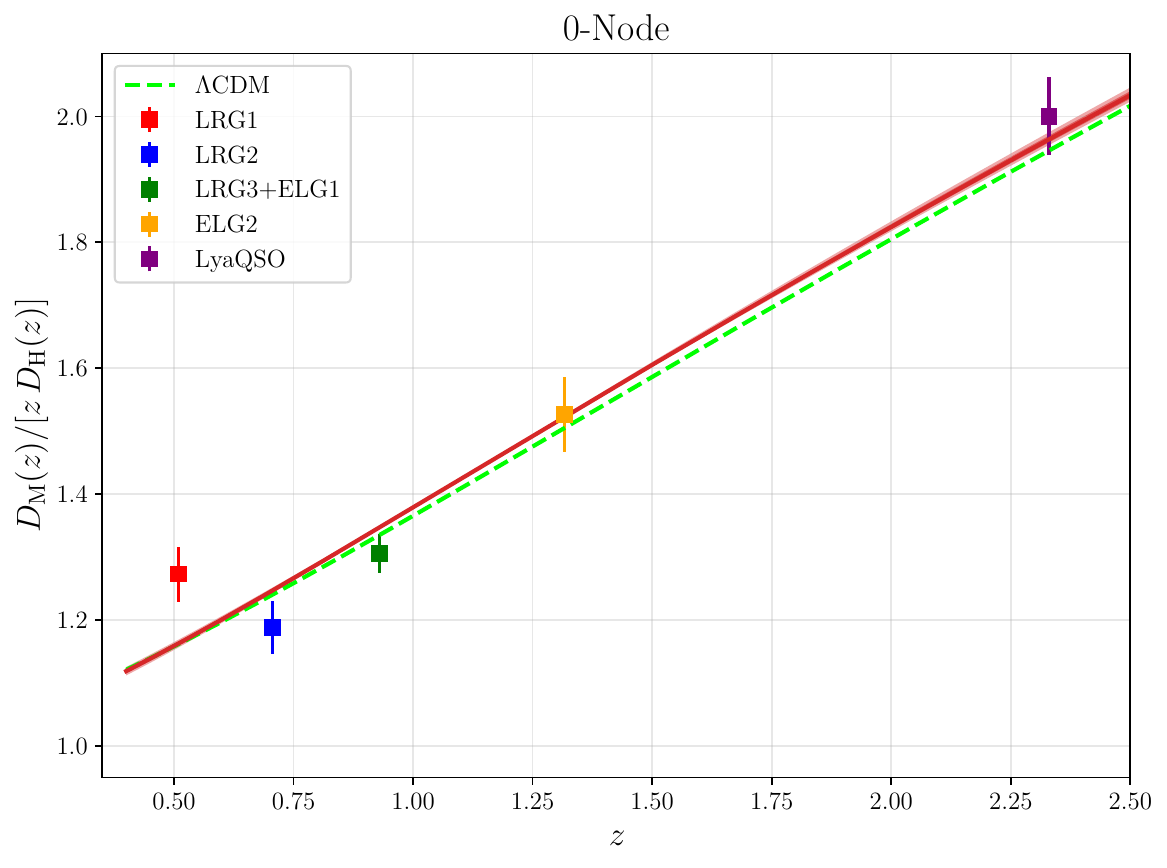}
   
    \includegraphics[trim=1mm 0mm 0mm 0mm, clip, width=0.35\textwidth]{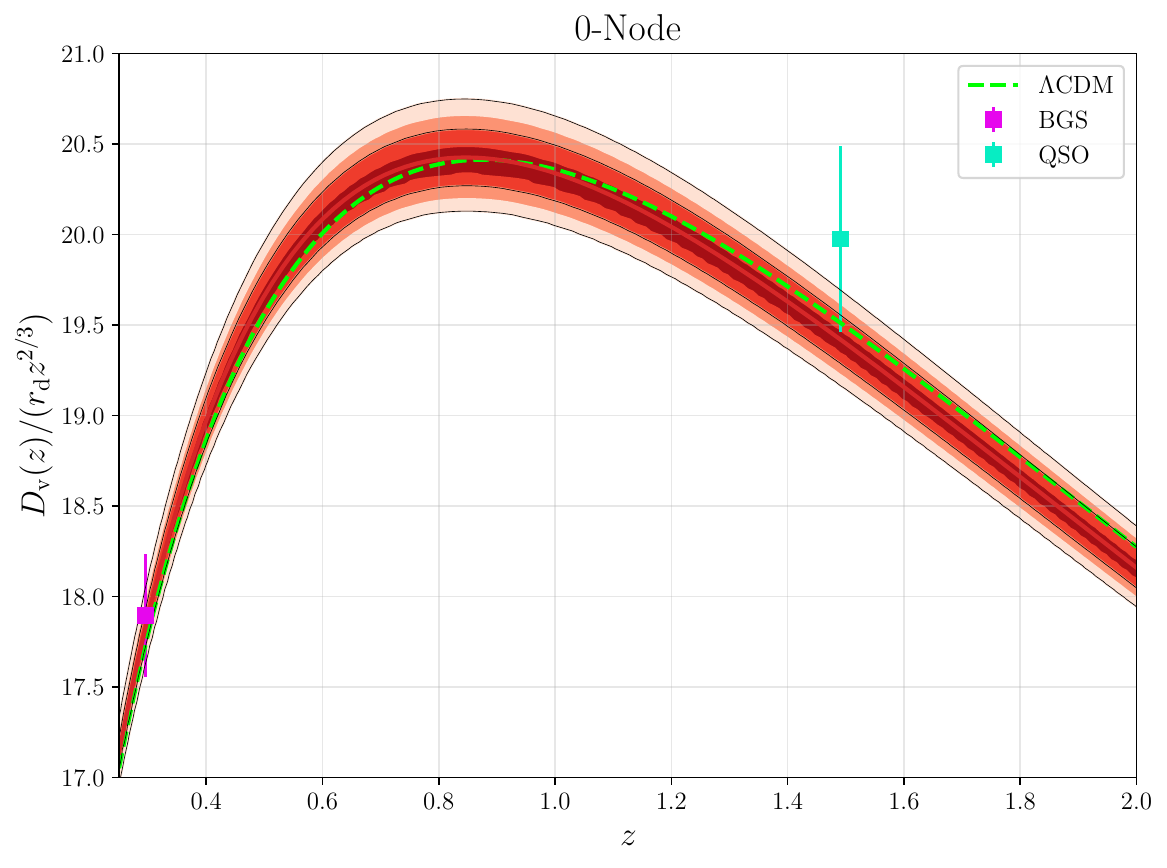}
     \includegraphics[trim=1mm 0mm 0mm 0mm, clip, width=0.33\textwidth]{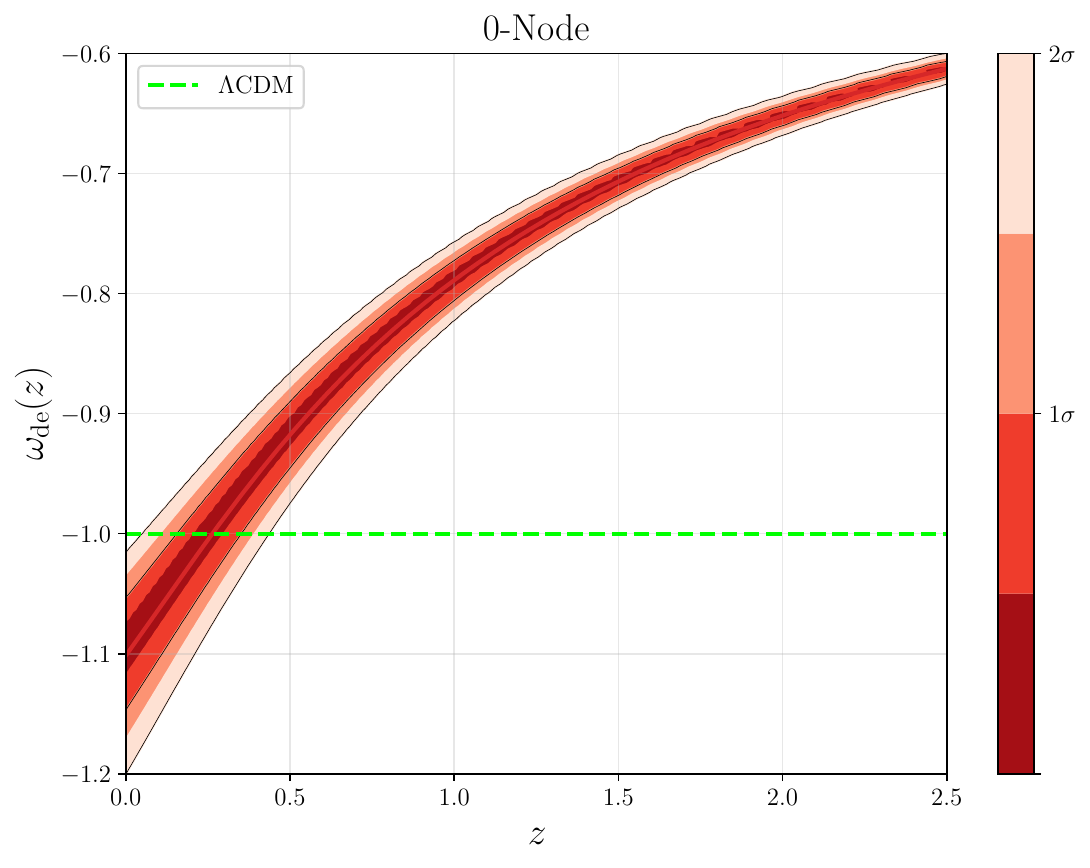}
}

         \makebox[\textwidth][c]{
    \includegraphics[trim=1mm 0mm 0mm 0mm, clip, width=0.35\textwidth]{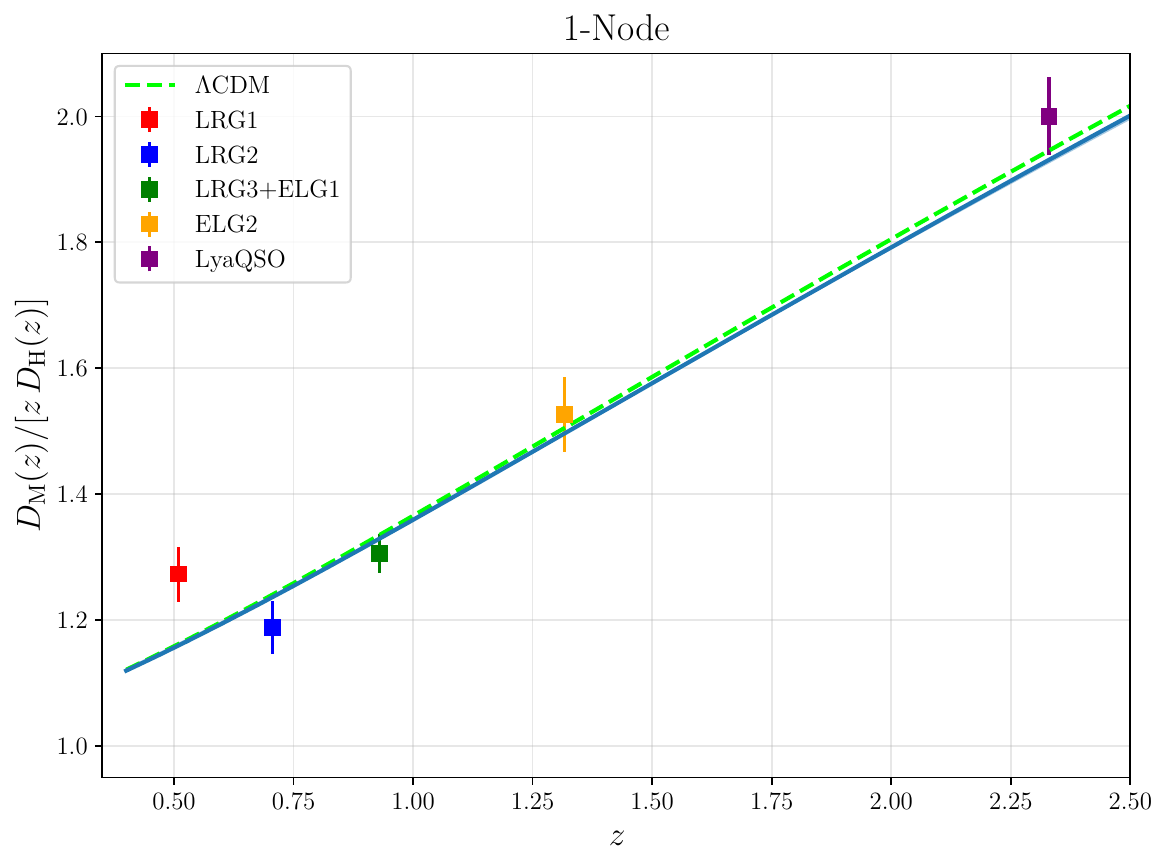}
   
    \includegraphics[trim=1mm 0mm 0mm 0mm, clip, width=0.35\textwidth]{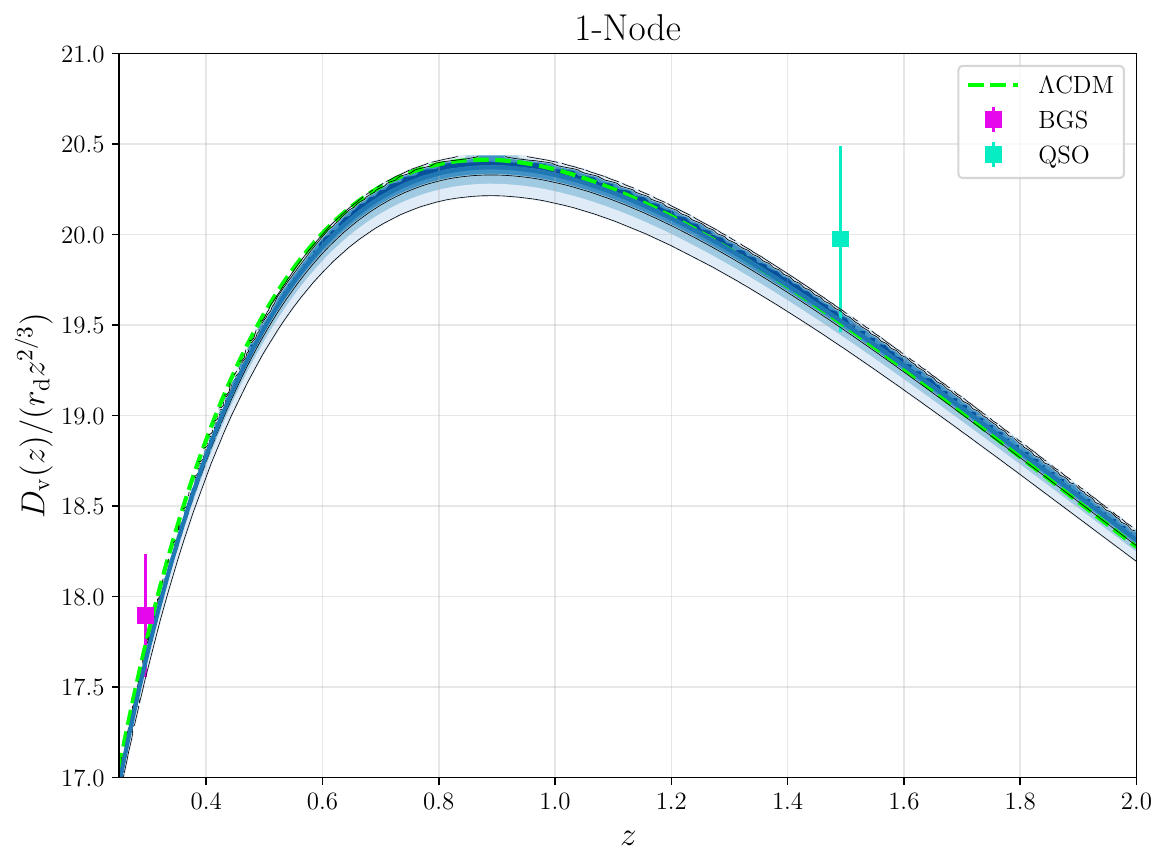}
     \includegraphics[trim=1mm 0mm 0mm 0mm, clip, width=0.33\textwidth]{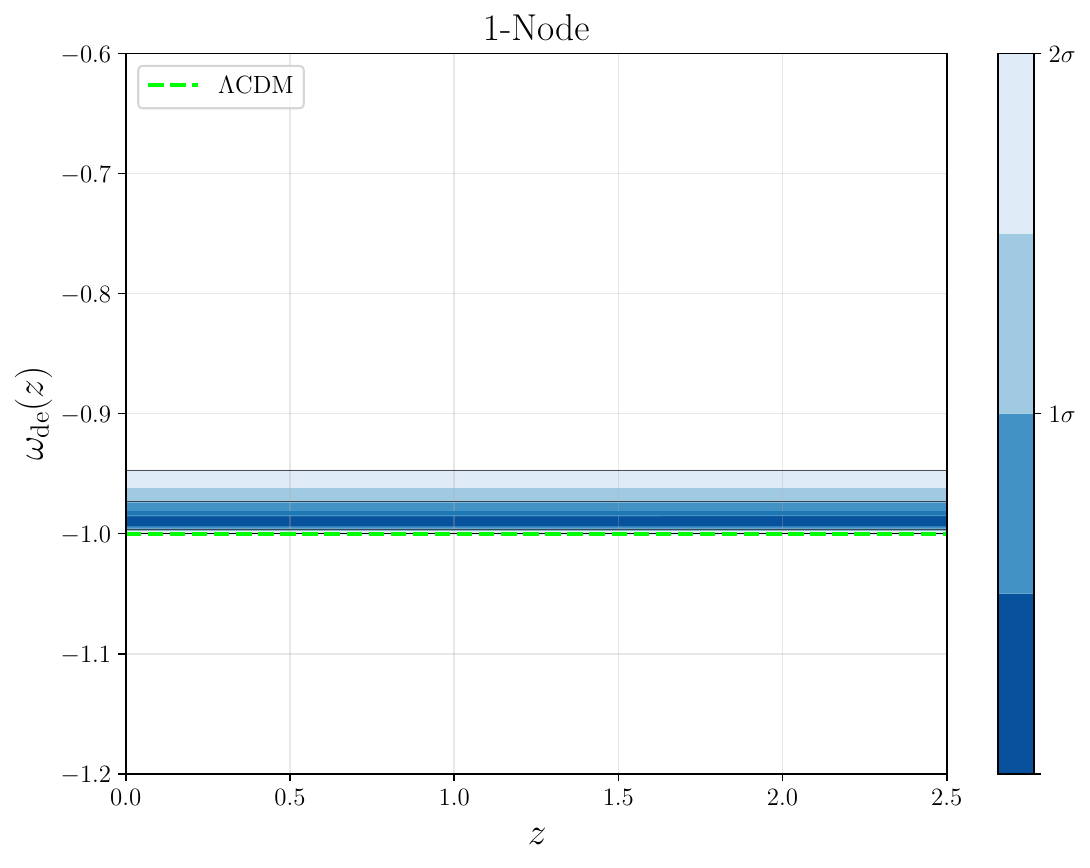}
}
    \makebox[\textwidth][c]{
    \includegraphics[trim=1mm 0mm 0mm 0mm, clip, width=0.35\textwidth]{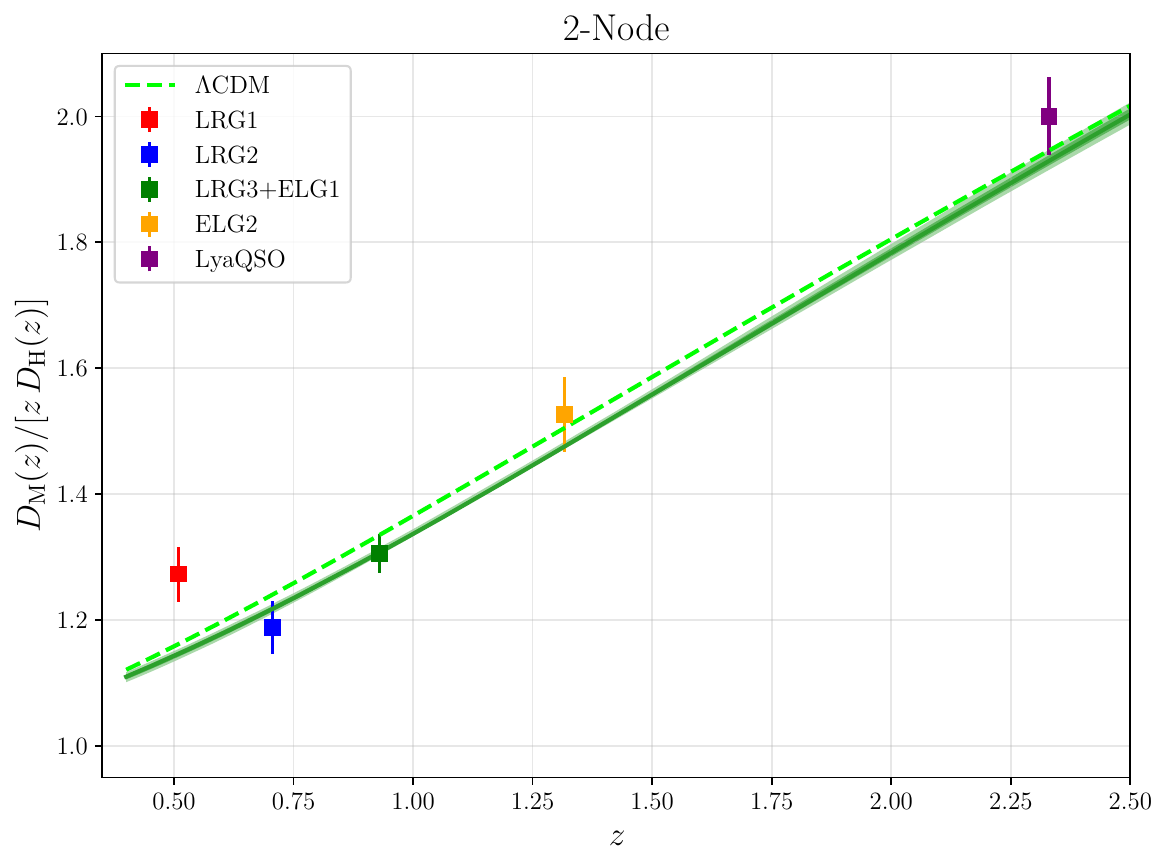}
   
    \includegraphics[trim=1mm 0mm 0mm 0mm, clip, width=0.35\textwidth]{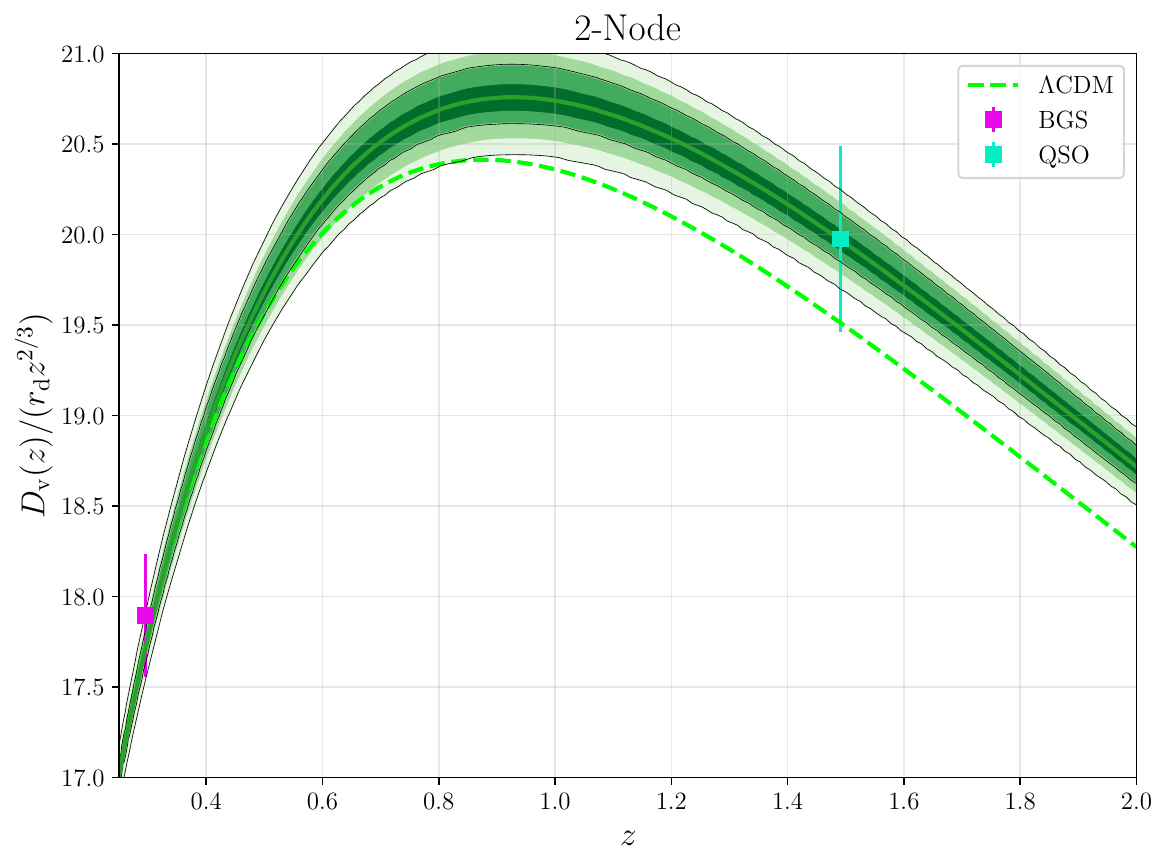}
     \includegraphics[trim=1mm 0mm 0mm 0mm, clip, width=0.33\textwidth]{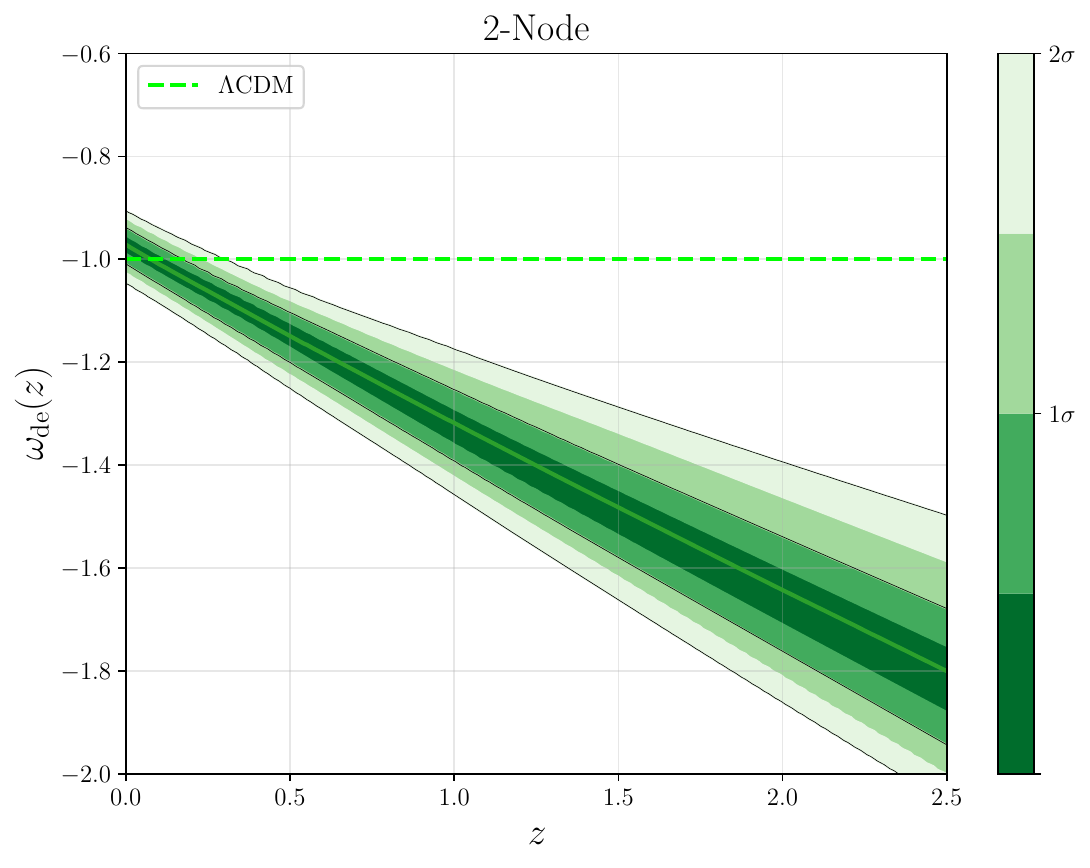}
}

    \makebox[\textwidth][c]{
    \includegraphics[trim=1mm 0mm 0mm 0mm, clip, width=0.35\textwidth]{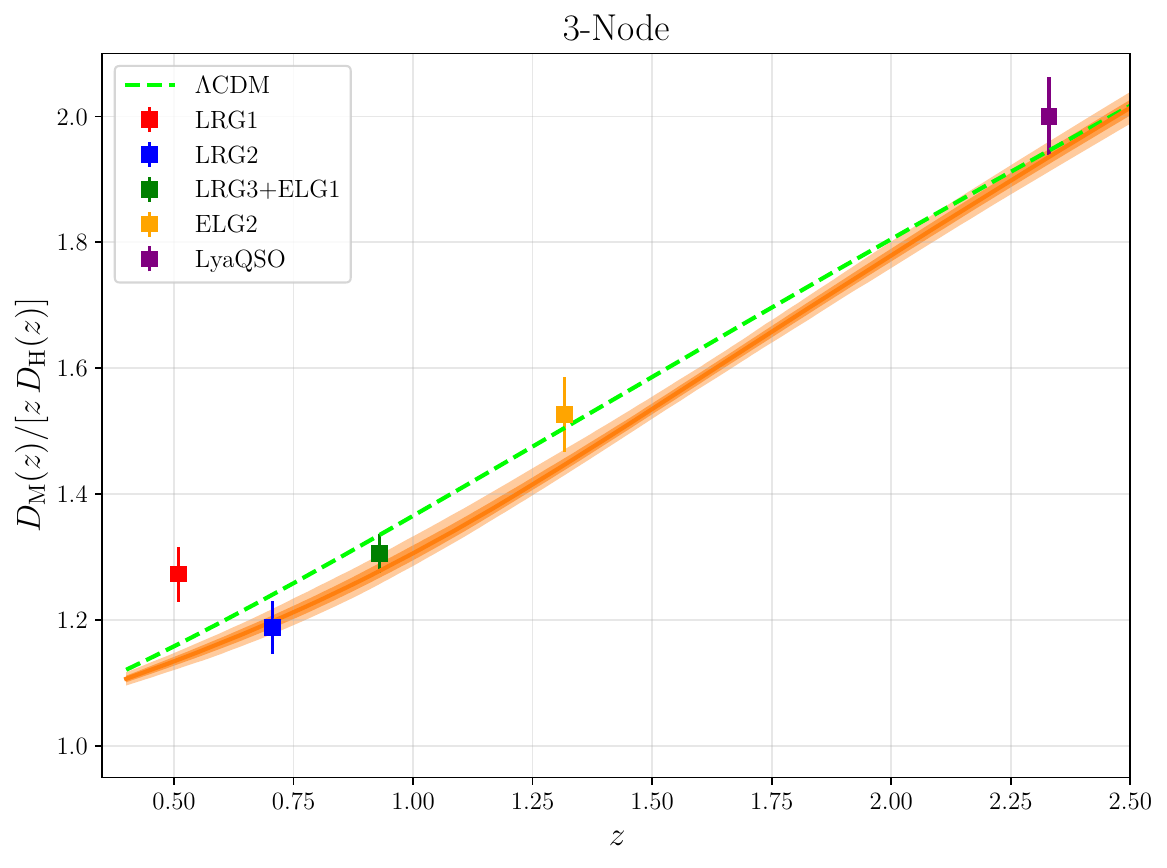}
   
    \includegraphics[trim=1mm 0mm 0mm 0mm, clip, width=0.35\textwidth]{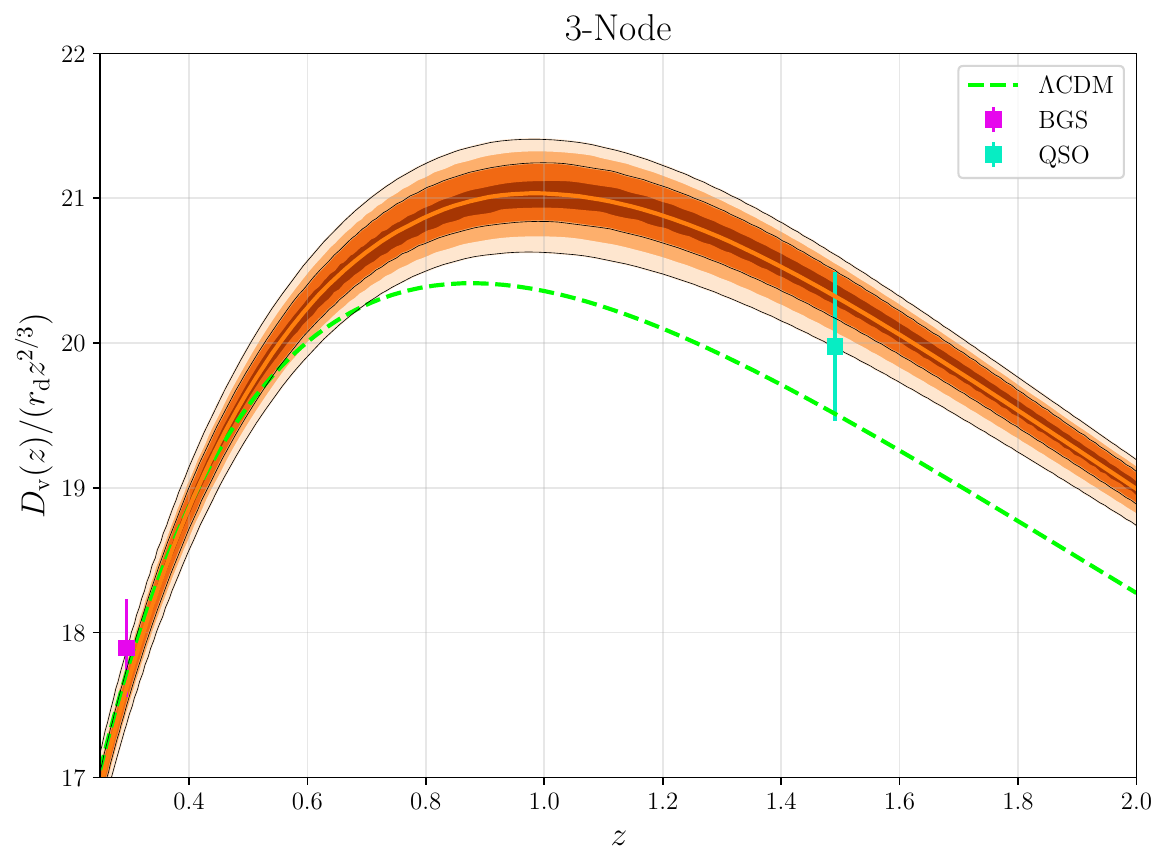}
\includegraphics[trim=1mm 0mm 0mm 0mm, clip, width=0.33\textwidth]{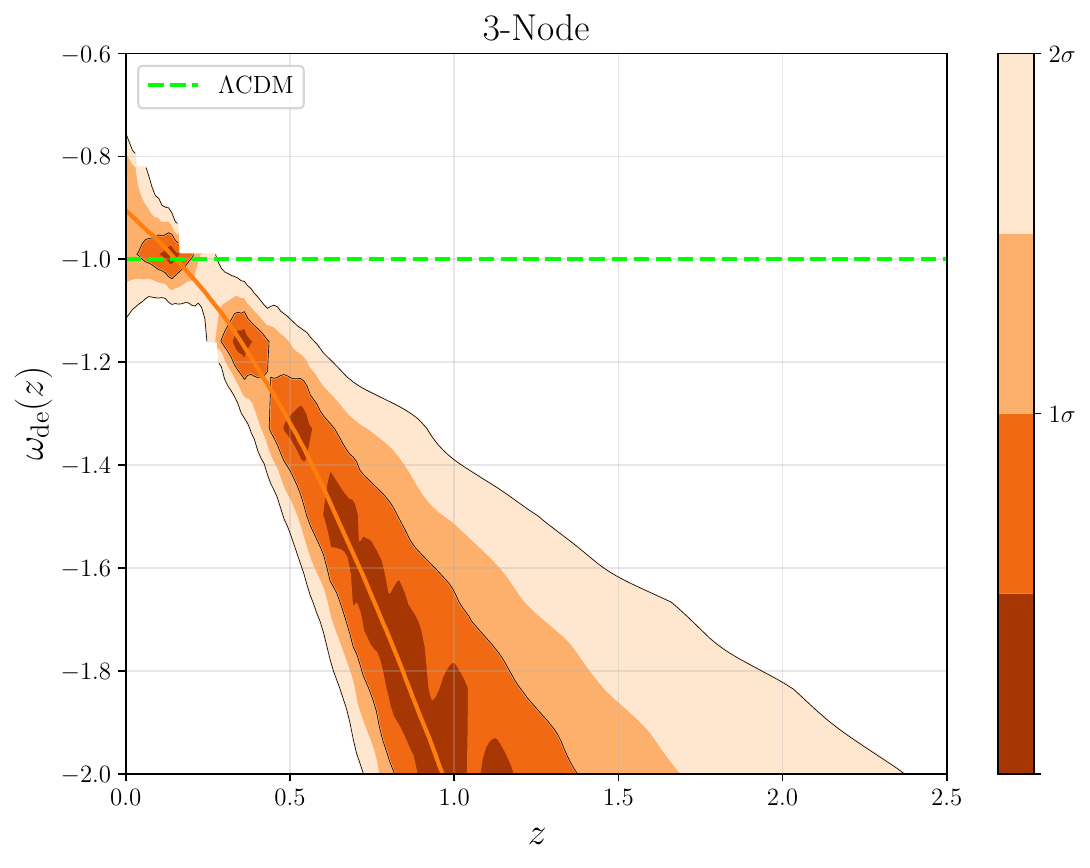}
}
  \captionsetup{singlelinecheck=off, justification=raggedright} 

 \caption{{2-$\sigma$} posterior reconstruction of $D_{\rm{M}}(z)/[z\,D_{\rm{H}}(z)]$ (left column), \(D_{\rm v}(z)/(r_{\rm d} z^{2/3})\) (middle column), and $\omega_{\rm{de}}(z)$ (right column) for the HDE models 0- to 3-Nodes, from top to bottom, obtained from the posterior chains using the DESI+PPS dataset. Shaded regions indicate different confidence levels, as indicated in the color bar on the right, and DESI data (Table~\ref{tab:DESI}) are included as reference points for comparison. }
    \label{fig:distances_pantheon}
\end{figure*}

In this section, we present and analyze the parameter inference results for both the $\Lambda$CDM model and the HDE reconstructions using the nodal approach, based on the two data combinations described earlier. For context, the 0-Node case corresponds to the standard HDE model with a fixed value, $f = -2$. The 1-Node reconstruction assumes a constant $f$ (as in the Barrow model), the 2-Node case models $f(z)$ as a linear function of redshift, and the 3-Node scenario adopts a quadratic spline for $f(z)$. The prior ranges used for each combination of models and data sets are summarized in Table~\ref{tab:prior}.

\subsection{Union3}

With respect to the DESI+SH0ES+Union3 dataset, the parameter constraints are summarized in the left panel of Figure~\ref{fig:triangle_both}, which displays the combined 1D and 2D posterior distributions for both cosmological and holographic parameters across the models from 0 to 3-Nodes. The corresponding mean parameter values are also listed in the top panel of Table~\ref{tab:results} with uncertainties reported at $68\%$ confidence level (c. l.). 

In the simplest 0-Node model, the dynamics of HDE is governed by the standard holographic parameter $c$, which is well constrained to ${c = 0.69\pm0.04}$ at $68\%$ c.l., consistent with previous constraints obtained from DESI DR1 data~\cite{li2024revisitingholographicdarkenergy}. For this model, the reconstructed EoS is shown in the first row of Figure~\ref{fig:distances_unions} and exhibits a phantom-like behavior at present (\(\omega_{\rm de} < -1\)), transitioning to a quintessence-like regime near \(z \approx 0.25\), in agreement with the trends shown in Figure \ref{fig:combined_omega_de}. 

Figure \ref{fig:distances_unions} also displays, in its left and middle panels, the posterior reconstruction of key cosmological distances, along with {1$\sigma$ and 2$\sigma$ confidence levels (c.l)}, enabling a direct comparison with BAO measurements, particularly those from DESI. This visualization allows identifying whether the data points were within the reconstructed contours at different confidence levels. {Although most BAO measurements are well accommodated, a notable exception is the LRG1 data point in the observable $D_{\rm M}(z)/[z\,D_{\rm H}(z)]$, whose discrepancy persists across all reconstruction cases. This behavior is particularly evident in the 0-Node configuration, which exhibits the weakest statistical performance. In contrast, for the QSO data, the reconstructed curves shift between the $1\sigma$ and $2\sigma$ regions, demonstrating the improvement in $\chi^{2}$ achieved by higher-node reconstructions.} Additionally,  as shown by both {p-value} and the evidence reported in ~\cref{tab:results}, the simplest HDE scenario is strongly disfavored over the standard $\Lambda$CDM model, based on the reference indicators summarized in~\cref{tab:aic}. Although theoretically well motivated, the limited flexibility of the 0-Node model reduces its ability to match current observations, suggesting that more elaborate extensions may be needed.

The 1-Node case provides an {insignificant} improvement fit over the standard 0-Node scenario, and also a \(\chi^2\) value slightly better than $\Lambda$CDM  with \(\Delta \chi^2_{\Lambda \rm CDM,\; 1\text{-Node}} = 0.48\), but not statistically significant. {This is further supported by the respective p-value and $\Delta \rm ln{\mathcal{Z}}$ that are $0.7866$ and $3.67$, respectively. } 
As shown in Figure~\ref{fig:triangle_both}, the posterior distribution of the amplitude parameter \(f_1\) lies close to zero, with a lower bound $f_1> -0.14$, suggesting a model's tendency to converge toward $\Lambda$CDM (recovered exactly when \(f_1 = 0\)), while still accommodating small deviations that may better fit specific features in the data. Despite this, Bayesian model comparison using Jeffreys’ scale still favors the standard $\Lambda$CDM model over the 1-Node scenario. Moreover, as illustrated in Figure~\ref{fig:distances_unions}, the 1-Node reconstruction reproduces the predictions of $\Lambda$CDM almost identically across all cosmological distances and the EoS. This behavior further suggests that the generalized HDE model effectively reduces to $\Lambda$CDM rather than exhibiting distinct BHDE features.

The advantages of the nodal reconstruction approach become particularly evident in the 2-Node HDE model, which produces a significant improvement in the fit quality with {p-value=0.013} compared to $\Lambda$CDM. In this case, the amplitude parameters $f_1$ and $f_2$, are correlated and deviate significantly from zero, with posterior distributions consistent with the dynamic HDE scenario, as shown in Figures~\ref{fig:triangle_both} and \ref{fig:distances_unions}. The preference is especially pronounced for the second amplitude, with a best-fit value of ${f_2=\rm{-}1.76_{-0.47}^{+0.28}}$. 
These parameters drive the evolution of the EoS, as illustrated by the green contours in Figure~\ref{fig:distances_unions}, where $\omega_{\rm de}(z)$ exhibits quintessence-like behavior to this day and crosses into the phantom regime at $z \approx 0.25$. However, at higher redshifts, the EoS drops to increasingly negative values and becomes poorly constrained, leading to a rapid growth of $\rho_{\rm de}$, marking a distinct departure from the previous cases. 
Despite these improvements, the LRG data points remain poorly fitted, consistent with the behavior seen in the 0- and 1-Node reconstructions. In general, the flexibility of the 2-Node model allows it to capture features in the data that the simpler $\Lambda$CDM framework cannot accommodate.

The 3-Node reconstruction yields the best statistical performance among all models considered, with an improvement of { p-value=0.0014} over the $\Lambda$CDM scenario. As shown in Figure~\ref{fig:triangle_both}, the amplitude parameters gradually shift away from the $\Lambda$CDM limit and the approach values characteristic of the standard HDE model. For example, the best-fit values are ${f_1=\rm{-} 1.71 _{-0.54}^{+0.32}}$, ${f_2 = \rm{-}1.91_{-0.56}^{+0.30}}$, and ${f_3 = \rm{-} 2.51_{-0.79}^{+0.48}}$ quoted at  68\% c.l., with $f_3$ being the least constrained due to its association with the highest redshift node.  
The corresponding reconstruction of cosmological distances is illustrated in Figure~\ref{fig:distances_unions} (blue contours), where the EoS undergoes a phase transition from quintessence-like to phantom-like behavior at $z \sim 0.25$. However, as in the 2-Node case, the constraints weaken at higher redshifts. Similar signatures of such transitions in the DE EoS have been discussed in~\cite{Escamilla_2023} for alternative reconstruction techniques, and more recently in~\cite{berti2025reconstructingdarkenergydensity} using DESI BAO data to reconstruct the DE density. 
The improvement in the statistic $\chi^2$ is driven by the greater agreement of the 3-Node model with the observational data, particularly $D_{\rm{M}}(z)/[z\,D_{\rm{H}}(z)]$ and \(D_{\rm v}(z)/(r_{\rm d} z^{2/3})\), as illustrated in Figure \ref{fig:distances_unions}. {Moreover, according to Jeffrey's scale, We find significant support for 3-node case over $\Lambda$CDM, consistent with the improved $\chi^2$ and p-value}.

Although Bayesian evidence improves for models with a higher number of nodes, some notable trends emerge. The holographic parameter $c$ is only well constrained in the 0- and 1-Node cases, while in higher-order models, its posterior saturates the prior range. Furthermore, models with more nodes tend to favor slightly higher values of both $\Omega_m$ and $h$, although the results remain compatible within the uncertainties.

For the CPL parameterization, we have found in addition to the cosmological values reported in Table~\ref{tab:results}, that ${w_0 =-0.704 \pm 0.13}$ and ${w_a = -2.05 \pm 0.716}$, which are in agreement with the values previously found in \cite{Malekjani:2024bgi} for a similar case.

\subsection{PantheonPlus}

In a manner analogous to the previous section, for DESI+PPS the 1D and 2D posterior distributions are shown in the right panel of Figure~\ref{fig:triangle_both}, and the values of the corresponding best fit parameters are listed in the bottom panel of Table~\ref{tab:results}.  For the 0-Node case, the holographic parameter is constrained to ${c = 0.73^{+0.05}_{-0.04}}$ at the $1\sigma$ level, again consistent with previous results \cite{li2024revisitingholographicdarkenergy} for this combination of datasets (excluding SH0ES). As in the + Union3 dataset, the standard HDE model performs worse than $\Lambda$CDM, exhibiting an increase of {p-value of 1} (see bottom panel of Table~\ref{tab:results}). This further supports the finding that the standard HDE model alone is insufficient to explain the data in a full way.

The functional posterior probabilities for the DESI observables and the DE EoS parameter $\omega_{\rm de}$ in this case are shown in Figure~\ref{fig:distances_pantheon}. The 0-Node reconstruction follows a similar trend as observed in the previous section. In turn, the 1-Node case also yields results consistent with the +Union3 dataset. The lower limit for the amplitude value ${f_1> -0.15 }$,  lies remarkably close to the cosmological constant case, with a reconstructed $\omega_{\rm de}$ shown in Figure~\ref{fig:distances_pantheon} ({second row}), which at different confidence levels is compatible with a constant EoS parameter whose value lies in the quintessence regime, suggesting minimal deviations from $\Lambda$CDM. 
Building on the discussion of the 2-Node results from the earlier Union3 dataset combination, we find essentially the same constraints on both cosmological and holographic parameters (see Table~\ref{tab:results}, Figures~\ref{fig:triangle_both}, \ref{fig:distances_unions}, \ref{fig:distances_pantheon}). The {p-value} and Bayesian evidence provide {significant} support for the 2-Node HDE model over $\Lambda$CDM~[\ref{tab:results}]. It is worth remarking that allowing the exponent $f(z)$ to vary with redshift results in a straight line, whose amplitudes remain closer to the standard HDE ($f = -2$) than to $\Lambda$CDM, which in turn changes the evolution of the reconstructed EoS with respect to the 0 and 1 node models (see Figure \ref{fig:distances_pantheon}). In this case, $\omega_{\rm de}$ stays in the quintessence regime only at very low redshifts before transitioning into the phantom regime. In particular, most of the DESI data points and their error bars now overlap with the reconstructed distances, indicating improved consistency with the observations.

For the three-node reconstruction, using the DESI+PPS dataset, our analysis indicates a clear statistical preference for the HDE model over the standard $\Lambda$CDM framework. As illustrated in Figure~\ref{fig:triangle_both}, this result is consistent with the trend observed in the DESI+SH0ES+Union3 dataset combination. The amplitude parameters, constrained at the $68\%$ confidence level, are:$
 {- 1.77_{-0.53    }^{+0.31} \quad f_2 =  - 1.94_{-0.55}^{+0.28}, \quad f_3 =  - 2.45_{-0.78}^{+0.48}}.
$
The cosmological implications of this HDE parameterization are further demonstrated in Figure~\ref{fig:distances_pantheon} (orange contours), where the reconstructed EoS exhibits a notable transition from quintessence-like behavior to phantom-like behavior at redshifts $z < 1$. Although constraints become weaker at higher redshifts ($z > 1$), this dynamical behavior is supported by independent DE reconstruction studies~\cite{berti2025reconstructingdarkenergydensity} and recent observational findings~\cite{Ormondroyd}. Quantitatively, the three-node HDE model shows a significantly better fit than $\Lambda$CDM, with a {p-value of $0.002$} in favor of the nodal-HDE scenario. {However, when the models are assessed using Bayesian evidence rather than solely through their $\chi^{2}$ and  p-values, we find that our model has statistically support in terms of overall preference when compared with $\Lambda$CDM. This outcome persists despite the fact that our model achieves a noticeably improved goodness-of-fit, as reflected in its lower $\chi^{2}$ and p-value. The Bayesian evidence naturally incorporates an Occam’s razor penalty for additional degrees of freedom, and therefore counterbalances any improvement in fit by accounting for the increased model complexity. This difference between a better p-value and the penalization of extra parameters is a well-known feature of cosmological model comparison. It signifies that an improved fit alone does not guaranty stronger model support when evaluated under a full Bayesian framework, which weighs both fit quality and model complexity simultaneously}.

In the same way as in the previous case, for the CPL parameterization, for this dataset we found ${w_0 =-0.806 \pm 0.0816}$ and ${w_a = -1.523 \pm 0.503}$ with a similar $\chi^2$ value to the 2-Node case.

Furthermore, to provide a more comprehensive test of a dynamical dark energy scenario, we also present the results for the same data set using the CPL parameterization, as shown in Table~\ref{tab:results}. 
Consistent with the findings of the DESI~\cite{DESI:2024mwx}, the CPL model outperforms the standard $\Lambda$CDM framework, a trend that is also supported by our analysis. Additionally, we observe that the 0- and 1-node cases perform worse than the CPL model, further indicating that the standard HDE scenario is not favored by current cosmological observations. In contrast, the 2-node reconstruction yields results comparable to the CPL parameterization, while the 3-node case provides an even better fit than CPL.

\begin{figure}[!htbp]
    \centering
    \includegraphics[ width=0.49 \linewidth]{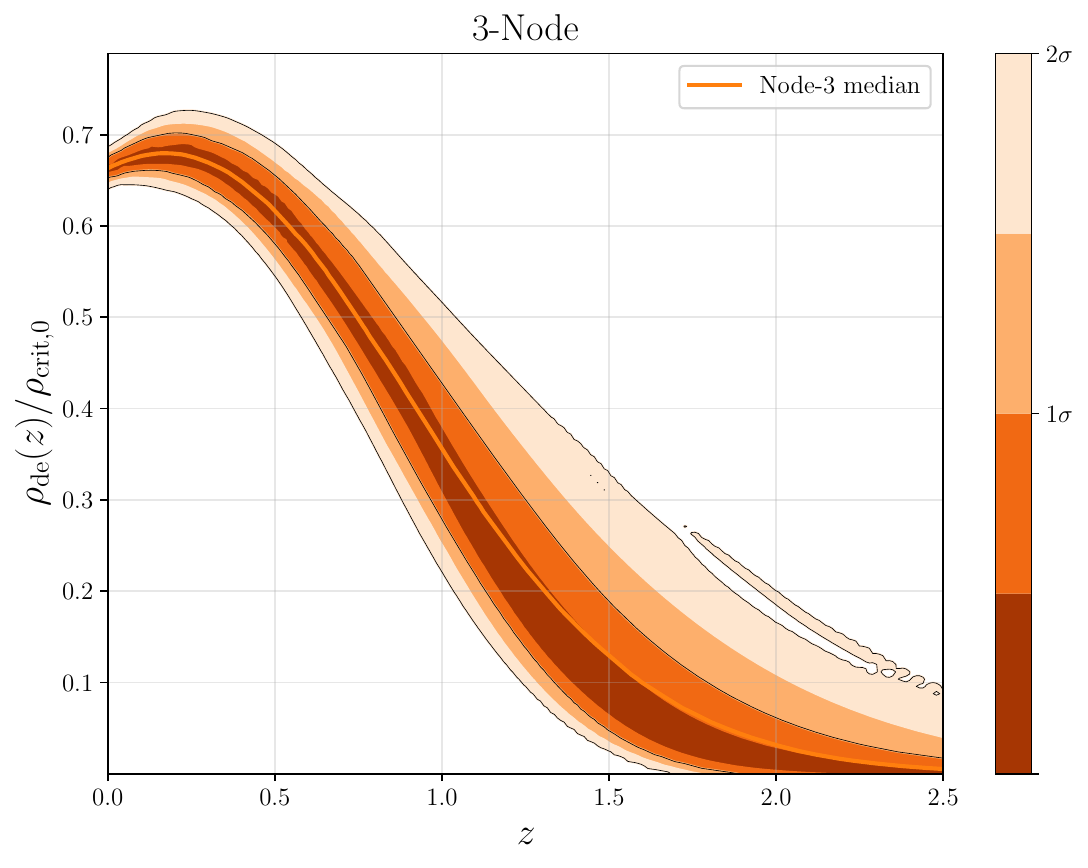}
    \includegraphics[width=0.49\linewidth]{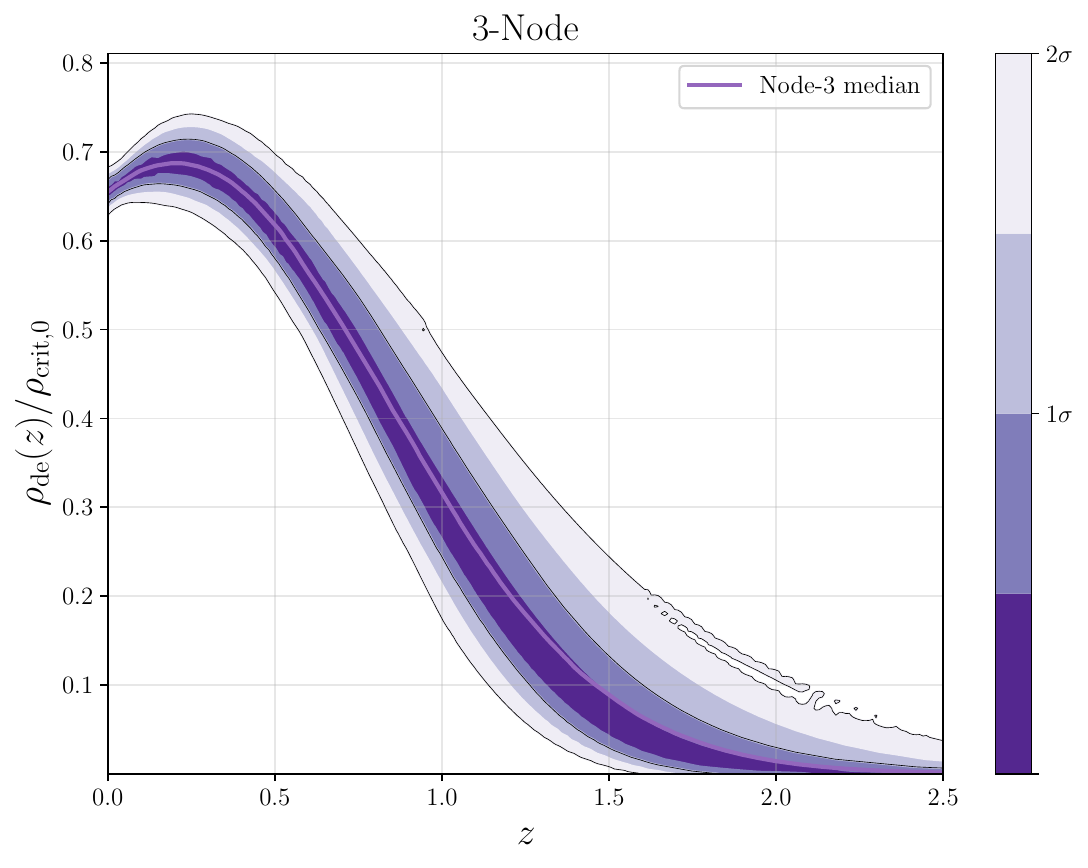}
    \captionsetup{singlelinecheck=off, justification=raggedright} 
    \caption{Functional posterior energy density ratio for the 3-Node case, for DESI+SH0ES+Union3 (purple contours) and DESI+PPS (orange contours).}
    \label{fig:function_all}
\end{figure}

Finally, in Figure~\ref{fig:function_all}, we present the corresponding evolution of DE density for the case of 3-nodes. The plots illustrate the dynamic behavior of the reconstructed function and its deviations from $\Lambda$CDM. Here, at low redshifts, the density remains nearly constant, but becomes dynamically evolving at higher redshifts for the Pantheon+ dataset. {A similar feature is observed in the DESI DR1 release [see figure 4 of \cite{DESI:2024aqx}] which is consistent with our findings.} The dynamical evolution is even more pronounced for the Union3 dataset, plateauing near by $\rho(z)\sim0$, by constructions, but resembling some behaviors previously found \cite{Akarsu:2019hmw, Escamilla_2023}.

\textcolor{black}{Moreover, we stress that the comparatively high value of $H_0$ inferred in our main analysis is largely driven by the inclusion of local distance ladder probes. For completeness, we provide a preliminary exploration of early time constraints in Appendix~\ref{sec:appendix}. }
\section{Conclusions} \label{sec:conclusion}

The holographic dark energy (HDE) model, rooted in the holographic principle, proposes that the universe's DE originates from quantum vacuum fluctuations constrained by horizon-scale entropy bounds. By identifying the energy density as $\rho_{\rm{de}} = 3c^2 M_{\rm{p}}^2 L^{-2}$ and choosing the future event horizon as the infrared cut-off scale $L$, the model successfully accounts for the universe's late-time acceleration. Although earlier proposals such as the Hubble radius or particle horizon failed to produce the observed acceleration, the future event horizon provides a consistent framework.
Building on this foundation, recent generalizations of the HDE model modify the entropy–area relation by introducing a redshift-dependent function $f(z)$ into the entropy exponent. This leads to a dynamic DE density $\rho_{\rm{de}} \propto L^{f(z)}$, encompassing standard HDE, $\Lambda$CDM, Barrow, and Tsallis models as special cases.

In this work, we adopt a robust nodal reconstruction approach that does not assume a specific parametric form of $f(z)$, but instead reconstructs it directly from the data using a piecewise-spline method. This adds significant flexibility to the analysis, allowing us to explore a broader class of models directly constrained by observations. The resulting framework captures potential deviations from the cosmological constant and offers a powerful tool for testing the nature of DE beyond standard assumptions.

We present a comprehensive analysis of generalized HDE models, using the latest cosmological datasets DESI+SH0ES+Union3 and DESI+PPS, which reveal several key insights into the viability of HDE as an alternative to the standard $\Lambda$CDM paradigm and how holographic DE could use a nodal with a reconstruction approach of up to 3 points of the exponent present in the expression \ref{eq:density_general}, which indirectly is a reconstruction of the future event horizon. 

Although theoretically attractive, we have seen that the conventional HDE model, or 0-Node, demonstrates statistically inferior performance compared to $\Lambda$CDM, as evidenced by higher values of {p-value} and Bayesian evidence metrics for both datasets. This suggests that the simplest formulation of HDE struggles to reconcile with current observational constraints. However, introducing nodal reconstructions significantly improves the model's flexibility and agreement with the data.

In contrast, for the case of 1-Node (comparable to Barrow and Tsallis HDE), we observe an {insignificant} improvement in fit quality ({p-value=0.75}), with the amplitude parameters tending toward $\Lambda$CDM-like behavior. This indicates that minimal extensions of HDE can partially accommodate observational data without substantially deviating from the standard model.

The most compelling results emerge in the 2- and 3-Node reconstructions. The 2-Node HDE model shows {significant} evidence over $\Lambda$CDM ({p-value=0.0101} {and $\Delta {\rm ln} \mathcal{Z}=-1.16$}), with amplitude parameters that deviate significantly from zero and capture DDE features. The reconstructed EoS exhibits a redshift-dependent transition from quintessence-like to phantom-like behavior, and recent studies \cite{DESI:2024mwx,Ormondroyd, abedin2025searchinteractiondarksector} have used different reconstruction methods and model approaches. 

The 3-Node case further enhances the fit {with the p-value of $0.0014$ and $0.002$ for the Union3 and Pantheon+ dataset}, although Bayesian evidence still only {significantly disfavors} $\Lambda$CDM due to increased model complexity. {Even when the 3-Node reconstruction is compared to $\Lambda$CDM but also to the CPL parametrization, it shows comparable performance, as reflected by their similar p-values across both datasets. Another statistically relevant observation from Table~\ref{tab:results} is that, for  both dataset combination, the $\chi^2$ values of the CPL parametrization and the 2-Node reconstruction are also closely aligned.}

Finally, we can see that in both cases of SNe Ia, there is a trend for $\Omega_{m}$ and $H_0$ from the cases of 0 to 3 Nodes. In this way, the last case agrees more with the value by SH0ES for $H_0$  (${71.60 \pm 0.70\,\rm{km/s/Mpc}}$ \cite{Riess:2021jrx}), which could be due to the 3-Node case being the most flexible case of them. In fact, the differences for $\Lambda$CDM and SH0ES are ${2.33 \, \sigma}$ and ${2.41\, \sigma}$ for the DESI+SH0ES+Union3 and the DESI+PPS, respectively, while using the nodal reconstruction in HDE we found ${1.12 \, \sigma}$ and ${1.11
\, \sigma }$, thus showing an improvement of this parameter. This trend could be useful in alleviating the Hubble tension, but for this, it is necessary to utilize the CMB data, which is considered in future work.

Notably, all multi-node reconstructions (except the 1-Node) satisfy fundamental thermodynamic requirements, with monotonically increasing amplitude functions preserving entropy bounds and density positivity of the DE for late times. Although when we start to consider 3 nodes, the reconstructed function becomes more negative, and there is a trend to $f(z)<-3$ for high redshifts, in that regime the energy density for DE could present a super decreasing behavior and then the entropy could become a decreasing function, as we can see in Figure~\ref{fig:function_all}. This is due to $\rho_{\rm{de}} \propto S_{\rm G}$, which may appear to be a violation of the second law of thermodynamics, this result is only for the DE component, and it is necessary to analyze the whole system, something similar to the study~\cite{Saridakis:2020cqq}, where the authors apply the Barrow's entropy to the evolution of the apparent horizon.

We may conclude that the standard HDE framework is facing serious problems when using current observations. However, the nodal reconstruction approach, particularly the 2-Node and 3-Node cases, shows promising consistency with the data by presenting an antagonistic behavior on its EoS, and also offering a compelling avenue to explore beyond $\Lambda$CDM physics and possible corrections to the original HDE model. These results suggest that flexible, data-driven reconstructions can better capture the underlying dynamics of DE. Looking ahead, future datasets and theoretical developments will be crucial to assess whether these HDE extensions can provide a robust alternative to the cosmological constant paradigm. In particular, the objective was to test the impact of DESI using extended nodal reconstructions and alternative methods to probe unresolved issues such as Hubble tension. Moreover, while this work relies primarily on local observations, future efforts will incorporate radiation and neutrino physics to enable joint analyzes with Planck and full-shape large-scale structure data, allowing a deeper exploration of neutrino and $S_8$ tensions~\cite{DESI:2025ejh}. Furthermore, upcoming data sets from LSST~\cite{Kumar:2024soe} and gamma-ray bursts~\cite{Adil:2024miw} may offer further avenues to validate our findings across different SNe Ia compilations.

\section{Data and Code Availability}
The complete codebase and data produced during our analysis are publicly available at \href{https://github.com/mazapataa/HDENodalRec}{HDENodalRec}.

\section*{Acknowledgment}
S.A.A. acknowledges the support of the DGAPA postdoctoral fellowship program at ICF-UNAM, Mexico. S.A.A. also acknowledges the use of High Performance Computing cluster Pegasus at IUCAA, Pune, India. M.A.Z. and G.G.-A. acknowledge the support of the SECIHTI. J.A.V. acknowledges support from FOSEC SEP-CONACYT Ciencia B\'asica A1-S-21925,  UNAM-DGAPA-PAPIIT IN117723, IN110325 and Cátedra de Investigación Marcos Moshinsky. Special thanks to Ing. Francisco Bustos and Lic. Reyes García who assisted considerably with the High Performance Computing at the ICF-UNAM.

\section{Appendix} \label{sec:appendix}

\begin{figure*}
    \centering
    \includegraphics[width=0.85\linewidth]{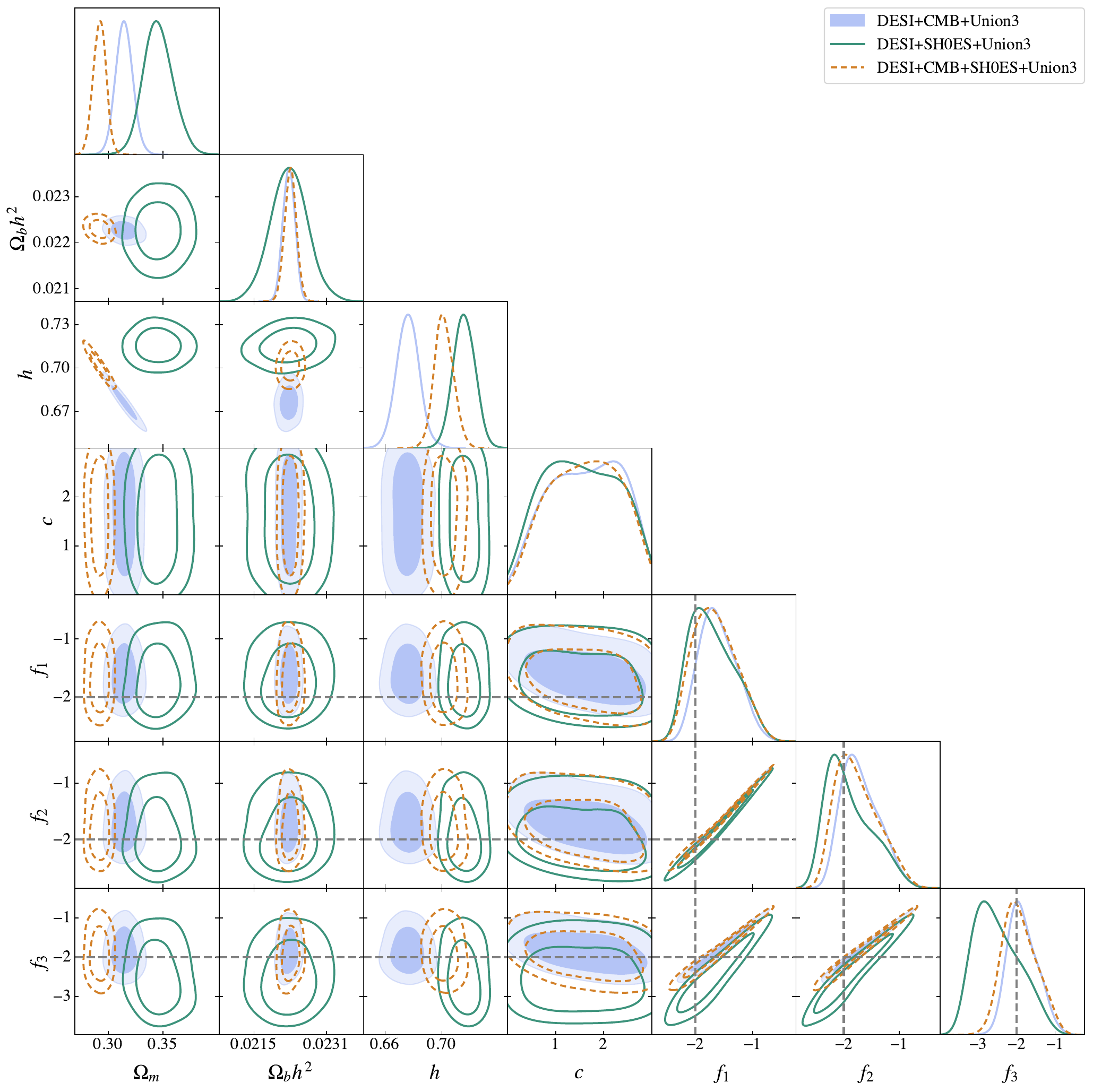}
    \caption{Triangle plot for marginalized posterior distributions using the DESI+CMB+Union3, DESI+SH0ES+Union3 and DESI+CMB+SH0ES+Union3 data combination at 1-$\sigma$ and 2-$\sigma$ confidence level. The dashed lines represent the case of standard HDE, i.e. the case with all nodes fixed to $f_i = -2$.}
    \label{fig:appendix}
\end{figure*}

\textcolor{black}{In this Appendix, we present a supplementary analysis aimed at clarifying the origin of the comparatively high values of the Hubble constant $H_0$ obtained in our nodal reconstruction of the holographic dark-energy scenario when only late-time distance probes are used. In Fig.~\ref{fig:appendix}, we show the posteriors for the three–node case and infer that the inclusion of CMB information substantially reduces the preferred value of $H_0$, bringing it into close agreement with the Planck determination within the standard cosmological framework~\cite{Planck:2018vyg}. This behavior indicates that the elevated $H_0$ values inferred in the absence of CMB data are primarily driven by the local-distance calibration entering through the \texttt{SH0ES} supernova sample, rather than reflecting an intrinsic prediction of the holographic nodal reconstruction itself. For the \texttt{DESI+CMB+Union3} data combination in the three–node scenario, we obtain the marginalized constraints $\Omega_m = 0.315 \pm 0.0079$, $h = 0.676 \pm 0.0081$, $c = 1.64_{-0.75}^{+0.91}$, $f_1 = -1.60_{-0.38}^{+0.29}$, $f_2 = -1.71_{-0.41}^{0.28}$, $f_3 = -1.84_{-0.45}^{+0.33} $.}

\textcolor{black}{In both the cases as shown in the two dimensional posteriors, the addition of CMB data drives the inferred Hubble constant toward the Planck-preferred range, while simultaneously tightening the constraints on the matter density and on the nodal amplitudes that parametrize departures from the reference holographic evolution. This supports the interpretation that background-only analyses including local distance-ladder information can artificially favor higher $H_0$ values, whereas CMB measurements impose a strong early-time anchor that propagates to late-time parameters.}

\textcolor{black}{At present, our treatment of the CMB information is restricted to a background-level implementation of compressed likelihood mentioned in the Appendix \cite{DESI:2025zgx}. In future work, we plan to extend this analysis by incorporating perturbation-level observables and the full Planck likelihood, including temperature, polarization, and lensing data, in order to perform a fully self-consistent assessment of the holographic nodal framework. Such an extension will allow us to test whether these models can simultaneously accommodate early- and late-Universe probes and to quantify their impact on cosmological tensions in a more rigorous manner. }

\bibliography{Bibliography}

@article{Trotta_2008,
   title={Bayes in the sky: Bayesian inference and model selection in cosmology},
   volume={49},
   ISSN={1366-5812},
   url={http://dx.doi.org/10.1080/00107510802066753},
   DOI={10.1080/00107510802066753},
   number={2},
   journal={Contemporary Physics},
   publisher={Informa UK Limited},
   author={Trotta, Roberto},
   year={2008},
   month=mar, pages={71–104} }

@article{TAVAYEF2018195,
title = {Tsallis holographic dark energy},
journal = {Physics Letters B},
volume = {781},
pages = {195-200},
year = {2018},
issn = {0370-2693},
doi = {https://doi.org/10.1016/j.physletb.2018.04.001},
url = {https://www.sciencedirect.com/science/article/pii/S0370269318302843},
author = {M. Tavayef and A. Sheykhi and Kazuharu Bamba and H. Moradpour}
}

@misc{Fischler:1998st,
    author = "Fischler, W. and Susskind, Leonard",
    title = "{Holography and cosmology}",
    eprint = "hep-th/9806039",
    archivePrefix = "arXiv",
    reportNumber = "SU-ITP-98-39A, UTTG-06-98",
    month = "6",
    year = "1998"
}

@article{Oliveros2022,
author={Oliveros, A.
and Sabogal, M. A.
and Acero, Mario A.},
title={Barrow holographic dark energy with Granda--Oliveros cutoff},
journal={The European Physical Journal Plus},
year={2022},
month={Jul},
day={07},
volume={137},
number={7},
pages={783},
issn={2190-5444},
doi={10.1140/epjp/s13360-022-02994-z},
url={https://doi.org/10.1140/epjp/s13360-022-02994-z}
}

@article{CHANG200614,
title = {Constraints on holographic dark energy from X-ray gas mass fraction of galaxy clusters},
journal = {Physics Letters B},
volume = {633},
number = {1},
pages = {14-18},
year = {2006},
issn = {0370-2693},
doi = {https://doi.org/10.1016/j.physletb.2005.10.095},
url = {https://www.sciencedirect.com/science/article/pii/S0370269305017120},
author = {Zhe Chang and Feng-Quan Wu and Xin Zhang}
}

@article{hdecurv,
    author = "Huang, Qing-Guo and Li, Miao",
    title = "{The Holographic dark energy in a non-flat universe}",
    eprint = "astro-ph/0404229",
    archivePrefix = "arXiv",
    doi = "10.1088/1475-7516/2004/08/013",
    journal = "JCAP",
    volume = "08",
    pages = "013",
    year = "2004"
}

@article{zhang,
  title={Revisit of the interaction between holographic dark energy and dark matter},
  author={Zhang, Zhenhui and Li, Song and Li, Xiao-Dong and Zhang, Xin and Li, Miao},
  journal={Journal of Cosmology and Astroparticle Physics},
  volume={2012},
  number={06},
  pages={009},
  year={2012},
  publisher={IOP Publishing}
}

@article{Li_2013,
   title={Holographic dark energy in a universe with spatial curvature and massive neutrinos: a full Markov Chain Monte Carlo exploration},
   volume={2013},
   ISSN={1475-7516},
   url={http://dx.doi.org/10.1088/1475-7516/2013/02/033},
   DOI={10.1088/1475-7516/2013/02/033},
   number={02},
   journal={Journal of Cosmology and Astroparticle Physics},
   publisher={IOP Publishing},
   author={Li, Yun-He and Wang, Shuang and Li, Xiao-Dong and Zhang, Xin},
   year={2013},
   month=feb, pages={033–033} }

@article{Jamil:2009sq,
    author = "Jamil, Mubasher and Saridakis, Emmanuel N. and Setare, M. R.",
    title = "{Holographic dark energy with varying gravitational constant}",
    eprint = "0906.2847",
    archivePrefix = "arXiv",
    primaryClass = "hep-th",
    doi = "10.1016/j.physletb.2009.07.048",
    journal = "Phys. Lett. B",
    volume = "679",
    pages = "172--176",
    year = "2009"
}

@article{chen,
author = {Chen, Bin and Li, Miao and Wang, Yi},
year = {2007},
month = {07},
pages = {256-267},
title = {Inflation with Holographic Dark Energy},
volume = {774},
journal = {Nuclear Physics B},
doi = {10.1016/j.nuclphysb.2007.04.007}
}

@article{Lu_2010,
   title={Observational constraints on holographic dark energy with varying gravitational constant},
   volume={2010},
   ISSN={1475-7516},
   url={http://dx.doi.org/10.1088/1475-7516/2010/03/031},
   DOI={10.1088/1475-7516/2010/03/031},
   number={03},
   journal={Journal of Cosmology and Astroparticle Physics},
   publisher={IOP Publishing},
   author={Lu, Jianbo and Saridakis, Emmanuel N and Setare, M.R and Xu, Lixin},
   year={2010},
   month=mar, pages={031–031} }

@article{Guberina_2005,
   title={Hint for quintessence-like scalars from holographic dark energy},
   volume={2005},
   ISSN={1475-7516},
   url={http://dx.doi.org/10.1088/1475-7516/2005/05/001},
   DOI={10.1088/1475-7516/2005/05/001},
   number={05},
   journal={Journal of Cosmology and Astroparticle Physics},
   publisher={IOP Publishing},
   author={Guberina, B and Horvat, R and Štefančić, H},
   year={2005},
   month=may, pages={001–001} }

@article{Gong_2004,
    author = "Gong, Yun-gui",
    title = "{Extended holographic dark energy}",
    eprint = "hep-th/0404030",
    archivePrefix = "arXiv",
    doi = "10.1103/PhysRevD.70.064029",
    journal = "Phys. Rev. D",
    volume = "70",
    pages = "064029",
    year = "2004"
}

@article{Li_2008,
   title={Some issues concerning holographic dark energy},
   volume={2008},
   ISSN={1475-7516},
   url={http://dx.doi.org/10.1088/1475-7516/2008/05/023},
   DOI={10.1088/1475-7516/2008/05/023},
   number={05},
   journal={Journal of Cosmology and Astroparticle Physics},
   publisher={IOP Publishing},
   author={Li, Miao and Lin, Chunshan and Wang, Yi},
   year={2008},
   month=may, pages={023} }

@article{PhysRevD.74.103505,
  title = {Dynamical vacuum energy, holographic quintom, and the reconstruction of scalar-field dark energy},
  author = {Zhang, Xin},
  journal = {Phys. Rev. D},
  volume = {74},
  issue = {10},
  pages = {103505},
  numpages = {7},
  year = {2006},
  month = {Nov},
  publisher = {American Physical Society},
  doi = {10.1103/PhysRevD.74.103505},
  url = {https://link.aps.org/doi/10.1103/PhysRevD.74.103505}
}

@article{articlecohen,
  title = {Effective Field Theory, Black Holes, and the Cosmological Constant},
  author = {Cohen, Andrew G. and Kaplan, David B. and Nelson, Ann E.},
  journal = {Phys. Rev. Lett.},
  volume = {82},
  issue = {25},
  pages = {4971--4974},
  numpages = {0},
  year = {1999},
  month = {Jun},
  publisher = {American Physical Society},
  doi = {10.1103/PhysRevLett.82.4971},
  url = {https://link.aps.org/doi/10.1103/PhysRevLett.82.4971}
}

@article{Brout:2022vxf,
    author = "Brout, Dillon and others",
    title = "{The Pantheon+ Analysis: Cosmological Constraints}",
    eprint = "2202.04077",
    archivePrefix = "arXiv",
    primaryClass = "astro-ph.CO",
    doi = "10.3847/1538-4357/ac8e04",
    journal = "Astrophys. J.",
    volume = "938",
    number = "2",
    pages = "110",
    year = "2022"
}

@article{cooray1999gravitational,
  title={Gravitational lensing as a probe of quintessence},
  author={Cooray, Asantha R and Huterer, Dragan},
  journal={The Astrophysical Journal},
  volume={513},
  number={2},
  pages={L95},
  year={1999},
  publisher={IOP Publishing}
}

@article{ASTIER20018,
title = {Can luminosity distance measurements probe the equation of state of dark energy?},
journal = {Physics Letters B},
volume = {500},
number = {1},
pages = {8-15},
year = {2001},
issn = {0370-2693},
doi = {https://doi.org/10.1016/S0370-2693(01)00072-7},
url = {https://www.sciencedirect.com/science/article/pii/S0370269301000727},
author = {P. Astier}
}

@article{quin1,
  title = {Cosmological Imprint of an Energy Component with General Equation of State},
  author = {Caldwell, R. R. and Dave, Rahul and Steinhardt, Paul J.},
  journal = {Phys. Rev. Lett.},
  volume = {80},
  issue = {8},
  pages = {1582--1585},
  numpages = {0},
  year = {1998},
  month = {Feb},
  publisher = {American Physical Society},
  doi = {10.1103/PhysRevLett.80.1582},
  url = {https://link.aps.org/doi/10.1103/PhysRevLett.80.1582}
}

@article{quin2,
  title = {Cosmological tracking solutions},
  author = {Steinhardt, Paul J. and Wang, Limin and Zlatev, Ivaylo},
  journal = {Phys. Rev. D},
  volume = {59},
  issue = {12},
  pages = {123504},
  numpages = {13},
  year = {1999},
  month = {May},
  publisher = {American Physical Society},
  doi = {10.1103/PhysRevD.59.123504},
  url = {https://link.aps.org/doi/10.1103/PhysRevD.59.123504}
}

@article{quin3,
  title = {Kinetically driven quintessence},
  author = {Chiba, Takeshi and Okabe, Takahiro and Yamaguchi, Masahide},
  journal = {Phys. Rev. D},
  volume = {62},
  issue = {2},
  pages = {023511},
  numpages = {8},
  year = {2000},
  month = {Jun},
  publisher = {American Physical Society},
  doi = {10.1103/PhysRevD.62.023511},
  url = {https://link.aps.org/doi/10.1103/PhysRevD.62.023511}
}

@article{phant1,
  author = {Caldwell, Robert R.},
  title = {A Phantom Menace? Cosmological consequences of a dark energy component with super-negative equation of state},
  journal = {Physics Letters B},
  volume = {545},
  number = {1-2},
  pages = {23--29},
  year = {2002},
  doi = {10.1016/S0370-2693(02)02589-3}
}

@article{phant2,
   title={Phantom Energy: Dark Energy with w < -1 Causes a Cosmic Doomsday},
   volume={91},
   ISSN={1079-7114},
   url={http://dx.doi.org/10.1103/PhysRevLett.91.071301},
   DOI={10.1103/physrevlett.91.071301},
   number={7},
   journal={Physical Review Letters},
   publisher={American Physical Society (APS)},
   author={Caldwell, Robert R. and Kamionkowski, Marc and Weinberg, Nevin N.},
   year={2003},
   month={08} }

@article{quintom1,
   title={Dynamical system analysis of quintom dark energy model},
   volume={78},
   ISSN={1434-6052},
   url={http://dx.doi.org/10.1140/epjc/s10052-018-6405-9},
   DOI={10.1140/epjc/s10052-018-6405-9},
   number={11},
   journal={The European Physical Journal C},
   publisher={Springer Science and Business Media LLC},
   author={Mishra, Sudip and Chakraborty, Subenoy},
   year={2018},
   month=nov }

@article{kessense1,
doi = {10.1088/1475-7516/2016/07/050},
url = {https://dx.doi.org/10.1088/1475-7516/2016/07/050},
year = {2016},
month = {jul},
publisher = {},
volume = {2016},
number = {07},
pages = {050},
author = {Bouhmadi-López, Mariam and Kumar, K. Sravan and Marto, João and Morais, João and Zhuk, Alexander},
title = {K-essence model from the mechanical approach point of view: coupled scalar field and the late cosmic acceleration},
journal = {Journal of Cosmology and Astroparticle Physics}
}

@article{kessense2,
  title = {Unified model of $k$-inflation, dark matter, and dark energy},
  author = {Bose, Nilok and Majumdar, A. S.},
  journal = {Phys. Rev. D},
  volume = {80},
  issue = {10},
  pages = {103508},
  numpages = {7},
  year = {2009},
  month = {Nov},
  publisher = {American Physical Society},
  doi = {10.1103/PhysRevD.80.103508},
  url = {https://link.aps.org/doi/10.1103/PhysRevD.80.103508}
}

@article{kessense3,
   title={Constraints on the speed of sound in the k-essence model of dark energy},
   volume={84},
   ISSN={1434-6052},
   url={http://dx.doi.org/10.1140/epjc/s10052-024-12547-6},
   DOI={10.1140/epjc/s10052-024-12547-6},
   number={2},
   journal={The European Physical Journal C},
   publisher={Springer Science and Business Media LLC},
   author={Dinda, Bikash R. and Banerjee, Narayan},
   year={2024},
   month=feb }

@article{early1,
  title = {Oscillating scalar fields and the Hubble tension: A resolution with novel signatures},
  author = {Smith, Tristan L. and Poulin, Vivian and Amin, Mustafa A.},
  journal = {Phys. Rev. D},
  volume = {101},
  issue = {6},
  pages = {063523},
  numpages = {25},
  year = {2020},
  month = {Mar},
  publisher = {American Physical Society},
  doi = {10.1103/PhysRevD.101.063523},
  url = {https://link.aps.org/doi/10.1103/PhysRevD.101.063523}
}

@article{early2,
  title = {Early Dark Energy can Resolve the Hubble Tension},
  author = {Poulin, Vivian and Smith, Tristan L. and Karwal, Tanvi and Kamionkowski, Marc},
  journal = {Phys. Rev. Lett.},
  volume = {122},
  issue = {22},
  pages = {221301},
  numpages = {7},
  year = {2019},
  month = {Jun},
  publisher = {American Physical Society},
  doi = {10.1103/PhysRevLett.122.221301},
  url = {https://link.aps.org/doi/10.1103/PhysRevLett.122.221301}
}

@misc{early3,
      title={Early Dark Energy Effects on the 21cm Signal}, 
      author={Tal Adi and Jordan Flitter and Ely D. Kovetz},
      year={2024},
      eprint={2410.22424},
      archivePrefix={arXiv},
      primaryClass={astro-ph.CO},
      url={https://arxiv.org/abs/2410.22424}, 
}

@misc{early4,
      title={Towards alleviating the $H_0$ and $S_8$ tensions with Early Dark Energy - Dark Matter drag}, 
      author={Théo Simon and Tal Adi and José Luis Bernal and Ely D. Kovetz and Vivian Poulin and Tristan L. Smith},
      year={2024},
      eprint={2410.21459},
      archivePrefix={arXiv},
      primaryClass={astro-ph.CO},
      url={https://arxiv.org/abs/2410.21459}, 
}

@article{escamilla_2025,
       author = {{Escamilla}, Luis A. and {Pan}, Supriya and {Di Valentino}, Eleonora and {Paliathanasis}, Andronikos and {V{\'a}zquez}, Jos{\'e} Alberto and {Yang}, Weiqiang},
        title = "{Testing an oscillatory behavior of dark energy}",
      journal = {\prd},
         year = 2025,
        month = jan,
       volume = {111},
       number = {2},
          eid = {023531},
        pages = {023531},
          doi = {10.1103/PhysRevD.111.023531},
archivePrefix = {arXiv},
       eprint = {2404.00181},
 primaryClass = {astro-ph.CO},
       adsurl = {https://ui.adsabs.harvard.edu/abs/2025PhRvD.111b3531E}
}

@article{Maldacena_1999,
    author = "Maldacena, Juan Martin",
    title = "{The Large N limit of superconformal field theories and supergravity}",
    eprint = "hep-th/9711200",
    archivePrefix = "arXiv",
    reportNumber = "HUTP-97-A097, HUTP-98-A097",
    doi = "10.4310/ATMP.1998.v2.n2.a1",
    journal = "Adv. Theor. Math. Phys.",
    volume = "2",
    pages = "231--252",
    year = "1998"
}

@article{susskind,
    author = {Susskind, Leonard},
    title = {The world as a hologram},
    journal = {Journal of Mathematical Physics},
    volume = {36},
    number = {11},
    pages = {6377-6396},
    year = {1995},
    month = {11},
    issn = {0022-2488},
    doi = {10.1063/1.531249},
    url = {https://doi.org/10.1063/1.531249},
}

@article{Dai_2020,
  title = {Reconciling Hubble constant discrepancy from holographic dark energy},
  author = {Dai, Wei-Ming and Ma, Yin-Zhe and He, Hong-Jian},
  journal = {Phys. Rev. D},
  volume = {102},
  issue = {12},
  pages = {121302},
  numpages = {6},
  year = {2020},
  month = {12},
  publisher = {American Physical Society},
  doi = {10.1103/PhysRevD.102.121302},
  url = {https://link.aps.org/doi/10.1103/PhysRevD.102.121302}
}

@misc{hooft,
      title={Dimensional Reduction in Quantum Gravity}, 
      author={G. 't Hooft},
      year={2009},
      eprint={gr-qc/9310026},
      archivePrefix={arXiv},
      primaryClass={gr-qc},
      url={https://arxiv.org/abs/gr-qc/9310026}, 
}

@article{AlbertoVazquez:2012ofj,
    author = "Alberto Vazquez, J. and Bridges, M. and Hobson, M. P. and Lasenby, A. N.",
    title = "{Reconstruction of the Dark Energy equation of state}",
    eprint = "1205.0847",
    archivePrefix = "arXiv",
    primaryClass = "astro-ph.CO",
    doi = "10.1088/1475-7516/2012/09/020",
    journal = "JCAP",
    volume = "09",
    pages = "020",
    year = "2012"
}

@article{Hee:2016nho,
    author = "Hee, S. and V{\'a}zquez, J. A. and Handley, W. J. and Hobson, M. P. and Lasenby, A. N.",
    title = "{Constraining the dark energy equation of state using Bayes theorem and the Kullback{\textendash}Leibler divergence}",
    eprint = "1607.00270",
    archivePrefix = "arXiv",
    primaryClass = "astro-ph.CO",
    doi = "10.1093/mnras/stw3102",
    journal = "Mon. Not. Roy. Astron. Soc.",
    volume = "466",
    number = "1",
    pages = "369--377",
    year = "2017"
}

@article{Akarsu:2019hmw,
    author = {Akarsu, {\"O}zg{\"u}r and Barrow, John D. and Escamilla, Luis A. and Vazquez, J. Alberto},
    title = "{Graduated dark energy: Observational hints of a spontaneous sign switch in the cosmological constant}",
    eprint = "1912.08751",
    archivePrefix = "arXiv",
    primaryClass = "astro-ph.CO",
    doi = "10.1103/PhysRevD.101.063528",
    journal = "Phys. Rev. D",
    volume = "101",
    number = "6",
    pages = "063528",
    year = "2020"
}

@article{Padilla:2019mgi,
    author = "Padilla, Luis E. and Tellez, Luis O. and Escamilla, Luis A. and Vazquez, Jose Alberto",
    title = "{Cosmological Parameter Inference with Bayesian Statistics}",
    eprint = "1903.11127",
    archivePrefix = "arXiv",
    primaryClass = "astro-ph.CO",
    doi = "10.3390/universe7070213",
    journal = "Universe",
    volume = "7",
    number = "7",
    pages = "213",
    year = "2021"
}

@article{Escamilla:2023shf,
    author = "Escamilla, Luis A. and Akarsu, Ozgur and Di Valentino, Eleonora and Vazquez, J. Alberto",
    title = "{Model-independent reconstruction of the interacting dark energy kernel: Binned and Gaussian process}",
    eprint = "2305.16290",
    archivePrefix = "arXiv",
    primaryClass = "astro-ph.CO",
    doi = "10.1088/1475-7516/2023/11/051",
    journal = "JCAP",
    volume = "11",
    pages = "051",
    year = "2023"
}

@article{Garcia-Arroyo:2024tqq,
    author = "Garcia-Arroyo, Gabriela and Ure{\~n}a-L{\'o}pez, L. Arturo and V{\'a}zquez, J. Alberto",
    title = "{Interacting scalar fields: Dark matter and early dark energy}",
    eprint = "2402.08815",
    archivePrefix = "arXiv",
    primaryClass = "astro-ph.CO",
    doi = "10.1103/PhysRevD.110.023529",
    journal = "Phys. Rev. D",
    volume = "110",
    number = "2",
    pages = "023529",
    year = "2024"
}

@article{Vazquez:2023kyx,
    author = "V{\'a}zquez, J. Alberto and Tamayo, David and Garcia-Arroyo, Gabriela and G{\'o}mez-Vargas, Isidro and Quiros, Israel and Sen, Anjan A.",
    title = "{Coupled multiscalar field dark energy}",
    eprint = "2305.11396",
    archivePrefix = "arXiv",
    primaryClass = "astro-ph.CO",
    doi = "10.1103/PhysRevD.109.023511",
    journal = "Phys. Rev. D",
    volume = "109",
    number = "2",
    pages = "023511",
    year = "2024"
}

@article{Vazquez:2020ani,
    author = "V{\'a}zquez, J. Alberto and Tamayo, David and Sen, Anjan A. and Quiros, Israel",
    title = "{Bayesian model selection on scalar $\epsilon$-field dark energy}",
    eprint = "2009.01904",
    archivePrefix = "arXiv",
    primaryClass = "gr-qc",
    doi = "10.1103/PhysRevD.103.043506",
    journal = "Phys. Rev. D",
    volume = "103",
    number = "4",
    pages = "043506",
    year = "2021"
}

@article{Akarsu:2015yea,
    author = {Akarsu, {\"O}zgur and Dereli, Tekin and Vazquez, J. Alberto},
    title = "{A divergence-free parametrization for dynamical dark energy}",
    eprint = "1501.07598",
    archivePrefix = "arXiv",
    primaryClass = "astro-ph.CO",
    doi = "10.1088/1475-7516/2015/06/049",
    journal = "JCAP",
    volume = "06",
    pages = "049",
    year = "2015"
}

@book{Burnham,
    title = {Model Selection and Multimodel Inference },
    author = {Kenneth P. Burnham, David R. Anderson},
    isbn = {978-0-387-95364-9},
    year = {2004},
    doi = {10.1007/b97636},
    publisher = {Springer New York}
}

@article{Escamilla_2023,
   title={Model selection applied to reconstructions of the Dark Energy},
   volume={83},
   ISSN={1434-6052},
   url={http://dx.doi.org/10.1140/epjc/s10052-023-11404-2},
   DOI={10.1140/epjc/s10052-023-11404-2},
   number={3},
   journal={The European Physical Journal C},
   publisher={Springer Science and Business Media LLC},
   pages = {17},
   author={Escamilla, Luis A. and Vazquez, J. Alberto},
   year={2023},
   month=mar }

@article{quintom2,
   title={Dynamical description of a quintom cosmological model nonminimally coupled with gravity},
   volume={80},
   ISSN={1434-6052},
   url={http://dx.doi.org/10.1140/epjc/s10052-020-08476-9},
   DOI={10.1140/epjc/s10052-020-08476-9},
   number={9},
   journal={The European Physical Journal C},
   publisher={Springer Science and Business Media LLC},
   author={Marciu, Mihai},
   year={2020},
   month=sep
}

@misc{abedin2025searchinteractiondarksector,
      title={In search of an interaction in the dark sector through Gaussian Process and ANN approaches}, 
      author={Mazaharul Abedin and Guo-Jian Wang and Yin-Zhe Ma and Supriya Pan},
      year={2025},
      eprint={2505.04336},
      archivePrefix={arXiv},
      primaryClass={astro-ph.CO},
      url={https://arxiv.org/abs/2505.04336}, 
}

@misc{Ormondroyd,
      title={Nonparametric reconstructions of dynamical dark energy via flexknots}, 
      author={A. N. Ormondroyd and W. J. Handley and M. P. Hobson and A. N. Lasenby},
      year={2025},
      eprint={2503.08658},
      archivePrefix={arXiv},
      primaryClass={astro-ph.CO},
      url={https://arxiv.org/abs/2503.08658}, 
}

@misc{berti2025reconstructingdarkenergydensity,
      title={Reconstructing the dark energy density in light of DESI BAO observations}, 
      author={Maria Berti and Emilio Bellini and Camille Bonvin and Martin Kunz and Matteo Viel and Miguel Zumalacarregui},
      year={2025},
      eprint={2503.13198},
      archivePrefix={arXiv},
      primaryClass={astro-ph.CO},
      url={https://arxiv.org/abs/2503.13198}, 
}

@misc{quintom3,
      title={Resolving Hubble Tension with Quintom Dark Energy Model}, 
      author={Sirachak Panpanich and Piyabut Burikham and Supakchai Ponglertsakul and Lunchakorn Tannukij},
      year={2020},
      eprint={1908.03324},
      archivePrefix={arXiv},
      primaryClass={gr-qc},
      url={https://arxiv.org/abs/1908.03324}, 
}

@article{Barboza_2008,
   title={A parametric model for dark energy},
   volume={666},
   ISSN={0370-2693},
   url={http://dx.doi.org/10.1016/j.physletb.2008.08.012},
   DOI={10.1016/j.physletb.2008.08.012},
   number={5},
   journal={Physics Letters B},
   publisher={Elsevier BV},
   author={Barboza, E.M. and Alcaniz, J.S.},
   year={2008},
   month=sep, pages={415–419} }

@article{jassal,
  title = {Observational constraints on low redshift evolution of dark energy: How consistent are different observations?},
  author = {Jassal, H. K. and Bagla, J. S. and Padmanabhan, T.},
  journal = {Phys. Rev. D},
  volume = {72},
  issue = {10},
  pages = {103503},
  numpages = {21},
  year = {2005},
  month = {Nov},
  publisher = {American Physical Society},
  doi = {10.1103/PhysRevD.72.103503},
  url = {https://link.aps.org/doi/10.1103/PhysRevD.72.103503}
}

@article{Efstathiou:1999tm,
    author = "Efstathiou, G.",
    title = "{Constraining the equation of state of the universe from distant type Ia supernovae and cosmic microwave background anisotropies}",
    eprint = "astro-ph/9904356",
    archivePrefix = "arXiv",
    doi = "10.1046/j.1365-8711.1999.02997.x",
    journal = "Mon. Not. Roy. Astron. Soc.",
    volume = "310",
    pages = "842--850",
    year = "1999"
}

@article{CPL,
  title = {Exploring the Expansion History of the Universe},
  author = {Linder, Eric V.},
  journal = {Phys. Rev. Lett.},
  volume = {90},
  issue = {9},
  pages = {091301},
  numpages = {4},
  year = {2003},
  month = {Mar},
  publisher = {American Physical Society},
  doi = {10.1103/PhysRevLett.90.091301},
  url = {https://link.aps.org/doi/10.1103/PhysRevLett.90.091301}
}

@article{chevallier:hal-00142125,
  TITLE = {{Accelerating Universes with Scaling Dark Matter}},
  AUTHOR = {Chevallier, M. and Polarski, D.},
  URL = {https://hal.science/hal-00142125},
  NOTE = {(uses Latex, 12 pages, 6 Figures) Minor corrections, Figures 4, 6 revised. Conclusions unchanged},
  JOURNAL = {{International Journal of Modern Physics D}},
  PUBLISHER = {{World Scientific Publishing}},
  VOLUME = {10},
  PAGES = {213-224},
  YEAR = {2001},
  HAL_ID = {hal-00142125},
  HAL_VERSION = {v1},
}

@article{PhysRevD.65.103512,
  title = {Future supernovae observations as a probe of dark energy},
  author = {Weller, Jochen and Albrecht, Andreas},
  journal = {Phys. Rev. D},
  volume = {65},
  issue = {10},
  pages = {103512},
  numpages = {21},
  year = {2002},
  month = {May},
  publisher = {American Physical Society},
  doi = {10.1103/PhysRevD.65.103512},
  url = {https://link.aps.org/doi/10.1103/PhysRevD.65.103512}
}

@article{Malekjani:2024bgi,
    author = "Malekjani, Mohammad and Davari, Zahra and Pourojaghi, Saeed",
    collaboration = "DESI",
    title = "{Cosmological constraints on dark energy parametrizations after DESI 2024: Persistent deviation from standard {\ensuremath{\Lambda}}CDM cosmology}",
    eprint = "2407.09767",
    archivePrefix = "arXiv",
    primaryClass = "astro-ph.CO",
    doi = "10.1103/PhysRevD.111.083547",
    journal = "Phys. Rev. D",
    volume = "111",
    number = "8",
    pages = "083547",
    year = "2025"
}

@article{Speagle_2020,
   title={dynesty: a dynamic nested sampling package for estimating Bayesian posteriors and evidences},
   volume={493},
   ISSN={1365-2966},
   url={http://dx.doi.org/10.1093/mnras/staa278},
   DOI={10.1093/mnras/staa278},
   number={3},
   journal={Monthly Notices of the Royal Astronomical Society},
   publisher={Oxford University Press (OUP)},
   author={Speagle, Joshua S},
   year={2020},
   month=feb, pages={3132–3158} }

@article{DESI:2024mwx,
    author = "Adame, A. G. and others",
    collaboration = "DESI",
    title = "{DESI 2024 VI: cosmological constraints from the measurements of baryon acoustic oscillations}",
    eprint = "2404.03002",
    archivePrefix = "arXiv",
    primaryClass = "astro-ph.CO",
    reportNumber = "FERMILAB-PUB-24-0154-PPD",
    doi = "10.1088/1475-7516/2025/02/021",
    journal = "JCAP",
    volume = "02",
    pages = "021",
    year = "2025"
}

@article{Bekenstein1994,
  title = {Entropy bounds and black hole remnants},
  author = {Bekenstein, Jacob D.},
  journal = {Phys. Rev. D},
  volume = {49},
  issue = {4},
  pages = {1912--1921},
  numpages = {0},
  year = {1994},
  month = {Feb},
  publisher = {American Physical Society},
  doi = {10.1103/PhysRevD.49.1912},
  url = {https://link.aps.org/doi/10.1103/PhysRevD.49.1912}
}

@article{HSU200413,
title = {Entropy bounds and dark energy},
journal = {Physics Letters B},
volume = {594},
number = {1},
pages = {13-16},
year = {2004},
issn = {0370-2693},
doi = {https://doi.org/10.1016/j.physletb.2004.05.020},
url = {https://www.sciencedirect.com/science/article/pii/S0370269304007658},
author = {Stephen D.H. Hsu}
}

@article{miao,
title = {A model of holographic dark energy},
journal = {Physics Letters B},
volume = {603},
number = {1},
pages = {1-5},
year = {2004},
issn = {0370-2693},
doi = {https://doi.org/10.1016/j.physletb.2004.10.014},
url = {https://www.sciencedirect.com/science/article/pii/S0370269304014388},
author = {Miao Li}
}

@article{Saridakis:2020cqq,
    author = "Saridakis, Emmanuel N. and Basilakos, Spyros",
    title = "{The generalized second law of thermodynamics with Barrow entropy}",
    eprint = "2005.08258",
    archivePrefix = "arXiv",
    primaryClass = "gr-qc",
    doi = "10.1140/epjc/s10052-021-09431-y",
    journal = "Eur. Phys. J. C",
    volume = "81",
    number = "7",
    pages = "644",
    year = "2021"
}

@article{Barrow,
    author = "Barrow, John D.",
    title = "{The Area of a Rough Black Hole}",
    eprint = "2004.09444",
    archivePrefix = "arXiv",
    primaryClass = "gr-qc",
    doi = "10.1016/j.physletb.2020.135643",
    journal = "Phys. Lett. B",
    volume = "808",
    pages = "135643",
    year = "2020"
}

@misc{Basilakos:2023seo,
    author = "Basilakos, Spyros and Lymperis, Andreas and Petronikolou, Maria and Saridakis, Emmanuel N.",
    title = "{Barrow holographic dark energy with varying exponent}",
    eprint = "2312.15767",
    archivePrefix = "arXiv",
    primaryClass = "gr-qc",
    month = "12",
    year = "2023"
}

@article{Saridakis:2020zol,
    author = "Saridakis, Emmanuel N.",
    title = "{Barrow holographic dark energy}",
    eprint = "2005.04115",
    archivePrefix = "arXiv",
    primaryClass = "gr-qc",
    doi = "10.1103/PhysRevD.102.123525",
    journal = "Phys. Rev. D",
    volume = "102",
    number = "12",
    pages = "123525",
    year = "2020"
}

@misc{li2024revisitingholographicdarkenergy,
      title={Revisiting holographic dark energy after DESI 2024}, 
      author={Tian-Nuo Li and Yun-He Li and Guo-Hong Du and Peng-Ju Wu and Lu Feng and Jing-Fei Zhang and Xin Zhang},
      year={2024},
      eprint={2411.08639},
      archivePrefix={arXiv},
      primaryClass={astro-ph.CO},
      url={https://arxiv.org/abs/2411.08639}, 
}

@article{Drepanou_2022,
title = {Cosmology of Tsallis and Kaniadakis holographic dark energy in Saez–Ballester theory and consideration of viscous van der Waals fluid},
journal = {Annals of Physics},
volume = {463},
pages = {169611},
year = {2024},
issn = {0003-4916},
doi = {https://doi.org/10.1016/j.aop.2024.169611},
url = {https://www.sciencedirect.com/science/article/pii/S0003491624000198},
author = {Khandro K. Chokyi and Surajit Chattopadhyay}
}

@misc{li2024comprehensivenumericalstudycategories,
      title={A comprehensive numerical study on four categories of holographic dark energy models}, 
      author={Jun-Xian Li and Shuang Wang},
      year={2024},
      eprint={2412.09064},
      archivePrefix={arXiv},
      primaryClass={astro-ph.CO},
      url={https://arxiv.org/abs/2412.09064}, 
}

@article{UNION3,
       author = {{Rubin}, David and {Aldering}, Greg and {Betoule}, Marc and {Fruchter}, Andy and {Huang}, Xiaosheng and {Kim}, Alex G. and {Lidman}, Chris and {Linder}, Eric and {Perlmutter}, Saul and {Ruiz-Lapuente}, Pilar and {Suzuki}, Nao},
        title = "{Union Through UNITY: Cosmology with 2,000 SNe Using a Unified Bayesian Framework}",
      journal = {arXiv e-prints},
         year = 2023,
        month = nov,
          eid = {arXiv:2311.12098},
        pages = {arXiv:2311.12098},
          doi = {10.48550/arXiv.2311.12098},
archivePrefix = {arXiv},
       eprint = {2311.12098},
 primaryClass = {astro-ph.CO},
       adsurl = {https://ui.adsabs.harvard.edu/abs/2023arXiv231112098R}
}

@article{DiValentino:2020vvd,
    author = "Di Valentino, Eleonora and others",
    title = "{Cosmology Intertwined III: $f \sigma_8$ and $S_8$}",
    eprint = "2008.11285",
    archivePrefix = "arXiv",
    primaryClass = "astro-ph.CO",
    reportNumber = "FERMILAB-PUB-20-495-AE",
    doi = "10.1016/j.astropartphys.2021.102604",
    journal = "Astropart. Phys.",
    volume = "131",
    pages = "102604",
    year = "2021"
}

@article{Planck:2018vyg,
    author = "Aghanim, N. and others",
    collaboration = "Planck",
    title = "{Planck 2018 results. VI. Cosmological parameters}",
    eprint = "1807.06209",
    archivePrefix = "arXiv",
    primaryClass = "astro-ph.CO",
    doi = "10.1051/0004-6361/201833910",
    journal = "Astron. Astrophys.",
    volume = "641",
    pages = "A6",
    year = "2020",
    note = "[Erratum: Astron.Astrophys. 652, C4 (2021)]"
}

@article{Riess:2021jrx,
    author = "Riess, Adam G. and others",
    title = "{A Comprehensive Measurement of the Local Value of the Hubble Constant with 1 km/s/Mpc Uncertainty from the Hubble Space Telescope and the SH0ES Team}",
    eprint = "2112.04510",
    archivePrefix = "arXiv",
    primaryClass = "astro-ph.CO",
    doi = "10.3847/2041-8213/ac5c5b",
    journal = "Astrophys. J. Lett.",
    volume = "934",
    number = "1",
    pages = "L7",
    year = "2022"
}

@article{Vagnozzi:2019ezj,
    author = "Vagnozzi, Sunny",
    title = "{New physics in light of the $H_0$ tension: An alternative view}",
    eprint = "1907.07569",
    archivePrefix = "arXiv",
    primaryClass = "astro-ph.CO",
    doi = "10.1103/PhysRevD.102.023518",
    journal = "Phys. Rev. D",
    volume = "102",
    number = "2",
    pages = "023518",
    year = "2020"
}

@article{Vagnozzi:2023nrq,
    author = "Vagnozzi, Sunny",
    title = "{Seven hints that early-time new physics alone is not sufficient to solve the Hubble tension}",
    eprint = "2308.16628",
    archivePrefix = "arXiv",
    primaryClass = "astro-ph.CO",
    doi = "10.3390/universe9090393",
    journal = "Universe",
    volume = "9",
    pages = "393",
    year = "2023"
}

@article{Nunes:2021ipq,
    author = "Nunes, Rafael C. and Vagnozzi, Sunny",
    title = "{Arbitrating the S8 discrepancy with growth rate measurements from redshift-space distortions}",
    eprint = "2106.01208",
    archivePrefix = "arXiv",
    primaryClass = "astro-ph.CO",
    doi = "10.1093/mnras/stab1613",
    journal = "Mon. Not. Roy. Astron. Soc.",
    volume = "505",
    number = "4",
    pages = "5427--5437",
    year = "2021"
}

@article{Abdalla:2022yfr,
    author = "Abdalla, Elcio and others",
    title = "{Cosmology intertwined: A review of the particle physics, astrophysics, and cosmology associated with the cosmological tensions and anomalies}",
    eprint = "2203.06142",
    archivePrefix = "arXiv",
    primaryClass = "astro-ph.CO",
    reportNumber = "FERMILAB-CONF-22-192-SCD",
    doi = "10.1016/j.jheap.2022.04.002",
    journal = "JHEAp",
    volume = "34",
    pages = "49--211",
    year = "2022"
}

@article{DiValentino:2021izs,
    author = "Di Valentino, Eleonora and Mena, Olga and Pan, Supriya and Visinelli, Luca and Yang, Weiqiang and Melchiorri, Alessandro and Mota, David F. and Riess, Adam G. and Silk, Joseph",
    title = "{In the realm of the Hubble tension\textemdash{}a review of solutions}",
    eprint = "2103.01183",
    archivePrefix = "arXiv",
    primaryClass = "astro-ph.CO",
    reportNumber = "IPPP/20/108",
    doi = "10.1088/1361-6382/ac086d",
    journal = "Class. Quant. Grav.",
    volume = "38",
    number = "15",
    pages = "153001",
    year = "2021"
}

@article{DiValentino:2022fjm,
    author = "Di Valentino, Eleonora",
    title = "{Challenges of the Standard Cosmological Model}",
    doi = "10.3390/universe8080399",
    journal = "Universe",
    volume = "8",
    number = "8",
    pages = "399",
    year = "2022"
}

@article{SupernovaSearchTeam:1998fmf,
    author = "Riess, Adam G. and others",
    collaboration = "Supernova Search Team",
    title = "{Observational evidence from supernovae for an accelerating universe and a cosmological constant}",
    eprint = "astro-ph/9805201",
    archivePrefix = "arXiv",
    doi = "10.1086/300499",
    journal = "Astron. J.",
    volume = "116",
    pages = "1009--1038",
    year = "1998"
}

@article{SupernovaCosmologyProject:1998vns,
    author = "Perlmutter, S. and others",
    collaboration = "Supernova Cosmology Project",
    title = "{Measurements of $\Omega$ and $\Lambda$ from 42 high redshift supernovae}",
    eprint = "astro-ph/9812133",
    archivePrefix = "arXiv",
    reportNumber = "LBNL-41801, LBL-41801",
    doi = "10.1086/307221",
    journal = "Astrophys. J.",
    volume = "517",
    pages = "565--586",
    year = "1999"
}

@article{Carroll:2000fy,
    author = "Carroll, Sean M.",
    title = "{The Cosmological constant}",
    eprint = "astro-ph/0004075",
    archivePrefix = "arXiv",
    reportNumber = "EFI-2000-13",
    doi = "10.12942/lrr-2001-1",
    journal = "Living Rev. Rel.",
    volume = "4",
    pages = "1",
    year = "2001"
}

@article{Perivolaropoulos:2021jda,
    author = "Perivolaropoulos, Leandros and Skara, Foteini",
    title = "{Challenges for \ensuremath{\Lambda}CDM: An update}",
    eprint = "2105.05208",
    archivePrefix = "arXiv",
    primaryClass = "astro-ph.CO",
    doi = "10.1016/j.newar.2022.101659",
    journal = "New Astron. Rev.",
    volume = "95",
    pages = "101659",
    year = "2022"
}

@article{Frieman:2008sn,
    author = "Frieman, Joshua and Turner, Michael and Huterer, Dragan",
    title = "{Dark Energy and the Accelerating Universe}",
    eprint = "0803.0982",
    archivePrefix = "arXiv",
    primaryClass = "astro-ph",
    reportNumber = "FERMILAB-PUB-08-613-A",
    doi = "10.1146/annurev.astro.46.060407.145243",
    journal = "Ann. Rev. Astron. Astrophys.",
    volume = "46",
    pages = "385--432",
    year = "2008"
}

@article{Nojiri:2017ncd,
    author = "Nojiri, S. and Odintsov, S. D. and Oikonomou, V. K.",
    title = "{Modified Gravity Theories on a Nutshell: Inflation, Bounce and Late-time Evolution}",
    eprint = "1705.11098",
    archivePrefix = "arXiv",
    primaryClass = "gr-qc",
    reportNumber = "PHYS.REPT.-692-(2017)-1-104, Phys.Rept. 692 (2017) 1-104",
    doi = "10.1016/j.physrep.2017.06.001",
    journal = "Phys. Rept.",
    volume = "692",
    pages = "1--104",
    year = "2017"
}

@article{Peebles:2002gy,
    author = "Peebles, P. J. E. and Ratra, Bharat",
    editor = "Hsu, Jong-Ping and Fine, D.",
    title = "{The Cosmological Constant and Dark Energy}",
    eprint = "astro-ph/0207347",
    archivePrefix = "arXiv",
    reportNumber = "KSUPT-02-3",
    doi = "10.1103/RevModPhys.75.559",
    journal = "Rev. Mod. Phys.",
    volume = "75",
    pages = "559--606",
    year = "2003"
}

@article{DES:2024jxu,
    author = "Abbott, T. M. C. and others",
    collaboration = "DES",
    title = "{The Dark Energy Survey: Cosmology Results with \ensuremath{\sim}1500 New High-redshift Type Ia Supernovae Using the Full 5 yr Data Set}",
    eprint = "2401.02929",
    archivePrefix = "arXiv",
    primaryClass = "astro-ph.CO",
    reportNumber = "FERMILAB-PUB-23-0821-PPD, DES-2023-805",
    doi = "10.3847/2041-8213/ad6f9f",
    journal = "Astrophys. J. Lett.",
    volume = "973",
    number = "1",
    pages = "L14",
    year = "2024"
}

@article{DESI:2025zgx,
    author = "Abdul Karim, M. and others",
    collaboration = "DESI",
    title = "{DESI DR2 results. II. Measurements of baryon acoustic oscillations and cosmological constraints}",
    eprint = "2503.14738",
    archivePrefix = "arXiv",
    primaryClass = "astro-ph.CO",
    reportNumber = "FERMILAB-PUB-25-0169-PPD",
    doi = "10.1103/tr6y-kpc6",
    journal = "Phys. Rev. D",
    volume = "112",
    number = "8",
    pages = "083515",
    year = "2025"
}

@article{DESI:2024aqx,
 archiveprefix = {arXiv},
 author = {Calderon, R. and others},
 collaboration = {DESI},
 doi = {10.1088/1475-7516/2024/10/048},
 eprint = {2405.04216},
 journal = {JCAP},
 pages = {048},
 primaryclass = {astro-ph.CO},
 title = {{DESI 2024: reconstructing dark energy using crossing statistics with DESI DR1 BAO data}},
 volume = {10},
 year = {2024}
}

@misc{DESI:2024jis,
    author = "Adame, A. G. and others",
    collaboration = "DESI",
    title = "{DESI 2024 V: Full-Shape Galaxy Clustering from Galaxies and Quasars}",
    eprint = "2411.12021",
    archivePrefix = "arXiv",
    primaryClass = "astro-ph.CO",
    reportNumber = "FERMILAB-PUB-24-0847-PPD",
    month = "11",
    year = "2024"
}

@misc{DESI:2025fii,
    author = "Lodha, K. and others",
    collaboration = "DESI",
    title = "{Extended Dark Energy analysis using DESI DR2 BAO measurements}",
    eprint = "2503.14743",
    archivePrefix = "arXiv",
    primaryClass = "astro-ph.CO",
    reportNumber = "FERMILAB-PUB-25-0164-PPD",
    month = "3",
    year = "2025"
}

@misc{DES:2025bxy,
    author = "Abbott, T. M. C. and others",
    collaboration = "DES",
    title = "{Dark Energy Survey: implications for cosmological expansion models from the final DES Baryon Acoustic Oscillation and Supernova data}",
    eprint = "2503.06712",
    archivePrefix = "arXiv",
    primaryClass = "astro-ph.CO",
    reportNumber = "DES-2024-0849, FERMILAB-PUB-25-0127-PPD",
    month = "3",
    year = "2025"
}

@article{BOSS:2014hhw,
    author = "Aubourg, \'Eric and others",
    collaboration = "BOSS",
    title = "{Cosmological implications of baryon acoustic oscillation measurements}",
    eprint = "1411.1074",
    archivePrefix = "arXiv",
    primaryClass = "astro-ph.CO",
    doi = "10.1103/PhysRevD.92.123516",
    journal = "Phys. Rev. D",
    volume = "92",
    number = "12",
    pages = "123516",
    year = "2015"
}

@article{Liddle:2007fy,
    author = "Liddle, Andrew R",
    title = "{Information criteria for astrophysical model selection}",
    eprint = "astro-ph/0701113",
    archivePrefix = "arXiv",
    doi = "10.1111/j.1745-3933.2007.00306.x",
    journal = "Mon. Not. Roy. Astron. Soc.",
    volume = "377",
    pages = "L74--L78",
    year = "2007"
}

@article{Cooke:2017cwo,
    author = "Cooke, Ryan J. and Pettini, Max and Steidel, Charles C.",
    title = "{One Percent Determination of the Primordial Deuterium Abundance}",
    eprint = "1710.11129",
    archivePrefix = "arXiv",
    primaryClass = "astro-ph.CO",
    doi = "10.3847/1538-4357/aaab53",
    journal = "Astrophys. J.",
    volume = "855",
    number = "2",
    pages = "102",
    year = "2018"
}

@article{Adil:2024miw,
    author = "Adil, Shahnawaz A. and Dainotti, Maria G. and Sen, Anjan A.",
    title = "{Revisiting the concordance \ensuremath{\Lambda}CDM model using Gamma-Ray Bursts together with supernovae Ia and Planck data}",
    eprint = "2405.01452",
    archivePrefix = "arXiv",
    primaryClass = "astro-ph.HE",
    doi = "10.1088/1475-7516/2024/08/015",
    journal = "JCAP",
    volume = "08",
    pages = "015",
    year = "2024"
}

@article{Kumar:2024soe,
    author = "Kumar, Dharmendra and Mitra, Ayan and Adil, Shahnawaz A. and Sen, Anjan A.",
    title = "{Exploring alternative cosmologies with the LSST: Simulated forecasts and current observational constraints}",
    eprint = "2406.06757",
    archivePrefix = "arXiv",
    primaryClass = "astro-ph.CO",
    reportNumber = "111.043503",
    doi = "10.1103/PhysRevD.111.043503",
    journal = "Phys. Rev. D",
    volume = "111",
    number = "4",
    pages = "043503",
    year = "2025"
}

@misc{DESI:2025ejh,
    author = "Elbers, W. and others",
    collaboration = "DESI",
    title = "{Constraints on Neutrino Physics from DESI DR2 BAO and DR1 Full Shape}",
    eprint = "2503.14744",
    archivePrefix = "arXiv",
    primaryClass = "astro-ph.CO",
    reportNumber = "FERMILAB-PUB-25-0168-PPD",
    month = "3",
    year = "2025"
}

@article{wilks1938,
  title = {The large-sample distribution of the likelihood ratio for testing composite hypotheses},
  author = {Wilks, Samuel S.},
  journal = {Annals of Mathematical Statistics},
  volume = {9},
  number = {1},
  pages = {60--62},
  year = {1938},
  doi = {10.1214/aoms/1177732360}
}

@article{Trotta:2008qt,
  author = {Trotta, Roberto},
  title = {Bayes in the sky: Bayesian inference and model selection in cosmology},
  journal = {Contemporary Physics},
  volume = {49},
  pages = {71--104},
  year = {2008},
  doi = {10.1080/00107510802066753},
  eprint = {0803.4089},
  archivePrefix = {arXiv},
  primaryClass = {astro-ph}
}


\end{document}